\documentclass[a4paper,11pt]{article}
\pdfoutput=1 % if your are submitting a pdflatex (i.e. if you have
\usepackage{jheppub} % for details on the use of the package, please
\usepackage[T1]{fontenc} % if needed
\usepackage{mathrsfs}
\usepackage{mathtools}
\usepackage{amsfonts}	% for mathematical fonts
\usepackage{bm}			% for bold math
\usepackage{bbm}		% for bold math
\usepackage{subfigure}	% for subfigures
\usepackage{tensor}		% for tensors
\usepackage{amsthm}		% for theorems, lemmas,...
\usepackage{braket}		% for braket notation
\usepackage{caption} 
\usepackage[normalem]{ulem}

\usepackage[nameinlink,capitalise]{cleveref}
\crefname{figure}{Figure}{Figures}

\newcommand{\D}{\mathrm{d}}				% for derivatives
\usepackage{xcolor}						% for colors
\title{\boldmath Black holes with synchronised Proca hair: \\ linear clouds and fundamental non-linear solutions}

\author{Nuno M. Santos$^1$,}
\affiliation{$^1$Centro de Astrof\'{i}sica e Gravita\c{c}\~{a}o -- CENTRA and \\  Departamento de F\'{i}sica, Instituto Superior T\'{e}cnico -- IST \\ Universidade de Lisboa -- UL, Avenida Rovisco Pais 1, 1049, Lisboa, Portugal}
\author{Carolina L. Benone$^2$,} 
\affiliation{$^2$Campus de Salin\'opolis, Universidade Federal do Par\'a, 68721-000, Salin\'opolis, Par\'a, Brazil}
\author{Lu\'\i s C. B. Crispino$^3$,}
\affiliation{$^3$Faculdade de F\'\i sica, Universidade Federal do Par\'a, 66075-110, Bel\'em, Par\'a, Brazil}
\author{Carlos A. R. Herdeiro$^4$ and Eugen Radu$^4$}
\affiliation{$^4$Centre for Research and Development  in Mathematics and Applications (CIDMA) and \\
Departamento de Matem\'atica da Universidade de Aveiro \\
Campus de Santiago, 3810-183 Aveiro, Portugal}

\emailAdd{nunomoreirasantos@tecnico.ulisboa.pt, benone@ufpa.br, crispino@ufpa.br, herdeiro@ua.pt, eugen.radu@ua.pt}
\abstract{Recent studies have made key progress on the black hole/solitonic solutions of the Einstein-Proca system. 
Firstly, fully non-linear dynamical evolutions of the Kerr black hole superradiant instability, triggered by a Proca field, have shown the formation of a new equilibrium state, a spinning black hole with synchronised Proca hair. Secondly, non-linear evolutions of spinning Proca stars have established that they are dynamically stable, unlike their scalar cousins. Thirdly, separability of the Proca equation on the Kerr background has been achieved. Motivated by these results, in this paper we reconsider Kerr black holes with synchronised Proca hair. The separability of the Proca equation on the Kerr background allows us to examine the stationary Proca clouds in greater detail, in particular their dependence on the different quantum numbers. These stationary clouds occur at a set of existence lines in the Kerr parameter space, from which the black holes with synchronised Proca hair bifurcate. We construct the domain of existence of these black holes, comparing the fundamental states missed in the original study with the first excited states and with the cousin scalar model, giving illustrative examples of Kerr-like and non-Kerr-like BHs. In the vanishing event horizon limit, these hairy black holes connect to the fundamental states of spinning Proca stars, which include the dynamically stable solutions.}

\date{\today}

\begin{document} 
\maketitle
\flushbottom

%N. Santos's comments: \Santos{red}
%
%C. Herdeiro's comments: \ch{orange}
%
%E. Radu's comments: \Radu{cyan}
%
%C. Benone's comments: \Benone{purple}\\
%
%L. Crispino's comments: \Crispino{brown}

%%%%%%%%%%%%%%%%%%%%%%%%%%%%%%%%%%%%%%%%%%%%%%%%%%%%%%%%%%%%%%%%%%%%%
%%%%%%%%%%%%%%%%%%%%%%%%%%%%%%%%%%%%%%%%%%%%%%%%%%%%%%%%%%%%%%%%%%%%%
\section{Introduction}
%%%%%%%%%%%%%%%%%%%%%%%%%%%%%%%%%%%%%%%%%%%%%%%%%%%%%%%%%%%%%%%%%%%%%
%%%%%%%%%%%%%%%%%%%%%%%%%%%%%%%%%%%%%%%%%%%%%%%%%%%%%%%%%%%%%%%%%%%%%

Successfully tested up to the $TeV$ scale, the Standard Model of particle physics turns out to describe but a tiny fraction of the matter-energy density of the Universe. It does not explain the phenomenological evidence for the existence of dark matter and dark energy, which are believed to make up about $95\%$ of the Cosmos~\cite{Aghanim:2018eyx}. Many models have been put forward to explain the dark side of the Universe.  Some, in particular, relate dark matter to hypothetical new, ultralight bosonic particles which are sufficiently weakly coupled to ordinary matter to have remained elusive to past and present experimental searches~\cite{Suarez:2013iw,Hui:2016ltb}.

Bosonic particles have an interesting interaction with Kerr black holes (BHs). They can extract  the BH's rotational energy through a radiation enhancement mechanism known as superradiance~\cite{Brito:2015oca}. 
For Kerr BHs, superradiance  occurs when the phase angular velocity of the boson, $\omega$, fulfills the condition
\begin{equation}
\frac{\omega}{m_j}<\Omega_\text{H}\equiv\frac{a}{2Mr_\text{H}}\ ,
\label{eq:superradiance}
\end{equation}
where $m_j$ is the boson's azimuthal total angular momentum and $\Omega_\text{H}$, $r_\text{H}=M+\sqrt{M^2-a^2}$ are, respectively, the BH's horizon angular velocity and event horizon (Boyer-Lindquist) radial coordinate,  in terms of the BHs's ADM mass $M$ and total angular momentum $J=Ma$. The enhancement is most efficient when the reduced Compton wavelength of the boson, $\lambda_\text{C}=\hbar/(\mu c)$, is comparable to the BH's gravitational radius, $R_\text{G}=GM/c^2$, $i.e.$
\begin{equation}
\alpha\equiv\frac{R_\text{G}}{\lambda_\text{C}}=\frac{GM\mu}{\hbar c}\approx0.15\left(\frac{M}{106M_\odot}\right)\left(\frac{\mu c^2}{10^{-17}~eV}\right)\sim 1 \ ,
\end{equation}
where $\mu$ is the boson's mass; $\alpha$ is the so-called gravitational fine-structure constant. For the known astrophysical BH masses, ranging between $1-10^{10}$ $M_\odot$, this implies that the bosonic particles are ultralight, with a mass range of roughly $10^{-10}-10^{-20}$ $eV$.

The non-vanishing bosonic field mass plays the role of a mirror, trapping the bosons in the vicinity of the BH and creating a recurrent energy/angular momentum enhancement of the bosonic state. At the linear level, $i.e.$ disregarding the bosons' backreaction on the background spacetime, the energy feeding of the particles fuels an exponential growth known as superradiant instability, or `BH bomb'~\cite{Press:1972zz}. At the non-linear level, the exponential superradiant growth stalls when the inequality \eqref{eq:superradiance}  saturates, $i.e.$
\begin{equation}
\frac{\omega}{m_j}=\Omega_\text{H} \ .
\label{eq:superradiance-sat}
\end{equation}
One may say the (phase angular velocity of the) cloud and the (horizon angular velocity of the) BH synchronise. A simple entropic estimate shows that up to about $29\%$ of the BH's energy could be mined, in an astrophysical timescale, by this process. The result is a classical condensate (often dubbed as \textit{cloud} but also as BH `\textit{hair}') which is stationary with respect to the slowed-down BH~\cite{Sanchis-Gual:2015lje,East:2017ovw,Herdeiro:2017phl}:
a Kerr BH with synchronised bosonic hair. These are stationary BH solutions of Einstein's gravity minimally coupled to complex bosons, first discussed in~\cite{Herdeiro:2014goa} for scalar and in~\cite{Herdeiro:2016tmi} for vector bosons. They challenge the \textit{no-hair} hypothesis~\cite{Ruffini:1971bza} (see also \cite{Herdeiro:2015waa,Cardoso:2016ryw}) \textit{even} in General Relativity. According to this hypothesis, BHs that could form dynamically in the presence of astrophysically (potentially) relevant generic matter-energy are fully characterised by global charges associated with Gauss laws, such as $M$ and $J$, and have no other degrees of freedom, broadly referred to as `hair'. 

The domain of existence of BHs with synchronised bosonic hair has two important boundaries. Firstly, for vanishing horizon size it yields the set of spinning bosonic stars, which have long been known  in the scalar case~\cite{Schunck:1996he,Yoshida:1997qf}, but only recently constructed in the vector case, \textit{a.k.a.} Proca stars~\cite{Brito:2015pxa}. Very recently, it has been shown that the spinning scalar stars suffer from a non-axisymmetric instability, whereas the spinning Proca stars are dynamically robust~\cite{Sanchis-Gual:2019ljs}. This suggests that the Proca case may be dynamically more interesting. Secondly, for vanishing bosonic field, the hairy BHs bifurcate from the Kerr family at the Kerr solutions that admit linear bound states of the corresponding massive bosonic field. These states exist at the threshold of superradience, $i.e.$ when Eq. \eqref{eq:superradiance-sat} holds, and are commonly known as \textit{stationary clouds}.

Stationary clouds around Kerr BHs were first found in the scalar case and around an extremal ($a=M$) BH~\cite{Hod:2012px}. Remarkably, in this particular case the radial function can be solved  analytically in terms of confluent hypergeometric functions. This analysis has then be extended, typically using numerical methods, to other regimes and other BHs -- see $e.g.$~\cite{Hod:2013zza,Hod:2014baa,Benone:2014ssa,Hod:2015goa,Siahaan:2015xna,Hod:2016lgi,Hod:2016yxg,Huang:2016qnk,Bernard:2016wqo,Sakalli:2016xoa,
Ferreira:2017cta,Richartz:2017qep,Huang:2017whw,Huang:2018qdl,Garcia:2018sjh,Delgado:2019prc,Kunz:2019bhm,Garcia:2019zla}.
Stationary clouds in the Kerr case are analogous to the atomic orbitals in the hydrogen atom~\cite{Baumann:2019eav} -- see also~\cite{Arvanitaki:2010sy}. They are finite on and outside the BH's event horizon, decay exponentially at spatial infinity and can be labeled by four quantum numbers: $n$, the number of nodes of the radial function; $\ell$, the orbital angular momentum; $j$, the total angular momentum; and $m_j$, the projection of the total angular momentum along the BH's axis of rotation. Similar configurations have also been obtained in analogue models of gravity such as the draining bathtub vortex~\cite{Benone:2014nla,Benone:2018xct}. 

Most studies of stationary clouds around rotating BHs have focused on the scalar case, whose equations of motion are separable on the Kerr spacetime. As for the massive vector bosons, it remained unclear for decades whether the Proca equation was separable or not on Kerr and the only study of clouds tackled the problem by solving the corresponding partial differential equations~\cite{Herdeiro:2016tmi} -- see also~\cite{East:2017mrj}. Recently, however, the separability of the Proca equation for a large family of spacetimes that includes the  Kerr BH was established using a proper ansatz [hereafter the Frolov-Krtou\v{s}-Kubiz\v{n}\'{a}k-Santos (FKKS) ansatz]~\cite{Frolov:2018ezx}. This development has allowed more detailed studies of the Proca superradiant instability -- see $e.g. $~\cite{Dolan:2018dqv,Cayuso:2019ieu}. The first goal of this paper is to  make use of this development to determine and characterize the stationary vector clouds around Kerr BHs in terms of $\{n,\ell,j,m_j\}$.

The stationary (scalar or vector) clouds define an existence line on the Kerr parameter space from which the BHs with synchronised hair bifurcate~\cite{Herdeiro:2014goa,Herdeiro:2016tmi}. There is a discrete set of families of BHs with synchronised hair, labelled by the parameters $(n,m_j)$; the parameters $(\ell,j)$ do not have significance when going from the linear to the non-linear theory. In the scalar case, the fundamental family of hairy BHs has \textit{nodeless} scalar field profiles, corresponding to $n=0$ and $m_j=1$~\cite{Herdeiro:2015gia}; nodeful solutions, $i.e.$ with $n\neq 0$, are excited states with higher energy~\cite{Wang:2018xhw}. The same holds for the solitonic limit. In the particular case of spherical, static scalar boson stars ($m_j=0$), it has been shown dynamically that the $n>0$ excited states decay into the fundamental $n=0$ ground state~\cite{Balakrishna:1997ej}.

In the original study of Proca stars~\cite{Brito:2015pxa} it was proved that, for static spherical Proca stars, one of the profile functions of the Proca potential must have at least one node; there are no nodeless solutions. In consistency with this observation, the spinning Proca stars reported in the same paper had one node for the corresponding function. Subsequently, the original study of BHs with synchronised Proca hair constructed BH solutions that also have one node of the same function~\cite{Herdeiro:2016tmi}. It was observed in~\cite{Herdeiro:2017phl} when interpreting the results in~\cite{East:2017ovw}, however, that in the spinning case (but not in the static case) there are nodeless Proca stars, and also hairy BHs with Proca hair, and  these are the true fundamental states. Nonetheless, the latter have not been studied in detail in the literature. The second goal of this paper is, therefore, to report a detailed study of the fundamental solutions of these hairy BHs. In particular, their solitonic limit corresponds precisely to the solutions that have been recently shown to be dynamically robust~\cite{Sanchis-Gual:2019ljs} -- see also~\cite{Herdeiro:2019mbz}.

This paper is organised as follows. In~\autoref{sec2} we consider the linear analysis of the stationary clouds on a fixed Kerr geometry. In  \autoref{sec:VectorBosons} the relativistic quantum-mechanical description of vector bosons is briefly addressed. The notation introduced therein will be useful to label the stationary clouds. \autoref{sec:ProcaEq} reviews the Proca equation on a curved spacetime, introduces the FKKS ansatz for the Proca field and presents the radial and angular equations it yields for the Kerr case, in Boyer-Lindquist coordinates. \autoref{sec:SCsKBHs} then sets the stage for the numerical integration of those equations and covers the results. ~\autoref{sec3} deals with the non-linear analysis. After briefly describing the setup in \autoref{sec31}, the domain of existence of the fundamental BHs with Proca hair is discussed in \autoref{sec32} and compared with that of the first excited states of hairy BHs and the cousin scalar model. In \autoref{sec33} we analyse illustrative solutions of both hairy BHs and spinning Proca stars.
Finally, a concise overview of the work is sketched in \autoref{sec:Conclusion}, together with some closing remarks on future prospects. \autoref{ap1} provides some illustrations of the vector spherical harmonics.

Natural units ($G=c=1$) are consistently used throughout the text. Additionally, the metric signature $(-,+,+,+)$ is adopted.

%%%%%%%%%%%%%%%%%%%%%%%%%%%%%%%%%%%%%%%%%%%%%%%%%%%%%%%%%%%%%%%%%%%%%
%%%%%%%%%%%%%%%%%%%%%%%%%%%%%%%%%%%%%%%%%%%%%%%%%%%%%%%%%%%%%%%%%%%%%
\section{Linear analysis: stationary clouds on a fixed Kerr geometry\label{sec2}}
%%%%%%%%%%%%%%%%%%%%%%%%%%%%%%%%%%%%%%%%%%%%%%%%%%%%%%%%%%%%%%%%%%%%%
%%%%%%%%%%%%%%%%%%%%%%%%%%%%%%%%%%%%%%%%%%%%%%%%%%%%%%%%%%%%%%%%%%%%%

%%%%%%%%%%%%%%%%%%%%%%%%%%%%%%%%%%%%%%%%%%%%%%%%%%%%%%%%%%%%%%%%%%%%%
\subsection{Vector bosons\label{sec:VectorBosons}}
%%%%%%%%%%%%%%%%%%%%%%%%%%%%%%%%%%%%%%%%%%%%%%%%%%%%%%%%%%%%%%%%%%%%%

In relativistic quantum mechanics, particles are  described by the orbital angular momentum $\hat{\boldsymbol{L}}$ and the intrinsic angular momentum $\hat{\boldsymbol{S}}$. The components of the individual operators satisfy the angular momentum commutation relations, i.e.
\begin{equation}
[\hat{L}_a,\hat{L}_b]=i\hbar\epsilon_{abc}\hat{L}_c\ ,\qquad 
[\hat{S}_a,\hat{S}_b]=i\hbar\epsilon_{abc}\hat{S}_c \ , \qquad [\hat{\boldsymbol{L}},\hat{\boldsymbol{S}}]=0 \ ,
\nonumber
\end{equation}
where $a,b,c=1,2,3$. The eigenstates of the operators $\hat{\boldsymbol{L}}^2$ (and $\hat{L}_z$) and $\hat{\boldsymbol{S}}^2$ (and $\hat{S}_z$), respectively denoted as $\ket{\ell,m_\ell}$ and $\ket{s,m_s}$, where $|m_\ell|\leqslant\ell$ and $|m_s|\leqslant s$, satisfy
\begin{align*}
&\hat{\boldsymbol{L}}^2\ket{\ell,m_\ell}=\hbar^2\ell(\ell+1)\ket{\ell,m_\ell} \ , \qquad ~\hat{L}_z\ket{\ell,m_\ell}=\hbar m_\ell\ket{\ell,m_\ell} \ ,\\
&\hat{\boldsymbol{S}}^2\ket{s,m_s}=\hbar^2s(s+1)\ket{s,m_s} \ , \qquad
\hat{S}_z\ket{s,m_s}=\hbar m_s\ket{s,m_s} \ .
\end{align*}

The total angular momentum $\hat{\boldsymbol{J}}$ is the sum of the orbital and intrinsic angular momenta, $i.e.$ $\hat{\boldsymbol{J}}=\hat{\boldsymbol{L}}+\hat{\boldsymbol{S}}$. Thus, according to the angular momentum addition theorem, the eigenstates of the operator $\hat{\boldsymbol{J}}^2$, here denoted by $\ket{\ell,s,j,m_j}$, can be expressed in terms of the eigenstates $\ket{\ell,m_\ell}$ and $\ket{s,m_s}$ as \cite{Griffiths:1995}
\begin{align*}
\ket{\ell,s,j,m_j}=\sum_{m_\ell+m_s=m_j}C_{m_\ell m_s m_j}^{\ell s j}\ket{\ell,m_\ell}\otimes\ket{s,m_s} \ ,
\end{align*}
where the coefficients $C_{m_\ell m_s m_j}^{\ell s j}$, with $j= | \ell-s|,  \dots  ,\ell+s-1, \ell+s$ and $|m_j|\leqslant j$, are the Clebsch-Gordan coefficients. These eigenstates satisfy
\begin{align*}
\hat{\boldsymbol{L}}^2\ket{\ell,s,j,m_j}&=\hbar^2\ell(\ell+1)\ket{\ell,s,j,m_j}\ , \qquad 
\hat{\boldsymbol{S}}^2\ket{\ell,s,j,m_j}=\hbar^2s(s+1)\ket{\ell,s,j,m_j}\ ,\\
\hat{\boldsymbol{J}}^2\ket{\ell,s,j,m_j}&=\hbar^2j(j+1)\ket{\ell,s,j,m_j} \ , \qquad 
\hat{J}_z\ket{\ell,s,j,m_j}=\hbar m_j\ket{\ell,s,j,m_j} \ ,
\end{align*}
and therefore $\{\ell,s,j,m_j\}$ are legitimate quantum numbers.

Vector bosons are characterized by $s=1$, which means that the quantum number $j$ can take a single value  when $\ell=0$ ($j=1$) and three different values when $\ell>0$ ($j=\ell-1,\ell,\ell+1$). In this case, the eigenstates of the operator $\hat{\boldsymbol{J}}^2$ in the spherical coordinate representation are the (`pure-orbital') vector spherical harmonics $\boldsymbol{Y}^\ell_{j,m_j}$, which can be expressed in terms of the scalar spherical harmonics $Y_{j,m_j}$ as
\begin{align}
\boldsymbol{Y}^{j-1}_{j,m_j}&=\frac{1}{\sqrt{j(2j+1)}}\left[r\boldsymbol{\nabla}+j\hat{\boldsymbol{e}}_{(r)}\right]Y_{j,m_j}\ ,\\
\boldsymbol{Y}^{j}_{j,m_j}&=-\frac{i}{\sqrt{j(j+1)}}[\boldsymbol{r}\times\boldsymbol{\nabla}]Y_{j,m_j}\ ,\\
\boldsymbol{Y}^{j+1}_{j,m_j}&=\frac{1}{\sqrt{(j+1)(2j+1)}}\left[r\boldsymbol{\nabla}-(j+1)\hat{\boldsymbol{e}}_{(r)}\right]Y_{j,m_j} \ ,
\label{eq:Vector-SHarm}
\end{align}
where
\begin{align*}
&\hat{\boldsymbol{e}}_{(r)}=\boldsymbol{\partial}_r \ ,
\quad\quad\hat{\boldsymbol{e}}_{(\theta)}=\frac{1}{r}\boldsymbol{\partial}_\theta \ ,
\quad\quad\hat{\boldsymbol{e}}_{(\varphi)}=\frac{1}{r\sin\theta}\boldsymbol{\partial}_\varphi\ ,\\
&\boldsymbol{\nabla}=\hat{\boldsymbol{e}}_{(r)}\partial_r+\hat{\boldsymbol{e}}_{(\theta)}\frac{1}{r}\partial_\theta+\hat{\boldsymbol{e}}_{(\varphi)}\frac{1}{r\sin\theta}\partial_\varphi \ .
\end{align*}
The vector spherical harmonics have parity $\hat{\mathbf{\Pi}}=(-1)^{\ell+1}$. Thus, upon a parity transformation, $\boldsymbol{Y}^\ell_{j,m_j}$ acquires a factor of $(-1)^j$, when $j=\ell\pm 1$, and of $(-1)^{j+1}$, when $j=\ell$. $\boldsymbol{Y}^{j\pm 1}_{j,m_j}$ ($\boldsymbol{Y}^j_{j,m_j}$) are said to have electric-type (magnetic-type) parity: they correspond to the magnetic (electric) field of electric multipole radiation and the electric (magnetic) field of magnetic multipole radiation \cite{Thorne:1980,Maggiore:2007}. The explicit form and a graphic representation of the first few `pure-orbital' vector harmonics are provided in \autoref{ap1}.

In curved spacetimes, $\{\ell,j,m_j\}$ are not in general legitimate quantum numbers, since curvature can break the conservation of angular momentum\footnote{Intrinsic angular momentum is expected to be conserved in curved spacetimes, though, otherwise curvature could induce transitions between particles or fields with different spins.}. However, in Schwarzschild spacetime, the total angular momentum $\hat{\boldsymbol{J}}$ is still conserved. This means that vector bosons only have definite total angular momentum. Such definiteness is broken in Kerr spacetime, in which the total angular momentum is no longer conserved. Nevertheless, choosing $\hat{J}_z$ to be aligned with the symmetry axis of Kerr spacetime at spatial infinity, $m_j$ remains a conserved quantity.  

It is convenient to use the quantum numbers $\{\ell,j,m_j\}$ to identify vector bosons, always bearing in mind that they are only physically meaningful in Minkowski spacetime. In particular, in the following, vector states will be labelled with $\ket{\ell,j,m_j}\equiv\ket{\ell,s=1,j,m_j}$.

%%%%%%%%%%%%%%%%%%%%%%%%%%%%%%%%%%%%%%%%%%%%%%%%%%%%%%%%%%%%%%%%%%%%%
\subsection{Proca equation\label{sec:ProcaEq}}
%%%%%%%%%%%%%%%%%%%%%%%%%%%%%%%%%%%%%%%%%%%%%%%%%%%%%%%%%%%%%%%%%%%%%

The Lagrangian density of a massive complex vector boson $A^\alpha$ reads
\begin{align}
\mathcal{L}_\text{M}=-\frac{1}{4}F_{\alpha\beta}\bar{F}^{\alpha\beta}-\frac{1}{2}\mu^2A_\alpha \bar{A}^\alpha \ ,
\label{pl}
\end{align}
where  $F_{\alpha\beta}=2 A_{[\beta;\alpha]}$. $F_{\alpha\beta}$ is the electromagnetic-field tensor, which is antisymmetric and gauge invariant, $\alpha,\beta=0,1,2,3$ and $\mu$ is the boson's mass. The variation of the action integral $\mathcal{S}=\int_V \D^4x\sqrt{-g}~\mathcal{L}_\text{M}$ with respect to the field $\bar{A}^\alpha$ leads to the Proca field equation~\cite{Proca:1938}:
\begin{align}
\nabla_\beta F^{\alpha\beta}+\mu^2A^\alpha=0 \ .
\label{pe}
\end{align}
Writing the equation in terms of the electromagnetic four-potential $A^\alpha$, 
\begin{align}
\nabla_\beta\nabla^\alpha A^\beta-\Box A^\alpha+\mu^2A^\alpha=0 \ ,
\label{eq:ProcaEq-1}
\end{align}
its four-divergence reads
\begin{align}
\nabla_\alpha\nabla_\beta\nabla^\alpha A^\beta-\nabla_\alpha(\Box A^\alpha)+\mu^2\nabla_\alpha A^\alpha=0 \ .
\label{eq:ProcaEq-4div-1}
\end{align}
Using the identities
\begin{align}
\nabla_{[\alpha}\nabla_{\beta]} A^\beta=\frac{1}{2}\tensor{R}{^\beta_\gamma_\alpha_\beta}A^\gamma=-\frac{1}{2}R_{\alpha\beta}A^\beta \ ,
\label{eq:CovD_commutator}\\
\nabla_{[\alpha}\nabla_{\beta]}(\nabla^\alpha A^\beta)=R_{\alpha\beta}\nabla^{[\alpha}A^{\beta]} \ ,
\end{align}
it follows that, for Ricci-flat spacetimes ($R_{\alpha\beta}=0$), such as the Kerr spacetime, Eq. \eqref{eq:ProcaEq-4div-1} reduces to
\begin{align}
\nabla_\alpha A^\alpha=0 \ .
\label{eq:LorenzCond}
\end{align}
This means that any massive complex vector boson minimally coupled to Einstein's gravity in a Ricci-flat spacetime satisfies the Lorenz condition. Moreover, under these conditions, the Proca equation~\eqref{eq:ProcaEq-1} simplifies to
\begin{align}
(\Box-\mu^2)A^\alpha=0 \ .
\label{eq:ProcaEq}
\end{align}
The dynamics of the divergenceless electromagnetic four-potential $A^\alpha$ is thus encoded in a set of four Klein-Gordon equations, one per component. The non-trivial separability of the Klein-Gordon equation in Kerr spacetime was first unveiled via variable separation by Carter \cite{Carter:1968ks}, shortly after noting the complete integrability of the Hamilton-Jacobi equation for Kerr geodesics \cite{Carter:1968rr}. Carter's seminal work broke down the original second-order partial differential equation (PDE) into two coupled second-order ordinary differential equations (ODEs), and paved the way for a thorough study of Kerr linear perturbations.  

Although the four equations of motion \eqref{eq:ProcaEq} are individually separable for a specific ansatz, the separability does not extend to the Lorenz condition \eqref{eq:LorenzCond}. In fact, the separability of these five second-order PDEs is not trivial and was only achieved recently via the FKKS ansatz~\cite{Frolov:2018ezx} (see also Ref.~\cite{Krtous:2018bvk}), following~\cite{Lunin:2017drx}. This separability has been established for the Kerr-NUT-(A)dS family of spacetimes.  The FKKS ansatz, which embodies the explicit and hidden symmetries of the metric, is
\begin{equation}
A^\alpha=B^{\alpha \beta}\nabla_\beta Z \ ,
\label{eq:Proca-ansatz}
\end{equation}
where $B^{\alpha\beta}$ is the polarisation tensor\footnote{In~\cite{Krtous:2018bvk}, the authors named $B^{\alpha\beta}$ \textit{polarisation tensor} without clarifying how the tensor encodes the different polarisations of massive vector bosons. It is worth pointing out that, as opposed to the polarisation tensors usually found in the literature, $B^{\alpha\beta}$ is not totally symmetric.} and $Z$ is an auxiliary complex scalar function for which a multiplicative separation of variables will hold -- \textit{cf.} Eq.\eqref{eq:Proca-ansatz-Z} below. The polarisation tensor $B^{\alpha\beta}$ is defined in terms of the principal tensor\footnote{The separability of the Hamilton-Jacobi, Klein-Gordon, and Dirac equations in Kerr-NUT-(A)dS spacetimes can be traced back to the existence of the principal tensor. For a review, see~\cite{Frolov:2017kze}.} $h_{\alpha\beta}$ as
\begin{align}
B^{\alpha\gamma}\left(g_{\gamma\beta}+i\frac{h_{\gamma\beta}}{\lambda}\right)=\delta^\alpha_\beta \ ,
\label{eq:PolTensor}
\end{align}
where $\lambda$ plays the role of a separation constant. 

To solve the Proca equation in the Kerr background with the ansatz \eqref{eq:Proca-ansatz} one proceeds as follows. In Boyer-Lindquist coordinates $(t,r,\theta,\varphi)$, the Kerr metric reads
\begin{align}
    \boldsymbol{g}&=-\frac{\Delta}{\Sigma}\left[\boldsymbol{\D} t-a\sin^2\theta\,\boldsymbol{\D}\varphi\right]^2+\frac{\Sigma}{\Delta}\,\boldsymbol{\D} r^2+\Sigma\,\boldsymbol{\D}\theta^2+\frac{\sin^2\theta}{\Sigma}\left[a\,\boldsymbol{\D} t-(r^2+a^2)\,\boldsymbol{\D}\varphi\right]^2\ ,
    \label{eq:Kerr-metric-BL}
\end{align}
where $\Sigma\equiv r^2+a^2\cos^2\theta$ and $\Delta\equiv r^2-2Mr+a^2$. The Kerr spacetime is stationary and axisymmetric; it has an event horizon at $r=r_\text{H}$, the largest root of $\Delta$. In Boyer-Lindquist coordinates, the Killing vectors associated with
these continuous symmetries are $\boldsymbol{\xi}_t=\boldsymbol{\partial}_t$ and $\boldsymbol{\xi}_\varphi=\boldsymbol{\partial}_\varphi$, respectively. Its principal tensor reads
\begin{align}
\boldsymbol{h}
=-(r\boldsymbol{\D}r+a^2\sin\theta&\cos\theta\boldsymbol{\D}\theta)\wedge\boldsymbol{\D}t\nonumber\\
&+a\sin\theta[r\sin\theta\boldsymbol{\D}r+(r^2+a^2)\cos\theta\boldsymbol{\D}\theta]\wedge\boldsymbol{\D}\varphi \ .
\label{eq:principal-tensor}
\end{align}
Using the metric and principal tensor, one constructs the polarization tensor. Its symmetric and antisymmetric parts are
\begin{align*}
&B^{(\alpha\beta)}=\frac{\lambda^2}{\Sigma}\left[\frac{1}{q_r}
\begin{pmatrix}
   -\frac{(r^2+a^2)^2}{\Delta} & 0 & 0 & -\frac{a(r^2+a^2)}{\Delta} \\
   0 & \Delta & 0 & 0 \\
   0 & 0 & 0 & 0 \\
   -\frac{a(r^2+a^2)}{\Delta} & 0 & 0 & -\frac{a^2}{\Delta}
\end{pmatrix}+
\frac{1}{q_\theta}
\begin{pmatrix}
   a^2\sin^2\theta & 0 & 0 & a \\
   0 & 0 & 0 & 0 \\
   0 & 0 & 1 & 0 \\
   a & 0 & 0 & \csc^2\theta
\end{pmatrix}
\right],\\
&B^{[\alpha\beta]}=-\frac{i\lambda}{\Sigma}\left[\frac{r}{q_r}
\begin{pmatrix}
   0 & -(r^2+a^2) & 0 & 0 \\
   (r^2+a^2) & 0 & 0 & a \\
   0 & 0 & 0 & 0 \\
   0 & -a & 0 & 0
\end{pmatrix}+
\frac{a\cos\theta}{q_\theta}
\begin{pmatrix}
   0 & 0 & -a\sin\theta & 0 \\
   0 & 0 & 0 & 0 \\
   a\sin\theta & 0 & 0 & \csc\theta \\
   0 & 0 & -\csc\theta & 0
\end{pmatrix}
\right],
\end{align*} 
respectively, where
\begin{equation}
q_r=r^2+\lambda^2 \ , \qquad q_\theta=\lambda^2-a^2\cos^2\theta \ .
\end{equation}
Note that the $r$-dependent terms are decoupled from the $\theta$-dependent terms, apart from the common factor $1/\Sigma$.

Additionally, we take the complex scalar function in Eq. \eqref{eq:Proca-ansatz} with the form
\begin{align}
Z(t,r,\theta,\varphi)=e^{-i\omega t}R(r)Q(\theta,\varphi)\ ,\qquad Q(\theta,\varphi)=S(\theta)e^{+im_j\varphi} \ ,
\label{eq:Proca-ansatz-Z}
\end{align}
where $R$ and $S$ are dubbed radial and angular functions, respectively, and $\omega$ and $m_j$ are the eigenvalues related to the aforementioned isometries.

With this construction, the ansatz \eqref{eq:Proca-ansatz} reduces the Proca equation to the two separated equations
\begin{align}
q_r\frac{\D}{\D r}\left[\frac{\Delta}{q_r}\frac{\D R}{\D r}\right]+\left[\frac{K_r^2}{\Delta}+\frac{2\lambda^2-q_r}{q_r}\sigma\lambda-q_r\mu^2\right]R=0 \ ,\label{eq:Kerr:RadEq}\\
\frac{q_\theta}{\sin\theta}\frac{\D}{\D\theta}\left[\frac{\sin\theta}{q_\theta}\frac{\D S}{\D \theta}\right]-\left[\frac{K_\theta^2}{\sin^2\theta}+\frac{2\lambda^2-q_\theta}{q_\theta}\sigma\lambda-q_\theta\mu^2\right]S=0 \ ,\label{eq:Kerr:AngEq}
\end{align}
where
\begin{equation}
K_r=(r^2+a^2)\omega-am_j\ ,\qquad K_\theta=m_j-a\omega\sin^2\theta \ ,  \qquad \sigma=\frac{a(m_j-a\omega)}{\lambda^2}+\omega \ .
\end{equation}

As mentioned in \autoref{sec:VectorBosons}, vector states can have electric-type or magnetic-type parity. The ansatz~\eqref{eq:Proca-ansatz} encodes all the electric-type states and the magnetic-type states with $j=|m_j|$ \cite{Dolan:2018dqv,Baumann:2019eav}. It remains unclear, however, whether all the magnetic-type states are captured by the FKKS ansatz or not. If so, further study of the Proca equation in Kerr spacetime may shed some light on how to recover all of them. If not, a new question arises:  whether it is possible to find a new ansatz which contains all the states.

In the following, all physical quantities will be expressed in terms of the boson's reduced Compton wavelength. It is therefore convenient to set $\lambda_\text{C}=\mu^{-1}=1$.

%%%%%%%%%%%%%%%%%%%%%%%%%%%%%%%%%%%%%%%%%%%%%%%%%%%%%%%%%%%%%%%%%%%%%
\subsection{Stationary vector clouds around Kerr black holes\label{sec:SCsKBHs}}
%%%%%%%%%%%%%%%%%%%%%%%%%%%%%%%%%%%%%%%%%%%%%%%%%%%%%%%%%%%%%%%%%%%%%

In general, Eqs. \eqref{eq:Kerr:RadEq} and \eqref{eq:Kerr:AngEq} form a non-standard coupled eigenvalue problem with an eigenvalue pair $\{\omega,\lambda\}$. To construct the stationary vector clouds around Kerr BHs, we need to find the bound states whose phase angular velocity fulfills the synchronization condition \eqref{eq:superradiance-sat}. Since $\omega$ is fixed \textit{a priori}, the eigenvalue pair can be chosen to be either $\{M,\lambda\}$ or $\{a,\lambda\}$. The existence of stationary clouds is only allowed for specific values of the background parameters $M$ and $a$. Such quantization follows from the regularity of the bound states and results in an \textit{existence line} in the two-dimensional Kerr parameter space defined by $(M,a)$ or, alternatively, $(M,\Omega_\text{H})$. 

The next subsections summarise the algorithm to solve the radial equation \eqref{eq:Kerr:RadEq} together with the synchronization condition and therefore determine the existence lines of stationary vector clouds around Kerr BHs. For convenience, Eqs. \eqref{eq:Kerr:RadEq} and \eqref{eq:Kerr:AngEq} will be considered as written in terms of $r_\text{H}$ and $a$ instead of $M$ and $a$. For this purpose, using the identity
\begin{align}
M=\frac{r_\text{H}^2+a^2}{2r_\text{H}} \ ,
\end{align}
the function $\Delta$ may be written as
\begin{align}
\Delta=r^2-(r_\text{H}^2+a^2)\frac{r}{r_\text{H}}+a^2 \ .
\end{align}

\subsubsection{Angular equation}

In the Minkowski limit, the phase angular velocity equals the inverse of the reduced Compton wavelength: $\omega=\mu=1$. Hence, Eq. \eqref{eq:Kerr:AngEq} can be written in the form\footnote{As opposed to the angular equation governing the dynamics of massless scalar bosons in the Kerr geometry, Eq. \eqref{eq:Kerr:AngEq} does not reduce to the spherical harmonic differential equation when $a\omega=0$ ($i.e.$ when $a=0$ or $\omega=0$), but only when $a=0$, $i.e.$ in the Schwarzschild limit. Note, however, that Eq. \eqref{eq:Kerr:AngEq-Minkowski} corresponds to the Minkowski limit ($M=0$, $a=0$) of Eq. \eqref{eq:Kerr:AngEq}, for which the condition $\omega=\mu$ holds.}
\begin{align}
\hat{\boldsymbol{J}}^2Q_0=\lambda^{\text{E}}_0(\lambda^{\text{E}}_0-1) Q_0 \ ,
\label{eq:Kerr:AngEq-Minkowski}
\end{align}
where $Q_0$ denotes the leading-order form of the function $Q$ at spatial infinity and the superscript `E' will become clear in the remainder of the present section. $\hat{\boldsymbol{J}}^2$ coincides with the square of the orbital angular momentum operator,
\begin{align}
\hat{\boldsymbol{J}}^2=-\frac{1}{\sin\theta}\frac{\partial}{\partial\theta}\left(\sin\theta\frac{\partial}{\partial \theta}\right)-\frac{1}{\sin^2\theta}\frac{\partial^2}{\partial\varphi^2} \ ,
\end{align}
whose eigenfunctions are the well-known scalar spherical harmonics of degree $j$ and order $m_j$, $Y_{j,m_j}$. In particular, $\hat{\boldsymbol{J}}^2Y_{j,m_j}(\theta,\varphi)=j(j+1)Y_{j,m_j}(\theta,\varphi)$. Since the Kerr spacetime is asymptotically flat, $j$ may be defined as the total angular momentum at spatial infinity.

The quadratic equation $\lambda^{\text{E}}_0(\lambda^{\text{E}}_0-1)=j(j+1)$ has two different solutions:
\begin{align}
\lambda^{\text{E}}_{0,-}=-j \ ,
\quad
\lambda^{\text{E}}_{0,+}=j+1\ .
\end{align}
This means that, at leading order, the electromagnetic four-potential $A^\alpha=(A^t,\boldsymbol{A})$ takes the form
\begin{align*}
A_{0}^t&=\frac{\lambda^{\text{E}}_0}{r^2+(\lambda^{\text{E}}_0)^2}\left(-\lambda^{\text{E}}_0\partial_t Z_0+ir\partial_r Z_0\right)=\frac{i\lambda^{\text{E}}_0}{r}\partial_rZ_0+\ldots=i\lambda^{\text{E}}_0\frac{e^{-i\omega t}}{r}\partial_rR_0^{[\infty]}Y_{j,m_j}+\ldots \ ,\\
A_{0}^r&=\frac{\lambda^{\text{E}}_0}{r^2+(\lambda^{\text{E}}_0)^2}\left(-ir\partial_t Z_0+\lambda^{\text{E}}_0\partial_r Z_0\right)=-\frac{i\lambda^{\text{E}}_0}{r}\partial_tZ_0+\ldots=-\lambda^{\text{E}}_0\frac{e^{-i\omega t}}{r}R_0^{[\infty]}Y_{j,m_j}+\ldots \ ,\\
A_{0}^\theta&=\frac{1}{r^2}\partial_\theta Z_0=\frac{e^{-i\omega t}}{r^2}R_0^{[\infty]}\partial_\theta Y_{j,m_j} \ ,\\
A_{0}^\varphi&=\frac{1}{r^2\sin^2\theta}\partial_\varphi Z_0=\frac{e^{-i\omega t}}{r^2\sin^2\theta}R_0^{[\infty]}\partial_\varphi Y_{j,m_j} \ ,
\end{align*}
where $R_0^{[\infty]}$ is the leading-order form of the function $R$ at spatial infinity. The spatial part of $A^\alpha$ can be written as

\begin{align}
\boldsymbol{A}_{0}&=
\frac{e^{-i\omega t}}{r}R_0^{[\infty]}\left[-\lambda^{\text{E}}_0Y_{j,m_j}\boldsymbol{\partial}_r+\frac{1}{r}(\partial_\theta Y_{j,m_j})\boldsymbol{\partial}_\theta+\frac{1}{r\sin^2\theta}(\partial_\varphi Y_{j,m_j})\boldsymbol{\partial}_\varphi\right]\nonumber\\
&=\frac{e^{-i\omega t}}{r}R_0^{[\infty]}\left[-\lambda^{\text{E}}_0Y_{j,m_j}\hat{\boldsymbol{e}}_{(r)}+(\partial_\theta Y_{j,m_j})\hat{\boldsymbol{e}}_{(\theta)}+\frac{1}{\sin\theta}(\partial_\varphi Y_{j,m_j})\hat{\boldsymbol{e}}_{(\varphi)}\right]\nonumber\\
&=\frac{e^{-i\omega t}}{r}R_0^{[\infty]}\left[r\boldsymbol{\nabla}-\lambda^{\text{E}}_0\hat{\boldsymbol{e}}_{(r)}\right]Y_{j,m_j}\nonumber\\
&=\sqrt{2j+1}\frac{e^{-i\omega t}}{r}R_0^{[\infty]}\times\left\{
\begin{array}{lr}
        \sqrt{j}~\boldsymbol{Y}^{j-1}_{j,m_j}, & \text{for } \lambda^{\text{E}}_0=\lambda^{\text{E}}_{0,-}\\
        \sqrt{j+1}~\boldsymbol{Y}^{j+1}_{j,m_j}, & \text{for } \lambda^{\text{E}}_0=\lambda^{\text{E}}_{0,+}\\
        \end{array}\right. \ .
\label{eq:MVB-A0}
\end{align}
The angular dependence of $\boldsymbol{A}_{0}$ is described by the `pure-orbital' vector spherical harmonics in flat space. Equation \eqref{eq:MVB-A0} with eigenvalues $\lambda^{\text{E}}_{0,\mp}$, corresponds to the $j=\ell\pm 1$ electric-type states of the vector field (cf. \autoref{sec:VectorBosons}). This explains the superscript `E'.

In the zero-angular-momentum (ZAMO) frame, characterized by the tetrad
\begin{align*}
	\begin{split}
		&\hat{\boldsymbol{e}}_{(t)}^\bullet=\frac{1}{\sqrt{\Xi\Sigma\Delta}}\left[\Xi\boldsymbol{\partial}_t+2Mar\boldsymbol{\partial}_\varphi\right] \ ,\\
		&\hat{\boldsymbol{e}}_{(\theta)}^\bullet=\frac{1}{\sqrt{\Sigma}}\boldsymbol{\partial}_\theta \ ,
	\end{split}
	\quad
	\begin{split}
		&\hat{\boldsymbol{e}}_{(r)}^\bullet=\sqrt{\frac{\Delta}{\Sigma}}\boldsymbol{\partial}_r\  ,\\	
		&\hat{\boldsymbol{e}}_{(\varphi)}^\bullet=\frac{\sqrt{\Sigma}}{\sqrt{\Xi}\sin\theta}\boldsymbol{\partial}_\varphi \ ,
	\end{split}
\end{align*}
where $\Xi\equiv (r^2+a^2)^2-a^2\Delta\sin^2\theta$, the electric field $\boldsymbol{E}$ and the magnetic field $\boldsymbol{B}$ have the following components:
\begin{align*}
	E_{(a)}=F_{\alpha\beta}\hat{e}_{(a)}^{\bullet \alpha} \hat{e}_{(t)}^{\bullet \beta} \ ,
	\qquad
	B_{(a)}=-\frac{1}{2}\epsilon_{\alpha\beta\gamma\delta}F^{\gamma\delta}~\hat{e}_{(a)}^{\bullet \alpha} \hat{e}_{(t)}^{\bullet \beta \ },
\end{align*}
with $\epsilon_{\alpha\beta\gamma\delta}\equiv\sqrt{-g}~[\alpha\beta\gamma\delta]$, where $[\alpha\beta\gamma\delta]$ is the four-dimensional Kronecker delta. The leading-order terms of $E_{(a)}$ and $B_{(a)}$ are given by
\begin{align*}
\begin{split}
E_{0(r)}&=-i\lambda^{\text{E}}_0\frac{e^{-i\omega t}}{r}R_0^{[\infty]}Y_{j,m_j} \ ,\\
E_{0(\theta)}&=i\frac{e^{-i\omega t}}{r}R_0^{[\infty]}\partial_\theta Y_{j,m_j} \ ,\\
E_{0(\varphi)}&=i\frac{e^{-i\omega t}}{r\sin\theta}R_0^{[\infty]}\partial_\varphi Y_{j,m_j} \ ,
\end{split}
\quad
\begin{split}
B_{0(r)}&=0 \ ,\\
B_{0(\theta)}&=-\frac{e^{-i\omega t}}{r\sin\theta}\frac{\D R_0^{[\infty]}}{\D r}\partial_\varphi Y_{j,m_j} \ ,\\
B_{0(\varphi)}&=\frac{e^{-i\omega t}}{r}\frac{\D R_0^{[\infty]}}{\D r}\partial_\theta Y_{j,m_j} \ ,
\end{split}
\end{align*}
or 
\begin{align*}
\boldsymbol{E}_0&=i\frac{e^{-i\omega t}}{r}R_0^{[\infty]}\left[r\boldsymbol{\nabla}-\lambda^{\text{E}}_0\hat{\boldsymbol{e}}_{(r)}\right]Y_{j,m_j}\\
&=i\sqrt{2j+1}\frac{e^{-i\omega t}}{r}R_0^{[\infty]}\times\left\{
\begin{array}{lr}
        \sqrt{j}~\boldsymbol{Y}^{j-1}_{j,m_j}, & \text{for } j-\ell=+1\\
        \sqrt{j+1}~\boldsymbol{Y}^{j+1}_{j,m_j}, & \text{for } j-\ell=-1\\
        \end{array}\right. \ ,\\
\boldsymbol{B}_0
&=\frac{e^{-i\omega t}}{r}\frac{\D R_0^{[\infty]}}{\D r}[\boldsymbol{r}\times\boldsymbol{\nabla}]Y_{j,m_j}=i\sqrt{j(j+1)}\frac{e^{-i\omega t}}{r}\frac{\D R_0^{[\infty]}}{\D r}\boldsymbol{Y}^j_{j,m_j} \ ,
\end{align*}
where we used the fact that $\hat{\boldsymbol{e}}_{(a)}^\bullet\rightarrow \hat{\boldsymbol{e}}_{(a)}$ as $r\rightarrow+\infty$. The electric field $\boldsymbol{E}_0$ depends on the difference $j-\ell$, whereas the magnetic field $\boldsymbol{B}_0$ does not.

When $\alpha\ll 1$, the angular eigenstates and eigenvalues for the electric-type states may be written as an expansion in $\alpha$. The next-to-leading order corrections to the angular eigenfunctions induce couplings to vector spherical harmonics of the same parity and thus the angular functions $Q$ have definite parity. For future reference, we present the expansion for the angular eigenvalues below \cite{Baumann:2019eav}:
\begin{align}
\lambda^{\text{E}}_\pm=\sum_{n=0}^\infty\lambda^{\text{E}}_{n,\pm}\alpha^n \ ,
\label{eq:AngEigen_E}
\end{align}
where the first terms of the series are given by
\begin{align}
\lambda^{\text{E}}_{1,\pm}&=-\frac{m_ja}{M\lambda^{\text{E}}_{0,\pm}} \ ,\\
\lambda^{\text{E}}_{2,\pm}&=-\frac{\lambda^{\text{E}}_{0,\pm}}{2\hat{n}^2(2\lambda^{\text{E}}_{0,\pm}-1)}+\frac{a^2(\lambda^{\text{E}}_{0,\pm}+1)[(\lambda^{\text{E}}_{0,\pm})^2-m_j^2]}{M^2(\lambda^{\text{E}}_{0,\pm})^3[2(\lambda^{\text{E}}_{0,\pm})^2+1]} \ ,\\
\lambda^{\text{E}}_{3,\pm}&=\frac{m_ja}{M}\left[\frac{1}{\hat{n}^2(2\lambda^{\text{E}}_{0,\pm}-1)}+\frac{a^2(\lambda^{\text{E}}_{0,\pm}+2)[(\lambda^{\text{E}}_{0,\pm})^2-m_j^2]}{M^2(\lambda^{\text{E}}_{0,\pm})^5(2\lambda^{\text{E}}_{0,\pm}+1)}\right] \ ,
\end{align}
with $\hat{n}\equiv n+\ell+1$. $\hat{n}\in\mathbb{N}$ is the principal quantum number and $n\in\mathbb{N}_0$ is the node number -- see \autoref{sec:Results}.

Equation \eqref{eq:Kerr:AngEq-Minkowski} allows us to recover the electric-type states of $A^a$ solely. The magnetic-type states with $j=\ell=|m_j|$, the only ones which are known to be captured by the FKKS ansatz, can be recovered considering the limits
\begin{align}
\lambda^{\text{M}}_0\equiv\lim_{\alpha\rightarrow 0} \lambda^{\text{M}}=0 \ , 
\qquad 
\chi\equiv\lim_{\alpha\rightarrow 0}\frac{a}{M}\frac{\alpha}{\lambda^{\text{M}}}= m_j\pm 1 \ , 
\end{align}
where the superscript `M' labels all quantities related to magnetic-type states with $j=|m_j|$. As first shown in~\cite{Baumann:2019eav}, the leading-order form of $\boldsymbol{A}$ is proportional to the vector spherical harmonic $\boldsymbol{Y}^{j}_{j,j}$. Unluckily, no expansion of $\lambda^{\text{M}}$ in powers of $\alpha$ is known. However, when considering marginally-bound states ($\omega^2=\mu^2=1$), the angular eigenvalue yields
\begin{align}
\lim_{\omega^2\rightarrow\mu^2}\lambda^{\text{M}}=\frac{2a}{m_j+1-a\omega+\sqrt{(m_j+1-a\omega)^2+4a\omega}}
\label{eq:AngEigen_M} \ ,
\end{align}
which vanishes in the Schwarzschild limit ($a\rightarrow 0$). 

Both the third-order expansion in $\alpha$ for $\lambda^\text{E}$ in Eq. \eqref{eq:AngEigen_E} and the limiting value of $\lambda^\text{M}$ (with $j=|m_j|$) in Eq. \eqref{eq:AngEigen_M} suffice to perform the numerical integration of the radial equation with great accuracy when $\alpha\ll 1$ -- see~\autoref{sec:Results}.

\subsubsection{Radial equation\label{sec:NumericalMethod}}

The integration of the radial equation is performed via the expansion 
\begin{align*}
\mathbb{R}(r)&=\sum_{n=0}^{N}c_n(r-r_\text{H})^n \ ,
\end{align*} 
for the radial function $R$ in Eq. \eqref{eq:Kerr:RadEq}. $N$ is the number of terms of the partial sum and the coefficients $\{c_n\}_{n=0,\ldots,N}$ are functions of\footnote{In general, the coefficients $\{c_n\}_{n=0,\ldots,N}$ also depend on $\omega$. However, here $\omega$ is fully defined via Eq. \eqref{eq:superradiance-sat}.} $r_\text{H}$, $a$, $\mu$, $m_j$ and
 $\lambda$, which, in turn, depends on $\ell$ and $j$. Plugging the expansion into Eq. \eqref{eq:Kerr:RadEq} and equating coefficients order by order, it is possible to write $\{c_n\}_{n=1,\ldots,N}$ in terms of $c_{0}$ . The latter is usually set to $1$. The choice of $N$ should be a trade-off between computational time and accuracy. Once the coefficients $\{c_n\}_{n=0,\ldots,N}$ are defined, one fixes the numerical values of $r_\text{H}$, $\mu$, $\ell$, $j$ and $m_j$, assigns a guess value to $a$ and computes the corresponding guess value for $\lambda$. 

The radial equation is then integrated from $r=r_\text{H}(1+\delta)$, with $\delta \ll 1$, to $r=r_{\infty}$, where $r_\infty$ stands for the numerical value of infinity. The solution must satisfy the boundary conditions 
\begin{align}
R(r=r_\text{H})=\mathbb{R}(r=r_\text{H}) \ ,
\qquad
R'(r=r_\text{H})=\mathbb{R}'(r=r_\text{H}) \ ,
\end{align}
where the prime denotes differentiation with respect to $r$.

The previous step is repeated for different guess values of $a$, until the solution satisfies the boundary conditions
\begin{align}
R(r=r_\infty)\rightarrow 0 \ ,
\qquad
R'(r=r_\infty)\rightarrow 0 \ .
\end{align}

\subsubsection{Results\label{sec:Results}}

When scanning the parameter space in search of stationary vector clouds with fixed quantum numbers $(\ell,j,m_j)$, solutions with different numbers of nodes $n$ ($n\in\mathbb{N}_0$) are found. Thus, each vector state may be labelled using the notation $\ket{n,\ell,j,m_j}$. Configurations with $n=0$ ($n\in\mathbb{N}$) are dubbed fundamental (excited) states. The greater the node number $n$, the more energetic the state.

The frequency spectra of massive vector \textit{quasi}-bound states, $i.e.$ with a complex frequency, can be written in the form
\begin{align}
\omega^{\text{(V)}}_{\ket{n,\ell,j,m_j}}&=1-\frac{\alpha^2}{2\hat{n}^2}-\frac{\alpha^4}{8\hat{n}^4}+\frac{f_{\text{V}}(n,\ell,j)}{\hat{n}^3}\alpha^4+\frac{h_{\text{V}}(\ell,j)}{\hat{n}^3}\frac{m_j a}{M}\alpha^5+\ldots \ ,
\label{eq:s1-freq-spectrum}
\end{align}
with $j=\ell\pm 1,\ell$ and where\footnote{Note that~\cite{Baumann:2019eav} $\omega^{\text{(V)}}_{\ket{n,j,j,m_j}}=\omega^{\text{(S)}}_{\ket{n,j,m_j}}$, where $\omega^{\text{(S)}}_{\ket{n,j,m_j}}$ denotes the frequency of a massive scalar quasi-bound state with quantum numbers $\{n,j,m_j\}$. This suggests that the magnetic-type vector states are equivalent to the scalar states with the same total angular momentum. If so, it should be possible to show that Eq. \eqref{eq:Kerr:RadEq} for magnetic-type states and its scalar counterpart are equivalent, at least in some limiting case.}
\begin{align*}
&f_\text{V}(n,\ell,j)=-\frac{4(6\ell j+3\ell+3j+2)}{(\ell+j)(\ell+j+1)(\ell+j+2)}+\frac{2}{\hat{n}} \ ,\\
&h_\text{V}(\ell,j)=\frac{16}{(\ell+j)(\ell+j+1)(\ell+j+2)} \ .
\end{align*}
The frequencies and corresponding instability rates were computed analytically in~\cite{Baumann:2019eav} via matched asymptotic expansions, except for the magnetic-type ($j=\ell$) vector states. The expression in Eq. \eqref{eq:s1-freq-spectrum} for $j=\ell$ is a conjecture. Nonetheless, the authors of \cite{Baumann:2019eav} confirmed that the conjectured frequencies do agree with those found numerically without relying on separability of the Proca equation. In fact, the analytic approximation is accurate when $\alpha\lesssim 0.2$, even for near-extremal Kerr BHs. 

Vector instability rates are proportional to the factor $(\omega-m_j\Omega_\text{H})$ and thus vanish whenever the synchronization condition holds. In that case, the contour lines for which 
\begin{align}
\omega^{\text{(V)}}_{\ket{n,\ell,j,m_j}}&=m_j\Omega_\text{H}
\label{eq:synchrono-P}
\end{align}
constitute an analytical approximation to the \textit{existence lines} of stationary vector clouds in the parameter space of Kerr BHs. For future reference, theses curves will be referred to as \textit{analytical existence lines} (AEL), whereas those obtained via the numerical algorithm laid out in \autoref{sec:NumericalMethod} will be named \textit{numerical existence lines} (NEL). Additionally, all existence lines will be presented in a ($M,\Omega_\text{H}$)-plane normalized to the boson's mass $\mu$, in which the domain of existence of Kerr BHs is shaded light green.  

The mass spectrum of Kerr BHs which support stationary vector clouds may be derived by solving Eq. \eqref{eq:synchrono-P} for $\alpha$. This yields 
\begin{align}
\alpha=\hat{n}(2\varpi)^{1/2}\left[1-\frac{1}{4}g_\text{V}\varpi+\frac{7}{32}g_\text{V}^2\varpi^2-\frac{33}{128}g_\text{V}^3\varpi^3+\ldots\right]\ ,
\label{eq:mass-spectrum}
\end{align}
where
\begin{align}
\varpi=1-m_j\Omega_\text{H}\ ,
\qquad
g_\text{V}=1-8\hat{n}f_\text{V} \ .
\end{align}
The first two terms in Eq. \eqref{eq:mass-spectrum} depend on $n$, $\ell$ and $m_j$, but not on $j$. The next-to-leading-order term, which depends on $j$ through $g_\text{V}$, must be taken into account to capture the leading-order behavior of stationary vector clouds. 

\begin{figure}[h]
  \centering
  \includegraphics[width=.6\linewidth]{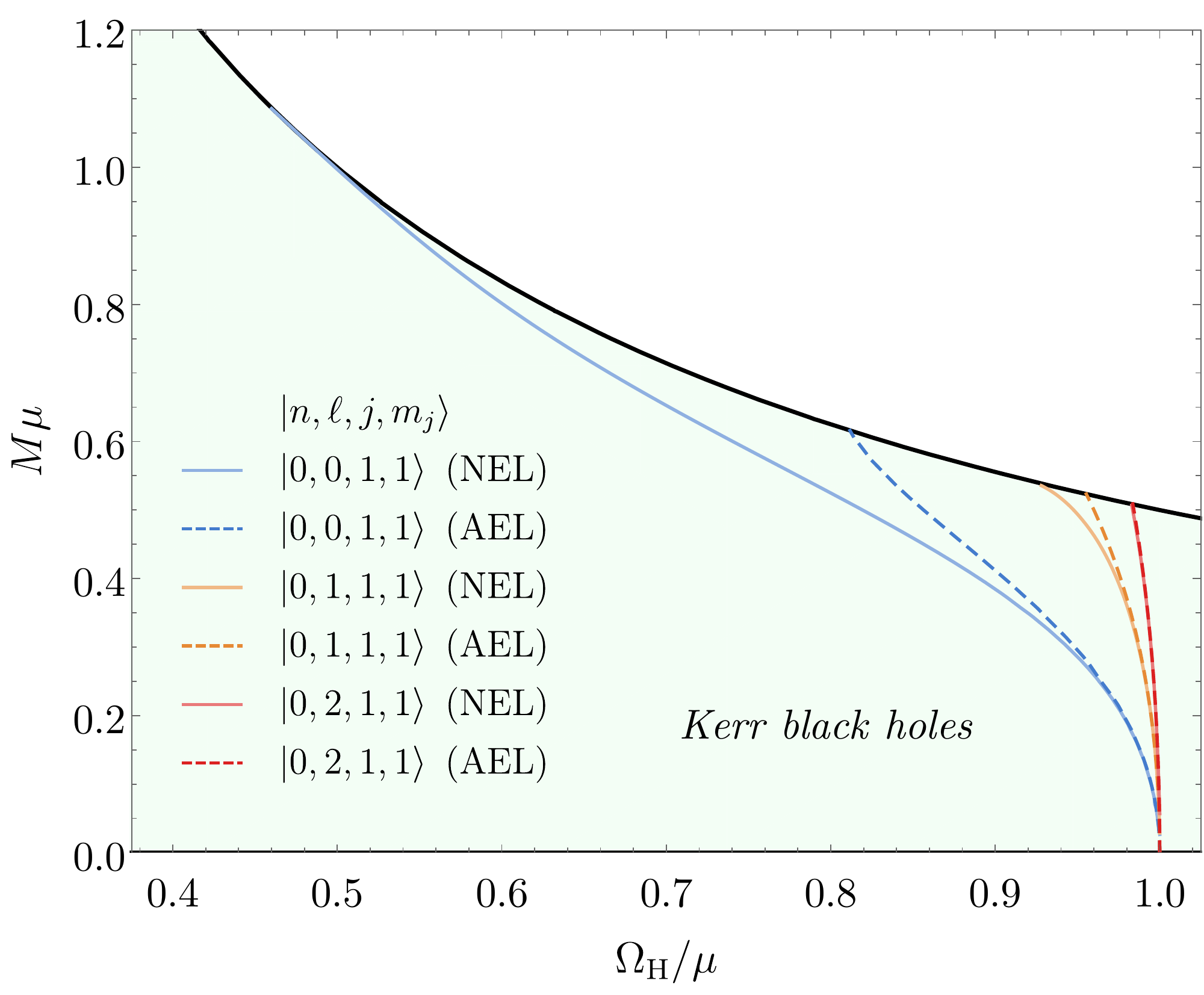}
 \includegraphics[width=.6\linewidth]{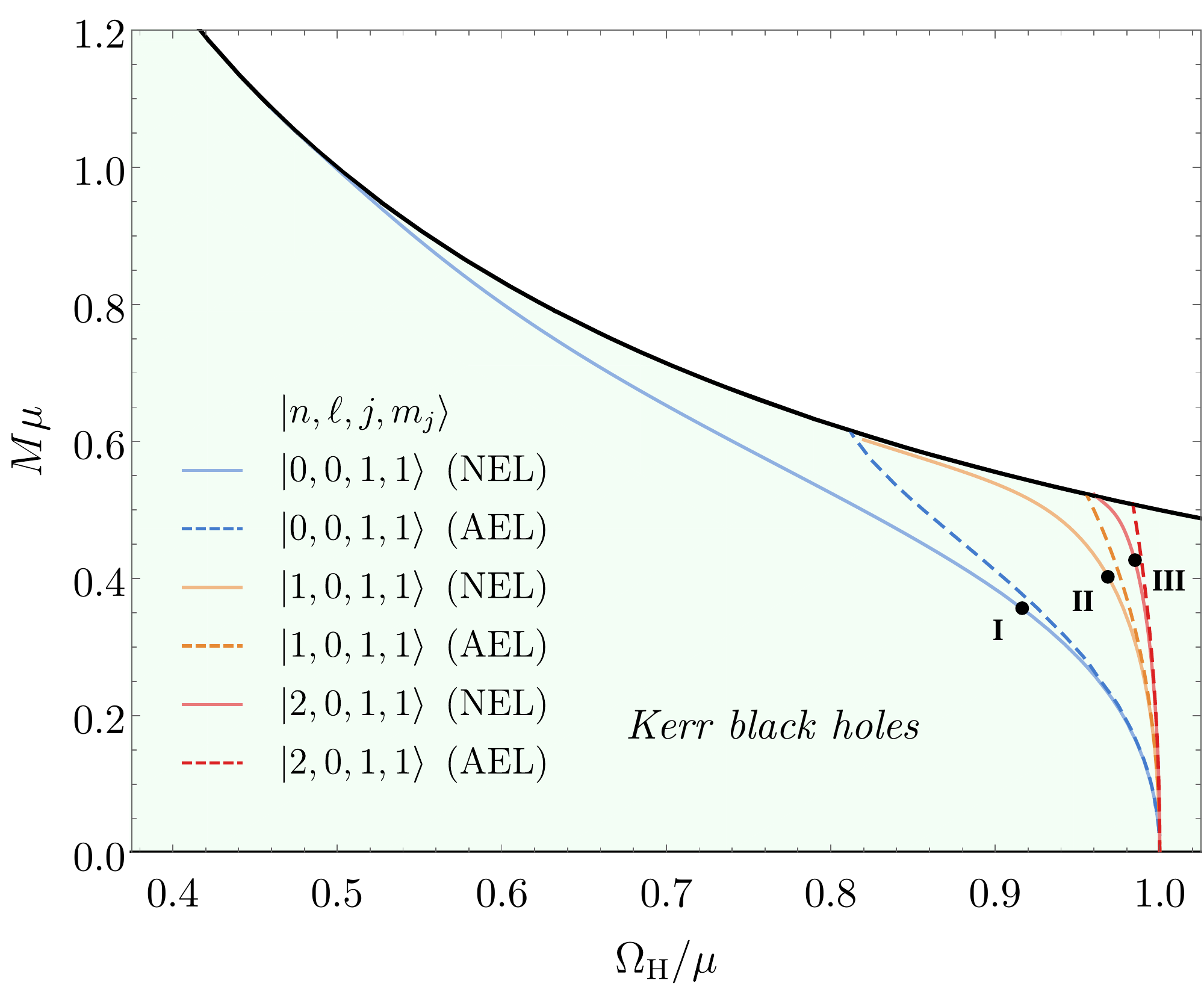}
  \caption{Existence lines for the vector stationary clouds in the ($M,\Omega_\text{H}$)-plane. The black solid line refers to extremal ($a=M$) Kerr BHs. (Top panel) $\ket{0,\ell,1,1}$ with $\ell\in\{0,1,2\}$.  The $\ell=0$ vector states are the least energetic, as they correspond to lower values of $\Omega_\text{H}$. The energy increases with $\ell$. (Bottom panel) $\ket{n,0,1,1}$ with $n\in\{0,1,2\}$. The $n=0$ vector states are the least energetic, as they correspond to lower values of $\Omega_\text{H}$. The energy increases with $n$.}
  \label{fig:SVC-j1m1}
\end{figure}

The existence lines for the vector states $\ket{0,\ell, 1,1}$ with\footnote{The existence of clouds with vanishing orbital angular momentum ($\ell=0$) is a distinctive feature of stationary vector clouds. This is intimately linked to a non-vanishing intrinsic angular momentum, as stationary scalar clouds with $\ell=0$ do not exist.} $\ell\in\{0,1,2\}$ are shown in \autoref{fig:SVC-j1m1} (top panel). When $m_j=1$, the line corresponding to the lowest values of $\Omega_\text{H}$ belongs to the electric-type state $\ket{0,0,1,1}$, which is therefore the fundamental mode with $m_j=1$. The analytical existence line for the electric-type state $\ket{0,0,1,1}$ is in agreement with its numerical counterpart when $\alpha\ll 1$. As $\alpha$ increases to values near the extremal case $a=M$ (black solid line), the two lines diverge from each other. This behavior appears to be a generic feature of existence lines corresponding to states for which $j=m_j$ and $\ell<j$ (see \autoref{fig:SVC-j1m1} -- bottom panel). The discrepancy, whose source remains unclear, suggests that higher-order corrections to the $\alpha$-expansion in Eq. \eqref{eq:s1-freq-spectrum} are needed when describing clouds around rapidly-rotating Kerr BHs. On the other hand, the analytical and numerical existence lines for the electric-type state $\ket{0,2,1,1}$, for which $j=m_j$ but $\ell>j$, appear to overlap over the full range of $\alpha$.

When $\{n,j,m_j\}$ are fixed, the existence lines move towards greater values of $\Omega_\text{H}$ as the orbital angular momentum $\ell$ increases. In fact, the larger the value of $\ell$, $i.e.$ the greater the energy of the state, the greater must the angular velocity $\Omega_\text{H}$ be for stationary equilibrium. Moreover, the existence lines converge in the limit of vanishing mass, $i.e.$ $M\rightarrow 0$, which reflects the fact that the spacetime becomes insensible to the cloud's features. These trends were also found for stationary scalar clouds around Kerr BHs \cite{Benone:2014ssa}.

\begin{figure}
\centering
\includegraphics[width=.6\linewidth]{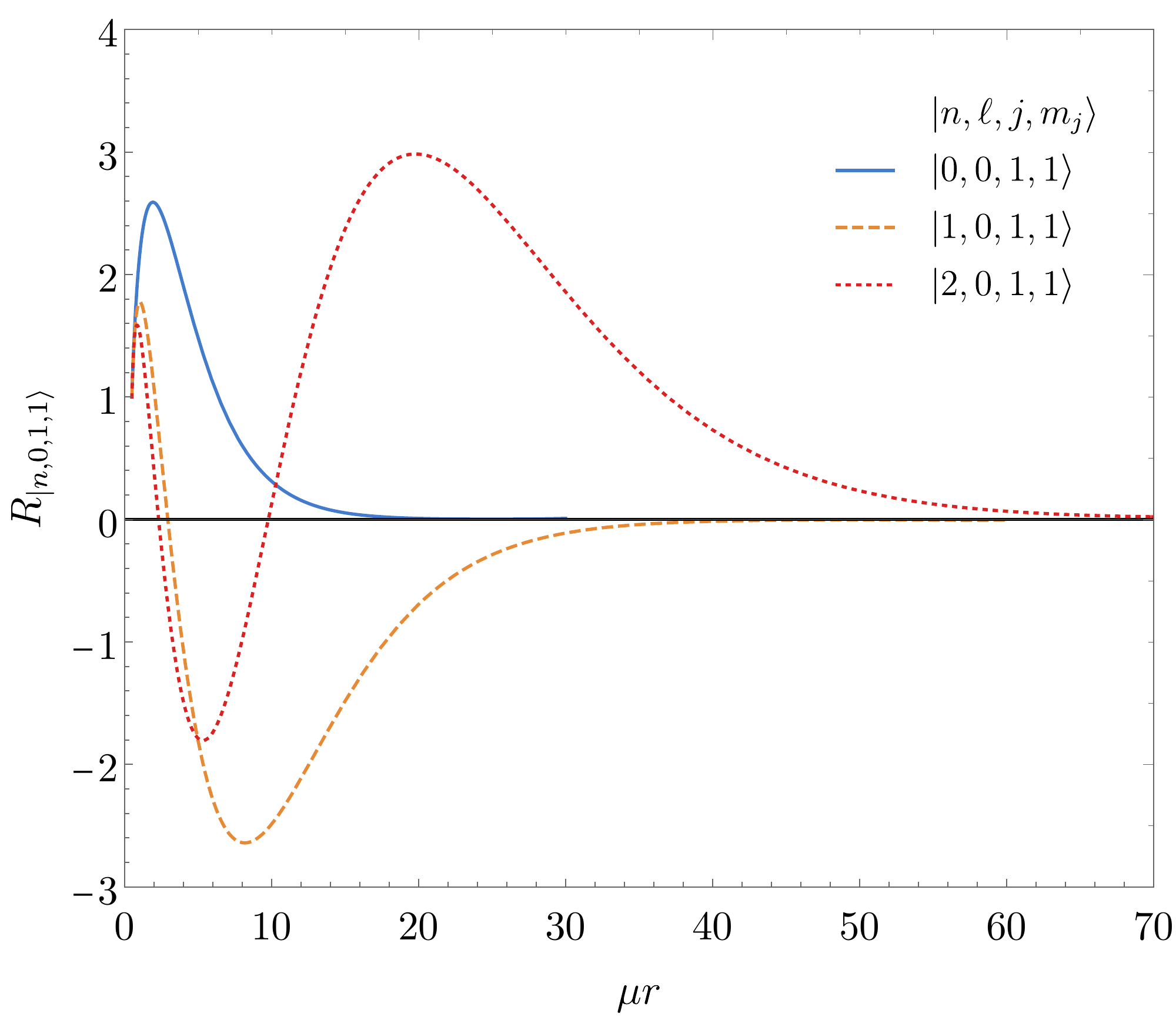}
\includegraphics[width=.6\linewidth]{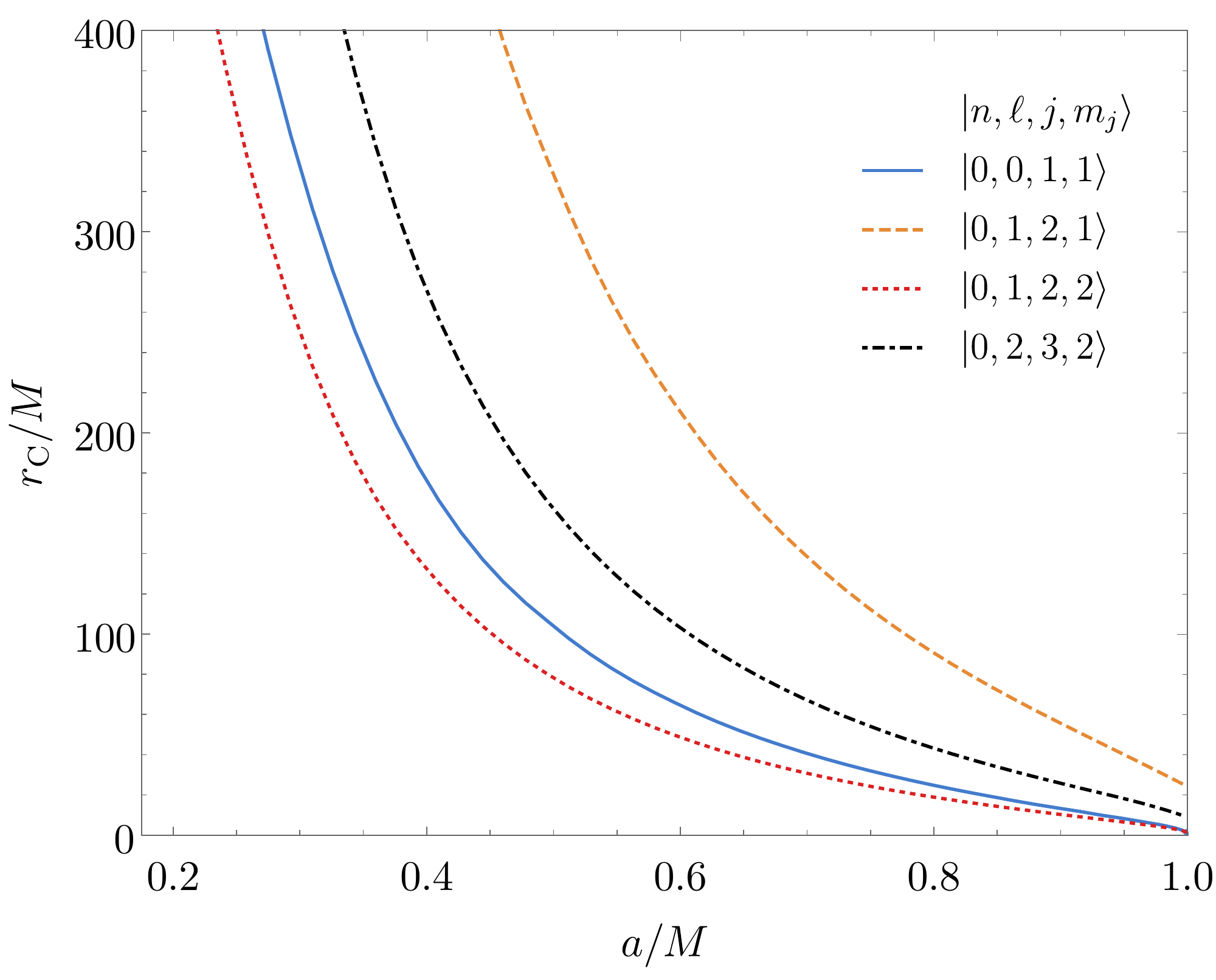}
\caption{(Top panel) Radial profiles of the bound states \textbf{I}, \textbf{II} and \textbf{III} in  \autoref{fig:SVC-j1m1} (bottom panel), characterized by $\mu r_\text{H}=0.5$. The radial functions are normalized so that $R(r_\text{H})=1$. The cloud is especially close to the event horizon when $n=0$. (Bottom panel) \textit{Radius} of different stationary vector clouds with $n=0$,  as a function of $a/M$.}
\label{fig:SPC-radial}
\end{figure}

The variation of the node number $n$ when $\{\ell,j,m_j\}$ are fixed yields identical behavior. The existence lines for the vector states $\ket{n,0,1,1}$ with $n\in\{0,1,2\}$ are plotted in \autoref{fig:SVC-j1m1} (bottom panel). The node number plays a similar role to that of the principal quantum number in the description of hydrogen's energy levels: the larger the node number $n$, the more energetic the state. Given two existence lines with the same $\{\ell,j,m_j\}$, the one with the largest node number $n$ lies to the right with respect to other in the $(M,\Omega_\text{H})$-plane. Additionally, they converge in the limit of vanishing $M$. 

The radial profile of the clouds \textbf{I}, \textbf{II} and \textbf{III} in \autoref{fig:SVC-j1m1} (bottom panel) are displayed in \autoref{fig:SPC-radial} (top panel). The function $R$ is finite over the whole $r$ domain outside the event horizon and vanishes (exponentially) as $r\rightarrow+\infty$, as required by asymptotic flatness. Besides, the local maximum closest to the event horizon decreases with increasing $n$. 

Finally, \autoref{fig:SPC-radial} (bottom panel) shows the dependence of the \textit{radius} of the cloud, hereafter denoted by $r_\text{C}$, on the rotation parameter $a$ for different vector stationary clouds with $n=0$. $r_\text{C}$ is defined as the value of $r$ closest to $r_\text{H}$ that locally maximizes the function $4\pi|R|^2$. Its value diverges in the Schwarzschild limit ($a\rightarrow 0$), in accordance with the fact that Schwarzschild BHs cannot carry stationary vector clouds. Moreover, the minimum of $r_\text{C}$, which occurs at $a=M$, is finite, which means that Kerr BHs do not support sufficiently tight clouds. Similar observations were already reported for stationary scalar clouds in~\cite{Benone:2014ssa}.

%%%%%%%%%%%%%%%%%%%%%%%%%%%%%%%%%%%%%%%%%%%%%%%%%%%%%%%%%%%%%%%%%%%%%
%%%%%%%%%%%%%%%%%%%%%%%%%%%%%%%%%%%%%%%%%%%%%%%%%%%%%%%%%%%%%%%%%%%%%
\section{Non-linear analysis: hairy black holes and Proca stars\label{sec3}}
%%%%%%%%%%%%%%%%%%%%%%%%%%%%%%%%%%%%%%%%%%%%%%%%%%%%%%%%%%%%%%%%%%%%%
%%%%%%%%%%%%%%%%%%%%%%%%%%%%%%%%%%%%%%%%%%%%%%%%%%%%%%%%%%%%%%%%%%%%%
We now address the fully non-linear solutions of the Einstein-complex-Proca model, described by the action
\begin{equation}
\mathcal{S}=\int \D^4x \sqrt{-g} \left(\frac{R}{16\pi }+\mathcal{L}_\text{M}\right) \ ,
\label{epaction}
\end{equation}
where $\mathcal{L}_\text{M}$ is the Proca Lagrangian density~\eqref{pl}. Varying this action one obtains the Proca equations~\eqref{pe} and the Einstein equations
\begin{equation}
R_{\alpha \beta}-\frac{1}{2}R g_{\alpha \beta}=8 \pi  T_{\alpha \beta} \ ,
\label{Einstein-eqs}
\end{equation}
where the Proca energy-momentum tensor is:
\begin{eqnarray}
T_{\alpha\beta}=\frac{1}{2}
( F_{\alpha \sigma }\bar{{F}}_{\beta \gamma}
+\bar{{F}}_{\alpha \sigma } {F}_{\beta \gamma}
)g^{\sigma \gamma}
-\frac{1}{4}g_{\alpha\beta}{F}_{\sigma\tau}\bar{{F}}^{\sigma\tau}+\frac{1}{2}\mu^2\left[  
{A}_{\alpha}\bar{{A}}_{\beta}
+\bar{{A}}_{\alpha}{A}_{\beta}
-g_{\alpha\beta} {A}_\sigma\bar{{A}}^\sigma\right]\ . \ \ \ \ \ \ \  \ 
\label{procaemt}
\end{eqnarray}
We follow the conventions of~\cite{Herdeiro:2016tmi}.   More details on the formalism can be found therein. If one linearises the model~\eqref{epaction} in the Proca field, one ends up with the vacuum Einstein equations and a test Proca field on a fixed curved background (that solves the vacuum Einstein equations). This corresponds precisely to the analysis of~\autoref{sec2}.

%%%%%%%%%%%%%%%%%%%%%%%%%%%%%%%%%%%%%%%%%%%%%%%%%%%%%%%%%%%%%%%%%%%%%
%%%%%%%%%%%%%%%%%%%%%%%%%%%%%%%%%%%%%%%%%%%%%%%%%%%%%%%%%%%%%%%%%%%%%
\subsection{The ansatz\label{sec31}}
%%%%%%%%%%%%%%%%%%%%%%%%%%%%%%%%%%%%%%%%%%%%%%%%%%%%%%%%%%%%%%%%%%%%%
%%%%%%%%%%%%%%%%%%%%%%%%%%%%%%%%%%%%%%%%%%%%%%%%%%%%%%%%%%%%%%%%%%%%%
To find the hairy BHs that bifurcate from the linear clouds that were studied in~\autoref{sec2} we use the metric ansatz\footnote{The Kerr metric in this coordinate system, together with the relation between $r$ in Eq.~\eqref{mansatz}, used in this section, and the radial Boyer-Lindquist coordinate used in~\autoref{sec2}, can be found in Appendix A of~\cite{Herdeiro:2015gia}.}
\begin{eqnarray}
  \boldsymbol{g}=-e^{2F_0}N \boldsymbol{\D} t^2+e^{2F_1}\left(\frac{\boldsymbol{\D} r^2}{N}+r^2 \boldsymbol{\D} \theta^2\right) + e^{2F_2}r^2 \sin^2\theta \left(\boldsymbol{\D} \varphi-W \boldsymbol{\D} t\right)^2 \ ,
 \label{mansatz}
\end{eqnarray}
where 
\begin{equation}
N\equiv 1 -\frac{r_\text{H}}{r} \ ,
\label{n}
\end{equation} 
and  $F_i,W$ are functions of the spheroidal coordinates $(r,\theta)$. The parameter $r_\text{H}$ is the radial coordinate of the event horizon, which is $\theta$-independent. 

For the Proca potential, we use an ansatz that depends on four functions $(V,H_a)$. All these functions depend on $(r,\theta)$. The ansatz has a harmonic time and azimuthal dependence, which introduces a (positive) frequency, $\omega>0$, and the azimuthal harmonic index, $m\in \mathbb{Z}$:
\begin{equation}
\bm{A}=e^{i(m\varphi-\omega t)}\left(
 iV \boldsymbol{\D} t  +H_1\boldsymbol{\D} r+H_2\boldsymbol{\D} \theta+i H_3 \sin \theta \boldsymbol{\D} \varphi 
\right) \ .
\label{procaa}
\end{equation}
Here, $m$ should be identified with $m_j$ of \autoref{sec2} and we shall focus on $m=1$. We follow closely~\cite{Herdeiro:2016tmi}, wherein all details can be found, namely: the explicit equations of motion for this ansatz (in Appendix B therein) and the boundary conditions at the horizon, spatial infinity and on the axis (in Section 4 therein). Details on the numerical method can be found in Section 3.3 of~\cite{Herdeiro:2015gia}.  The key feature for the existence of these BHs is the synchronisation condition~\eqref{eq:superradiance-sat}, where $\Omega_\text{H}=W(r_\text{H})$, the non-diagonal metric function in Eq.~\eqref{mansatz}, which on the horizon is independent of $\theta$, and $m,\omega$ are the parameters in the Proca ansatz~\eqref{procaa}. 

%%%%%%%%%%%%%%%%%%%%%
\subsection{Domain of existence}
\label{sec32}
%%%%%%%%%%%%%%%%%%%%%
When finding solutions via a relaxation method, such as the Newton-Raphson method used for this work, the initial guess plays a key role to guarantee convergence to the desired solutions. In~\cite{Herdeiro:2015gia}, the construction of hairy BHs started from the spinning Proca stars in~\cite{Brito:2015pxa}, which have one node for the temporal component of the Proca potential, $V$. Consequently, the hairy BHs reported in~\cite{Herdeiro:2015gia} also have one node in $V$. At that point, it was found no evidence for nodeless solutions of either spinning Proca stars or BHs with Proca hair, even though it was stated in~\cite{Herdeiro:2015gia} that no proof for the inexistence of nodeless solutions could be established (except for spherical Proca stars).

\begin{figure}[h!]
\centering
\includegraphics[width=.69\linewidth]{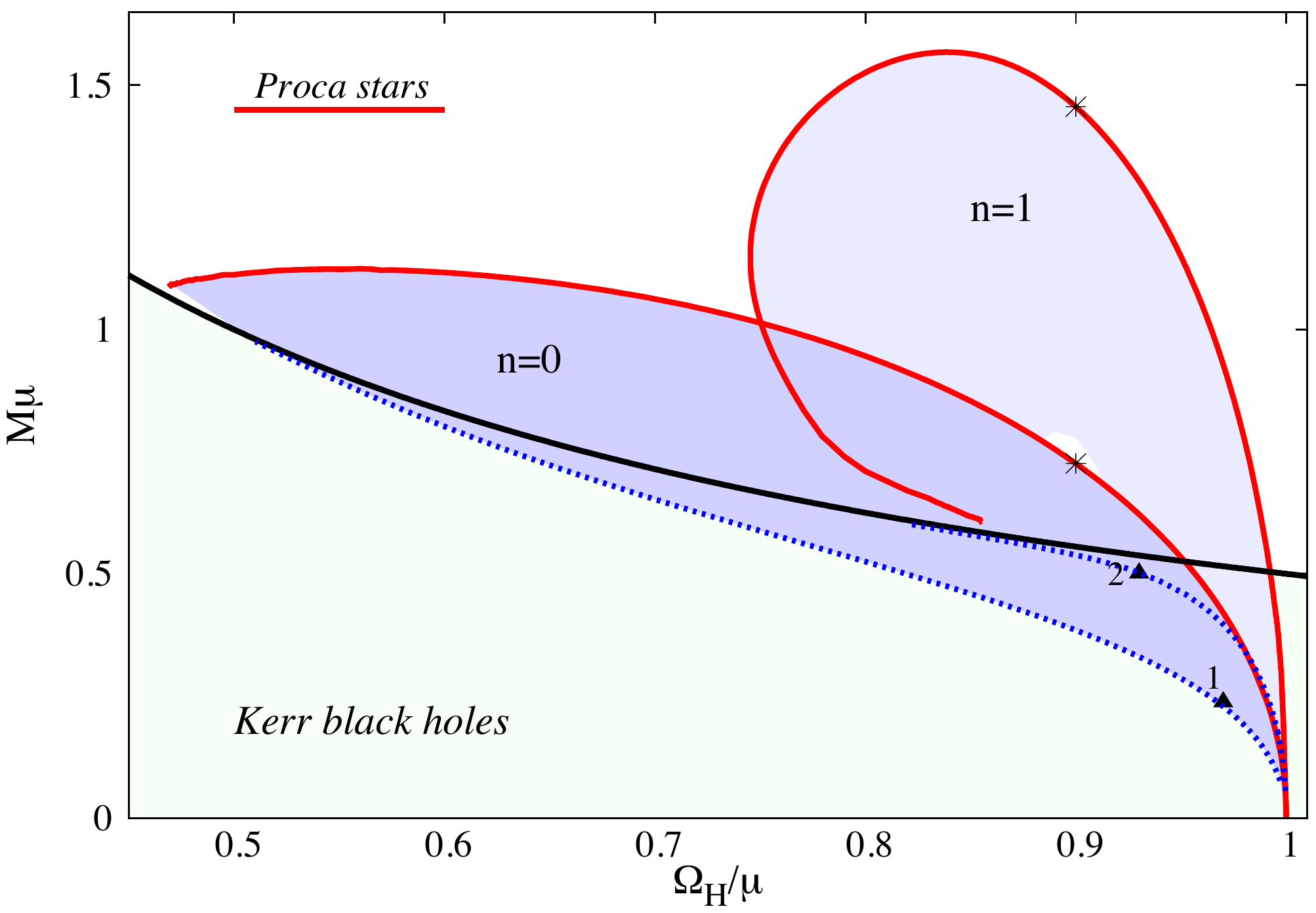}
\caption{Domain of existence of the $n=0$ (fundamental state) and $n=1$ (excited state) BHs with synchronised Proca hair with $m=1$ in an ADM mass $vs.$ angular velocity diagram, in units of $\mu$. The black solid line corresponds to extremal Kerr BHs; non-extremal Kerr BHs exist below that line, in the light green shaded region. Two Proca star solutions and two hairy BH solutions were highlighted (as stars and triangles), to be analysed in \autoref{sec33}.}
\label{fig3}
\end{figure}

These results were reconsidered after the numerical evolutions of the Kerr superradiant instability have been reported~\cite{East:2017ovw}. The data describing the equilibrium points attained in these evolutions matched spinning BHs with Proca hair and a nodeless Proca potential temporal component $V$, first constructed in~\cite{Herdeiro:2017phl}, wherein their domain of existence was exhibited. This domain of existence is shown in  \autoref{fig3}, together with the domain of existence of the nodeful $(n=1)$ solutions reported in~\cite{Herdeiro:2015gia}. The hairy BHs exist in the blue shaded regions. In each case ($n=0$ or $n=1$) the domain of existence is bounded by the solitonic limit (red solid lines) wherein the hairy BHs become spinning Proca stars with the same $n$, and by the bald limit (blue dotted lines), wherein they meet the Kerr parameter space at the corresponding existence line, with $m_j=m$, the same $n$ and $(\ell,j)=(0,m_j)$. Thus, the two blue dotted lines plotted in  \autoref{fig3} correspond to the blue ($n=0$) and yellow ($n=1$) numerical existence lines plotted in \autoref{fig:SVC-j1m1} (bottom panel). 

The existence line from which the \textit{fundamental} non-linear solutions bifurcate follows a similar rationale to that observed for the scalar case~\cite{Herdeiro:2014goa}. For a given $m=m_j$, the existence line with $\ell=0$ and $j=m_j$ is the leftmost one in the Kerr parameter space plotted in \autoref{fig:SVC-j1m1}. Thus it represents the threshold between the Kerr BHs that are stable against all modes with that $m_j$ and the ones that are unstable against at least one such mode. Since $m_j$ is the only of the three quantum numbers $(\ell,j,m_j)$ that remains significant in the non-linear theory -- it is associated to an isometry --, for each $m_j$ the existence line whence the hairy BHs bifurcate is the one with  $(\ell,j)=(0,m_j)$. BHs emerging from the other existence lines with $m=1$ are likely to exist but are excited states, 
with either more radial or angular nodes.

\begin{figure}[h!]
\centering
\includegraphics[width=.7\linewidth]{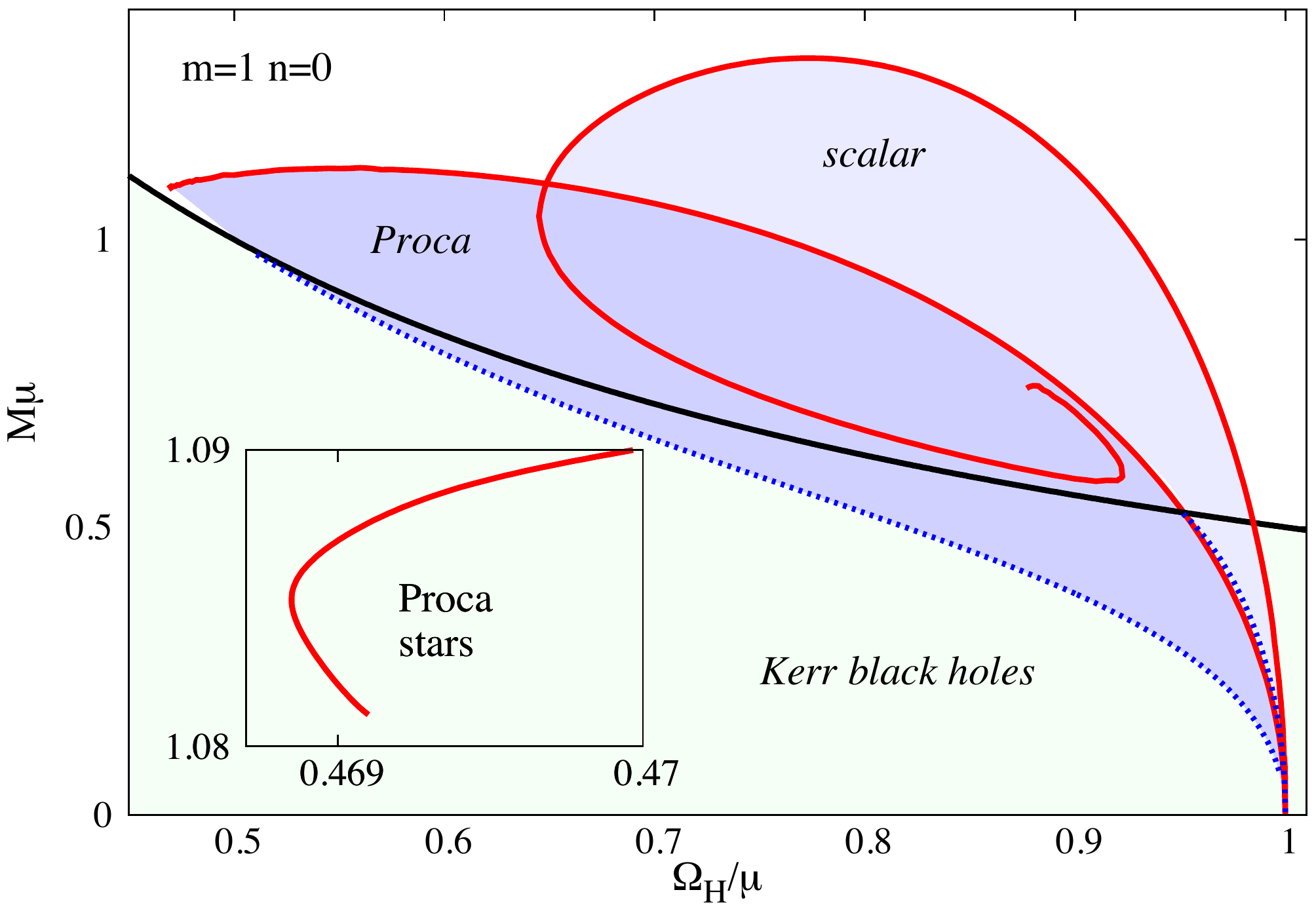}
\caption{Similar representation of the domain of existence as in \autoref{fig3}, but now comparing  the $(n,m)=(0,1)$ (fundamental states) Kerr BHs with synchronised scalar and Proca hair. The inset shows a detail of the backbending of the fundamental Proca stars line, as it attains the minimal frequency.}
\label{fig4}
\end{figure}

Inspection of \autoref{fig3} reveals two main features. Firstly, as expected, the excited states ($n=1$) can attain a larger ADM mass; secondly, the fundamental states of the BHs with Proca hair exist for a larger $\omega$-range; this also seems intuitive: excited states require a larger minimum angular velocity.\footnote{A similar trend can be observed in the scalar case, comparing $n=0$ with $n=1$ solutions~\cite{Wang:2018xhw}.} The same trends are observed when comparing the fundamental states $(n,m)=(0,1)$ of the scalar and the Proca hairy BHs -- \autoref{fig4}, with the scalar case playing the role of the excited Proca family, in this comparison. This had already been observed for the solitonic limit in~\cite{Herdeiro:2019mbz}. In the bald limit, this means that the fundamental Proca existence line spans lower $\Omega_\text{H}$ BHs. This is a manifestation of the well known fact that the superradiant instability is stronger for the vector case~\cite{Press:1972zz}. 

\begin{figure}[h!]
\centering
\includegraphics[width=.7\linewidth]{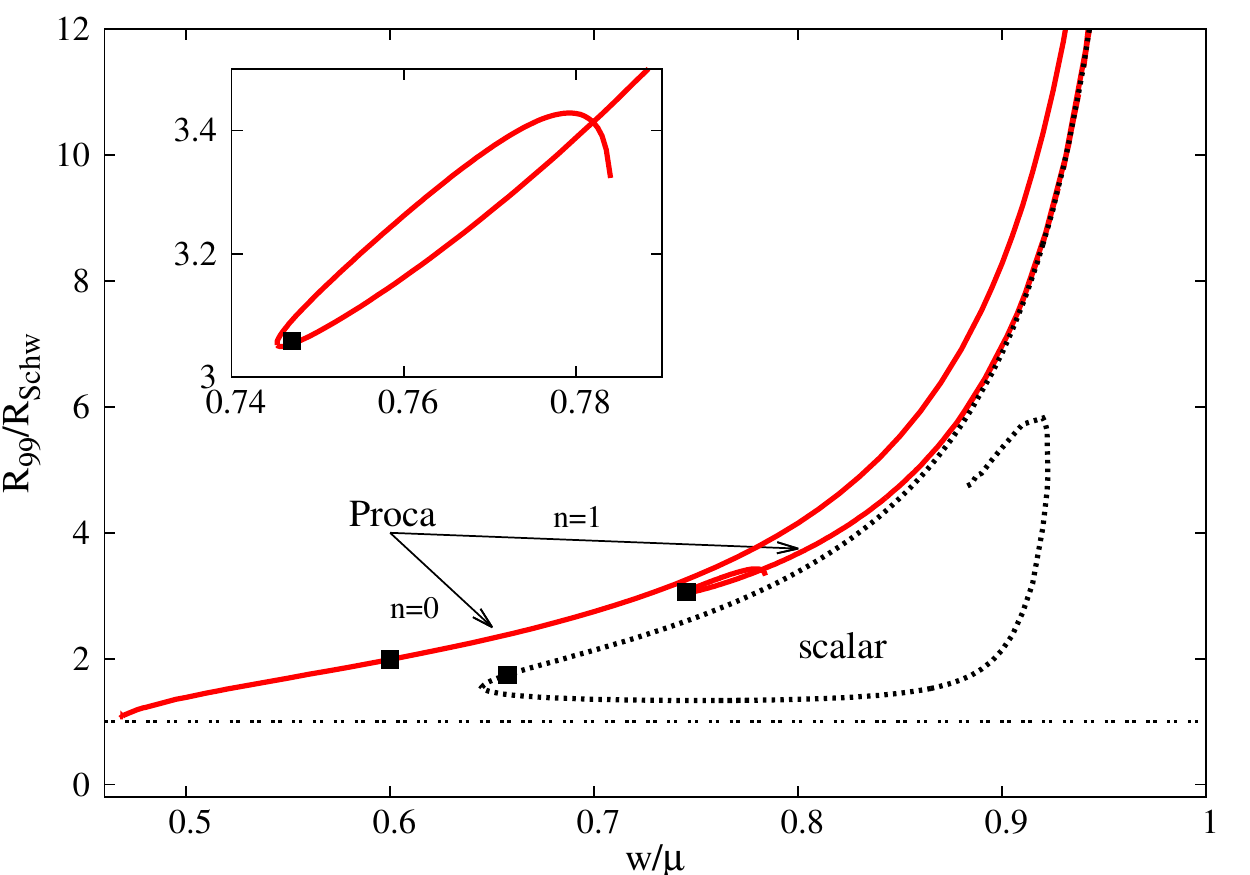}
\caption{Inverse compactness of the three families of solutions: $n=0$ and $n=1$ Proca stars (red solid lines) and $n=0$ scalar boson stars (black dotted line), all with $m=1$. The inset exhibits a detail of the $n=1$ Proca stars line. The squares mark the first occurrence of an ergo-region along the family of bosonic stars.}
\label{fig5}
\end{figure}

Unlike the scalar or the excited Proca case, in the case of $n=0$ spinning Proca stars, that compose the (red solid line) boundary of the domain of existence, it was not possible to explore the domain of solutions after the backbending, $i.e.$ when the minimum frequency is attained -- see inset in~\autoref{fig4}. The reason is that these solutions become rather compact and hence strong gravity configurations, making their computation numerically challenging. To assess this, we have used the same measure of compactness as, $e.g.$ in~\cite{AmaroSeoane:2010qx,Herdeiro:2015gia}, namely:
\begin{equation}
{\rm Compactness}^{-1}\equiv  \frac{R_{99}}{2M_{99}} \ ,
\label{compactness}
\end{equation}
where $R_{99}$ is the perimetral radius that contains 99\% of the star's mass, $M_{99}$. We recall that bosonic stars do not have a surface where a discontinuity of the energy density occurs; rather, they decay exponentially, vanishing only at infinity. The perimetral radius is a geometrically meaningful radial coordinate $R$: a circumference along the equatorial plane has perimeter $ 2\pi R$. 
The inverse compactness of  the $n=0,1$ Proca stars and $n=0$ scalar boson stars, all with $m=1$, is shown in~\autoref{fig5}. One observes that the inverse compactness is always greater than unity, meaning that all these stars are less compact than a BH. Moreover, the fundamental Proca stars become the most compact ones, precisely at the backbending, where they attain an inverse compactness $\lesssim 1.1$.

Another token of strong field gravity is the formation of ergo-regions in spinning spacetimes. In the solitonic limit, both the scalar and vector spinning stars do not have ergo-regions when $\omega\rightarrow \mu$ (see~\cref{fig3,fig4}), corresponding to the dilute regime where the stars are not compact and not strongly relativistic. Moving along the spiral and away from this dilute regime, in all cases the ergo-region appears in the first branch, $i.e.$, before the first backbending and for quite compact stars. The first occurrence of an ergo-region along the sequence of bosonic stars is marked with a square in~\autoref{fig5}.  The comparison between the three different cases shows that compactness is not the only factor determining the existence of an ergo-region. For all three cases (Proca with $n=0,1$ and scalar with $n=0$) the ergo-region of these stars is toroidal. In the family of the hairy BHs, this toroidal region adds up to the ergo-sphere around the spinning horizon. We have not scanned in detail the parameter space but one will get a rich ergo-region structure, including ergo-Saturns, analogous to those found for BHs with synchronised scalar hair $(n=0)$~\cite{Herdeiro:2014jaa}, Proca hair $(n=1)$~\cite{Herdeiro:2016tmi} and other cousin models, $e.g.$~\cite{Herdeiro:2018djx,Delgado:2019prc,Kunz:2019bhm,Kunz:2019sgn}.

%%%%%%%%%%%%%%%%%%%%%%%%%%%%%%%%%%%%%%%%%%%%%%%%%%%%%%%%%%%%%%%%%%%%%
%%%%%%%%%%%%%%%%%%%%%%%%%%%%%%%%%%%%%%%%%%%%%%%%%%%%%%%%%%%%%%%%%%%%%
\subsection{Analysis of specific solutions\label{sec33}}
%%%%%%%%%%%%%%%%%%%%%%%%%%%%%%%%%%%%%%%%%%%%%%%%%%%%%%%%%%%%%%%%%%%%%
%%%%%%%%%%%%%%%%%%%%%%%%%%%%%%%%%%%%%%%%%%%%%%%%%%%%%%%%%%%%%%%%%%%%%
In order to get a better intuition on the impact of the node number on the solutions let us consider a comparative study between  the profile functions of two spinning Proca stars, one with $n=0$ and another with $n=1$, and both with the same frequency $\omega/\mu=0.9$. These two configurations are highlighted as two stars in \autoref{fig3}. 

\begin{figure}
\centering
\includegraphics[width=.485\linewidth]{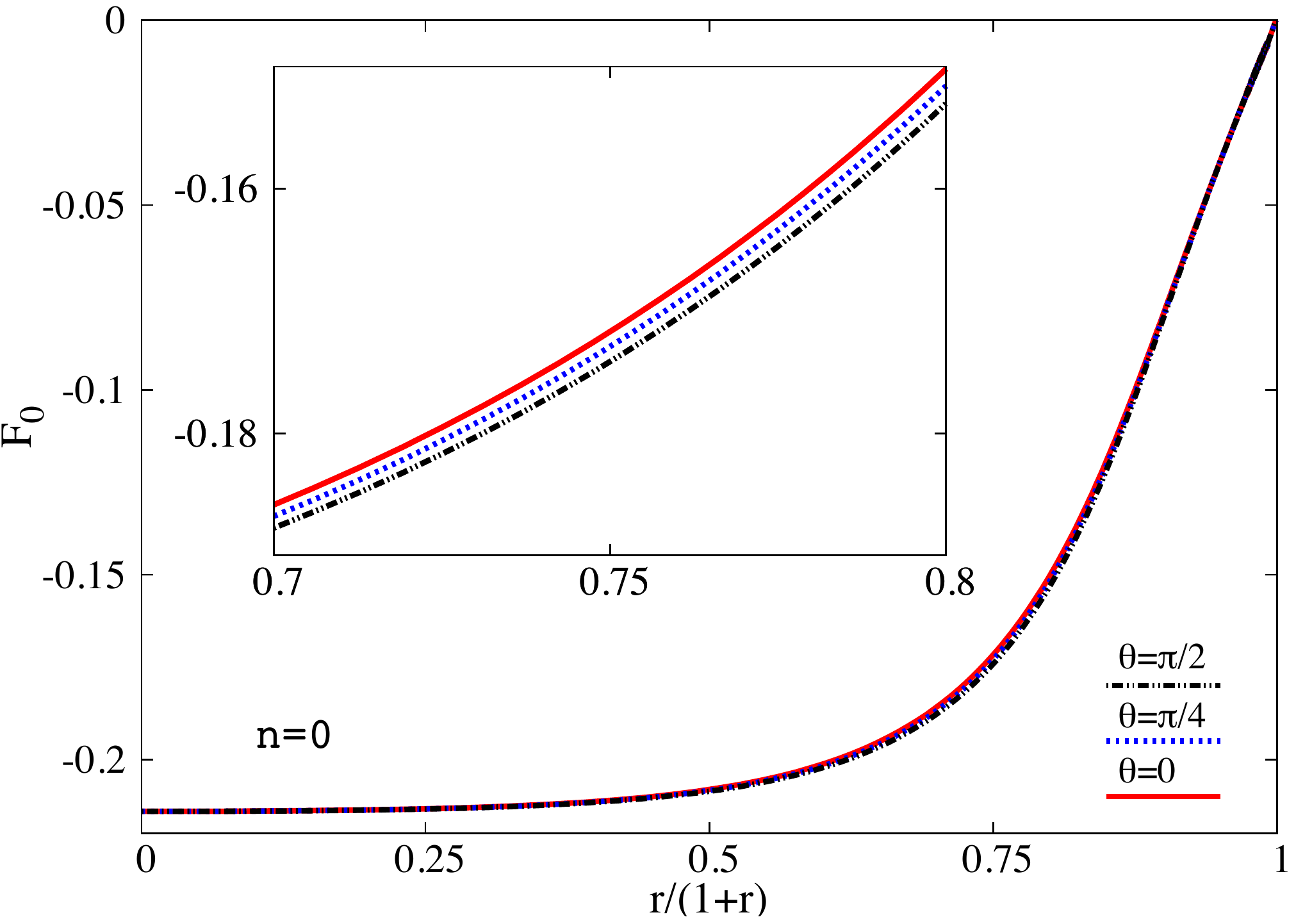}
\includegraphics[width=.485\linewidth]{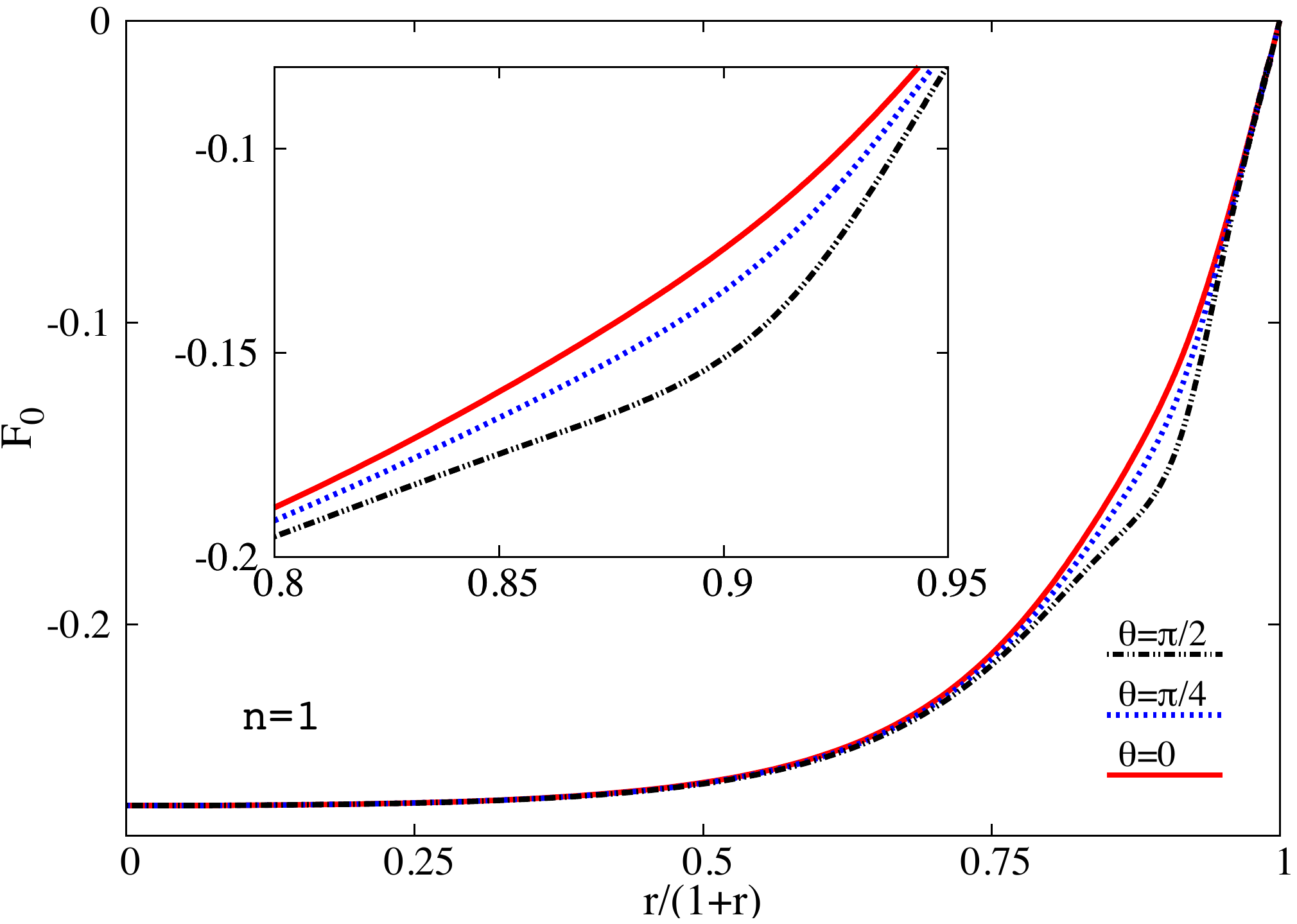}
\includegraphics[width=.485\linewidth]{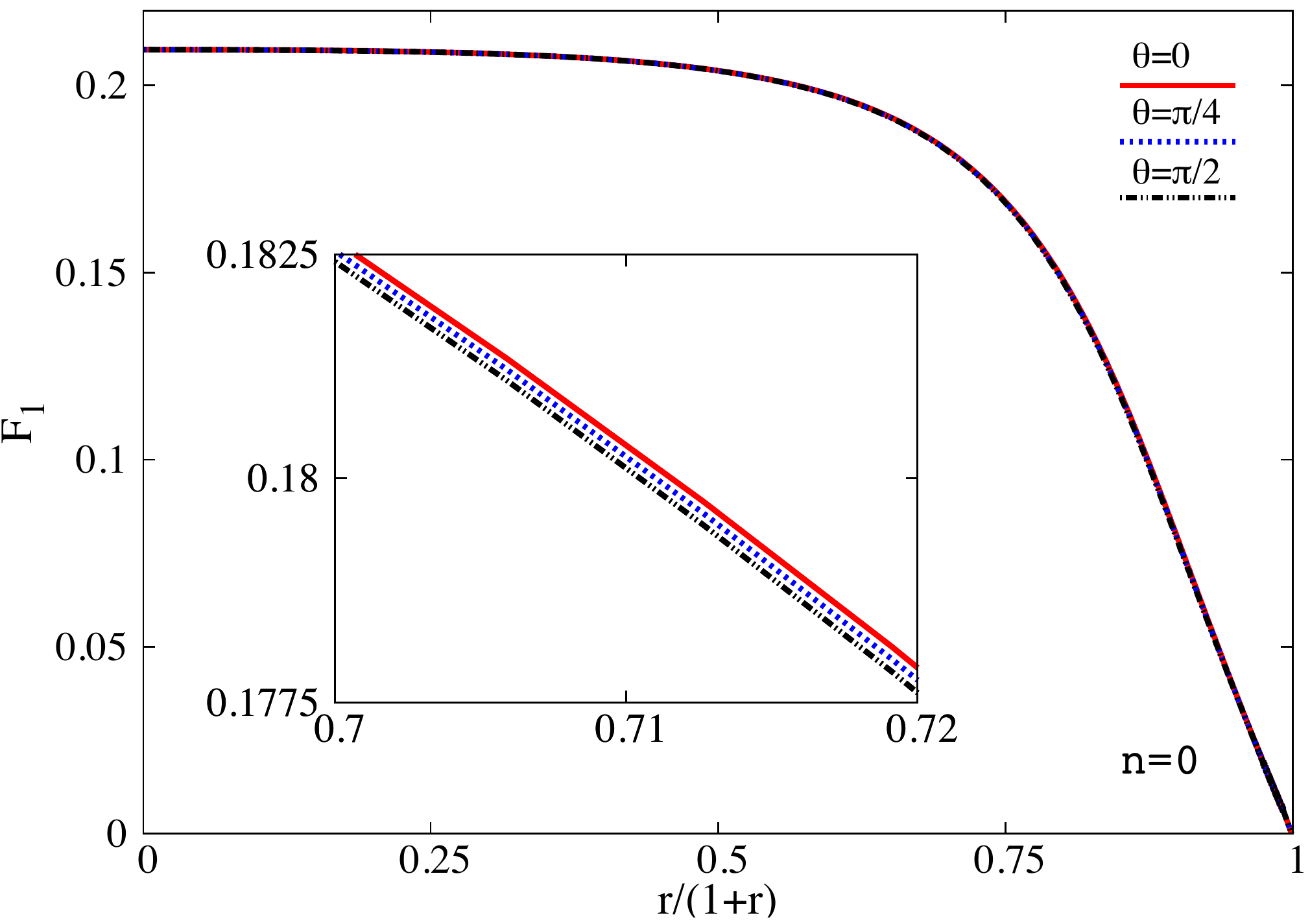}
\includegraphics[width=.485\linewidth]{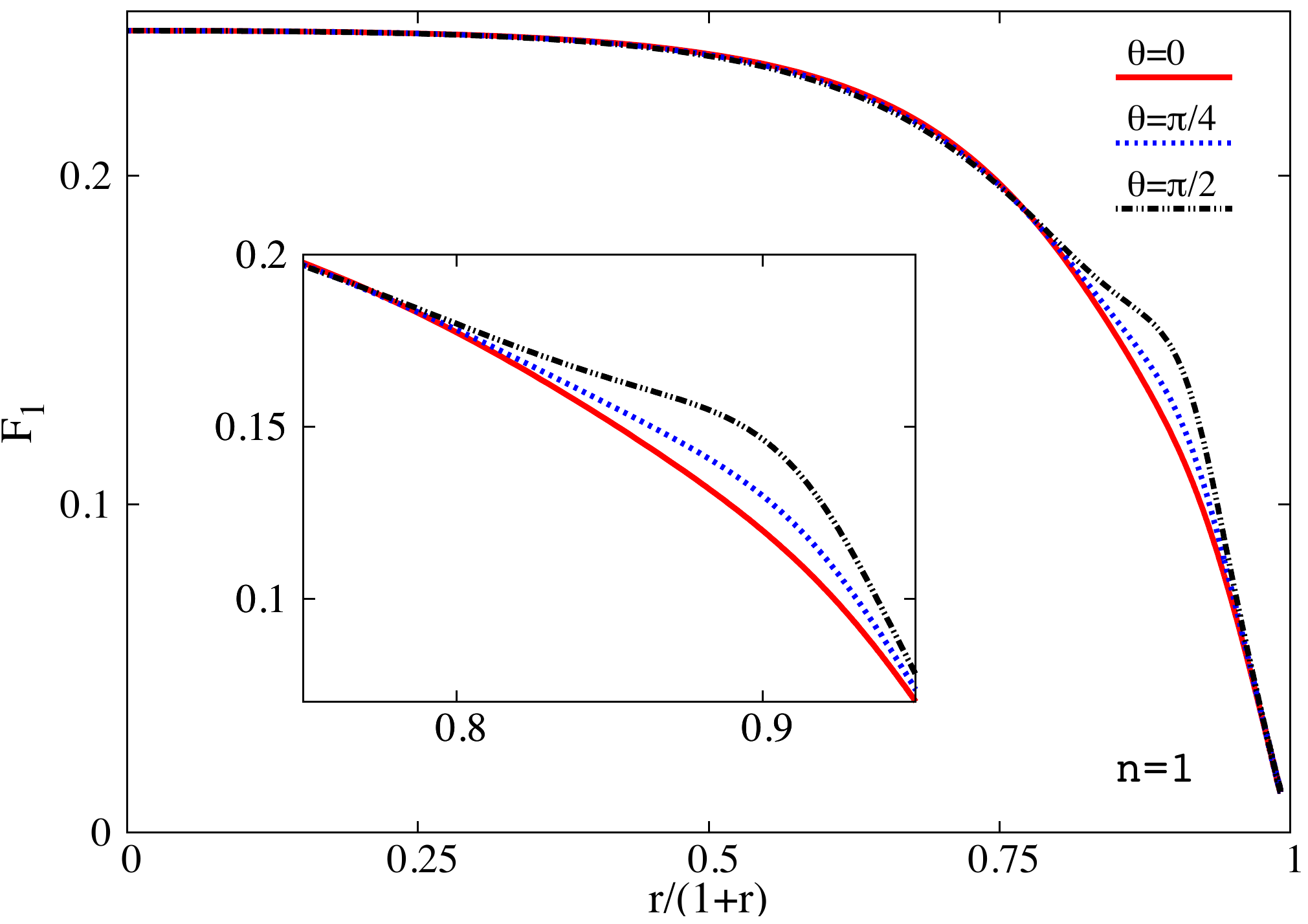}
\includegraphics[width=.485\linewidth]{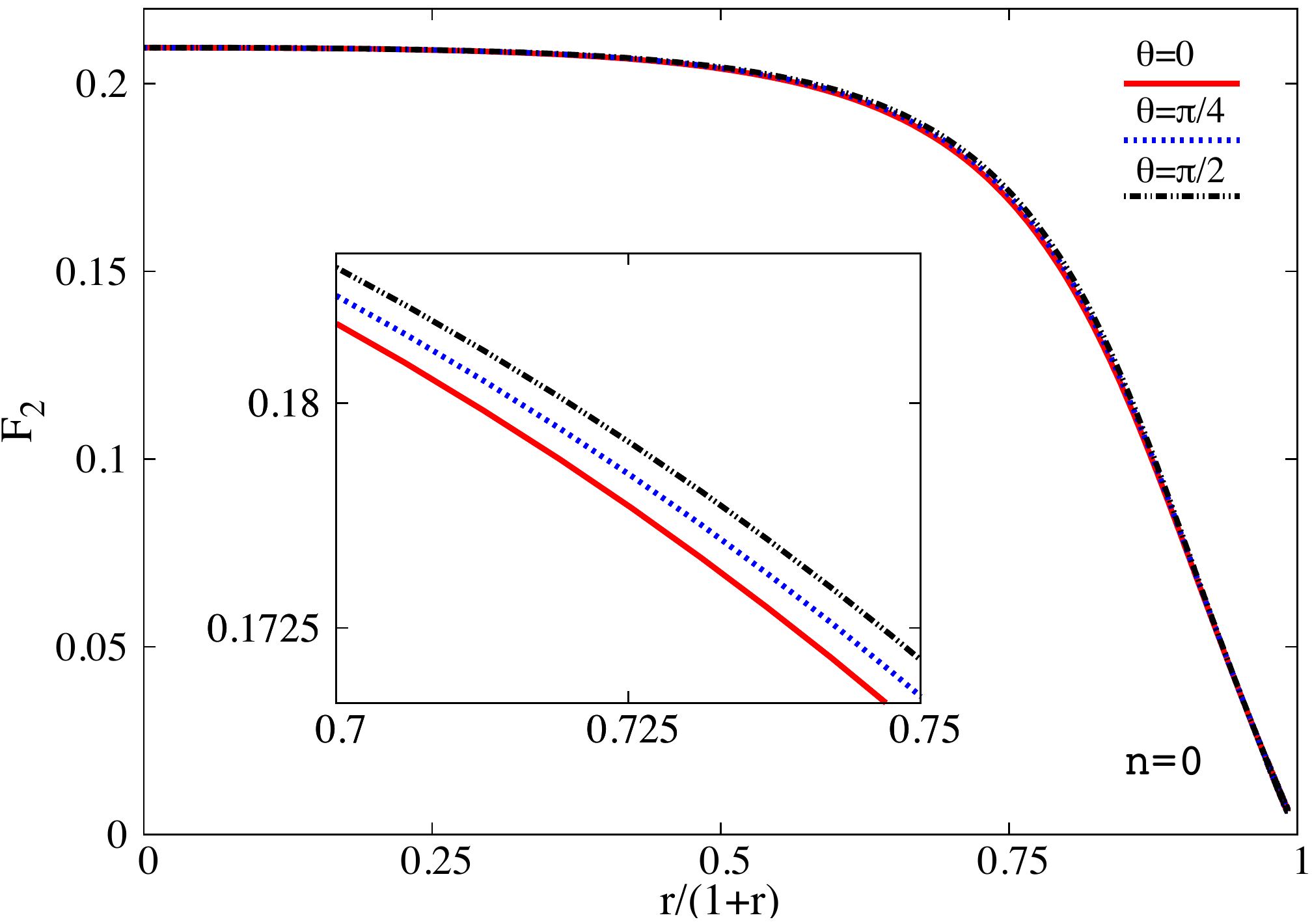}
\includegraphics[width=.485\linewidth]{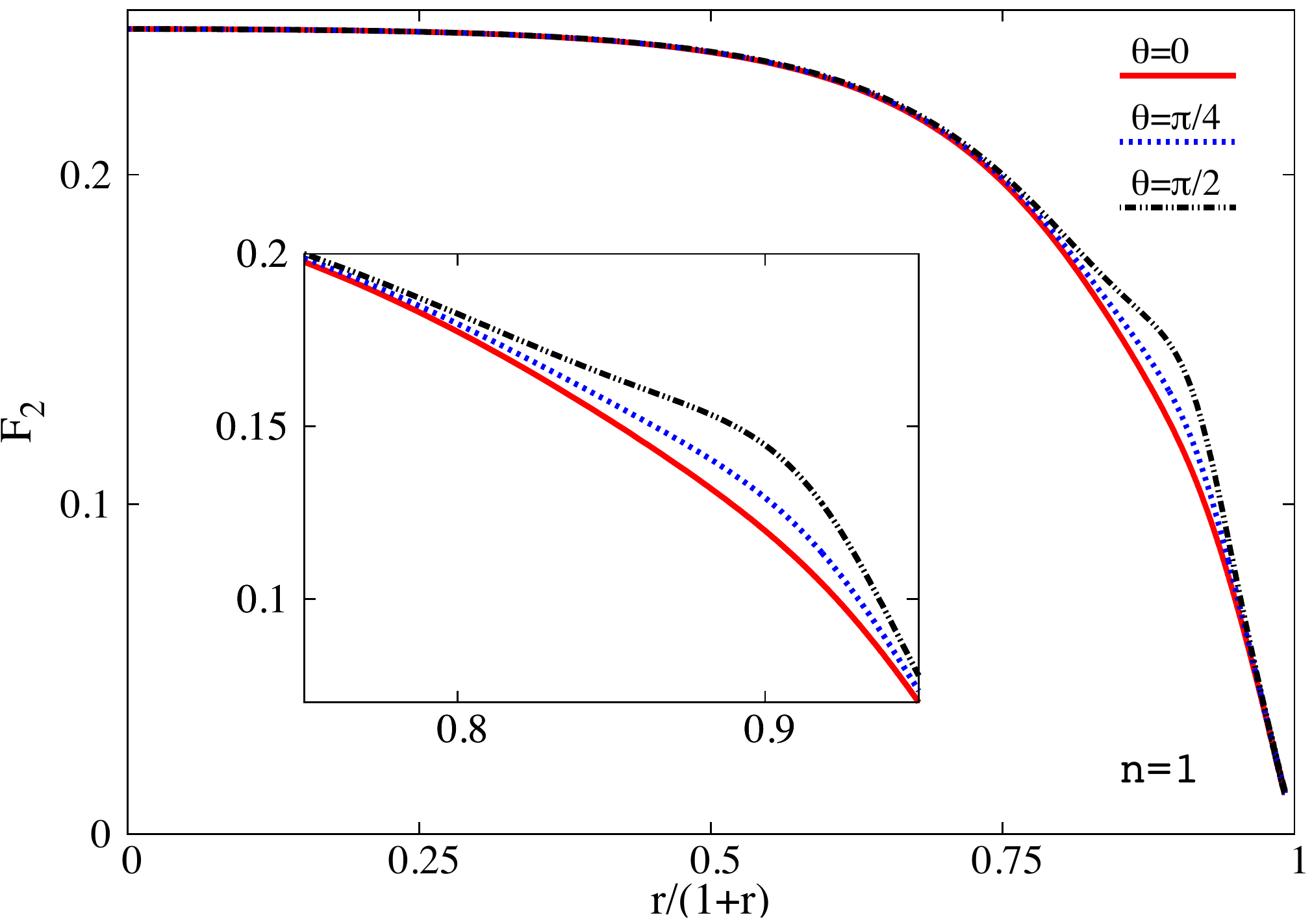}
\includegraphics[width=.485\linewidth]{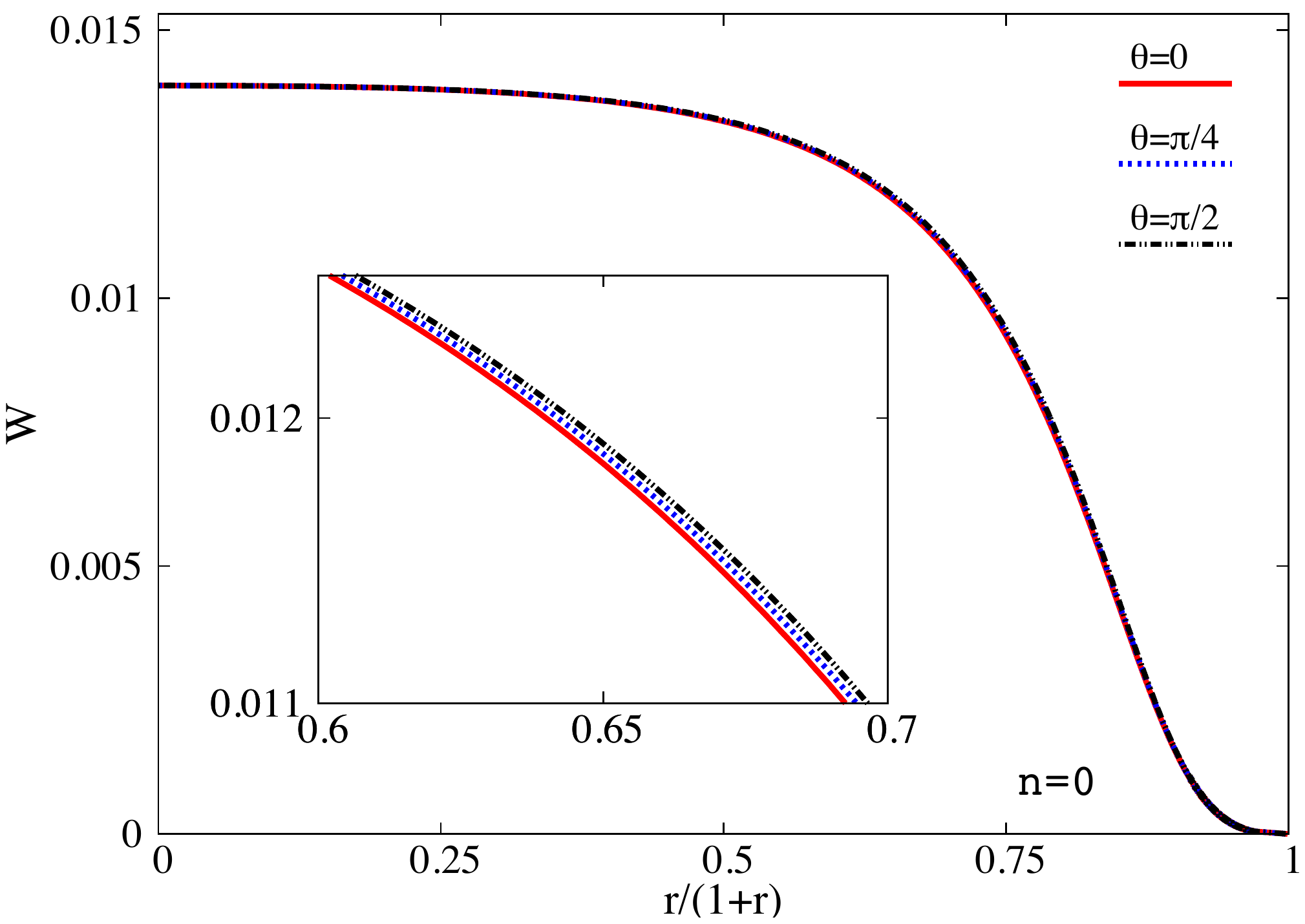}
\includegraphics[width=.485\linewidth]{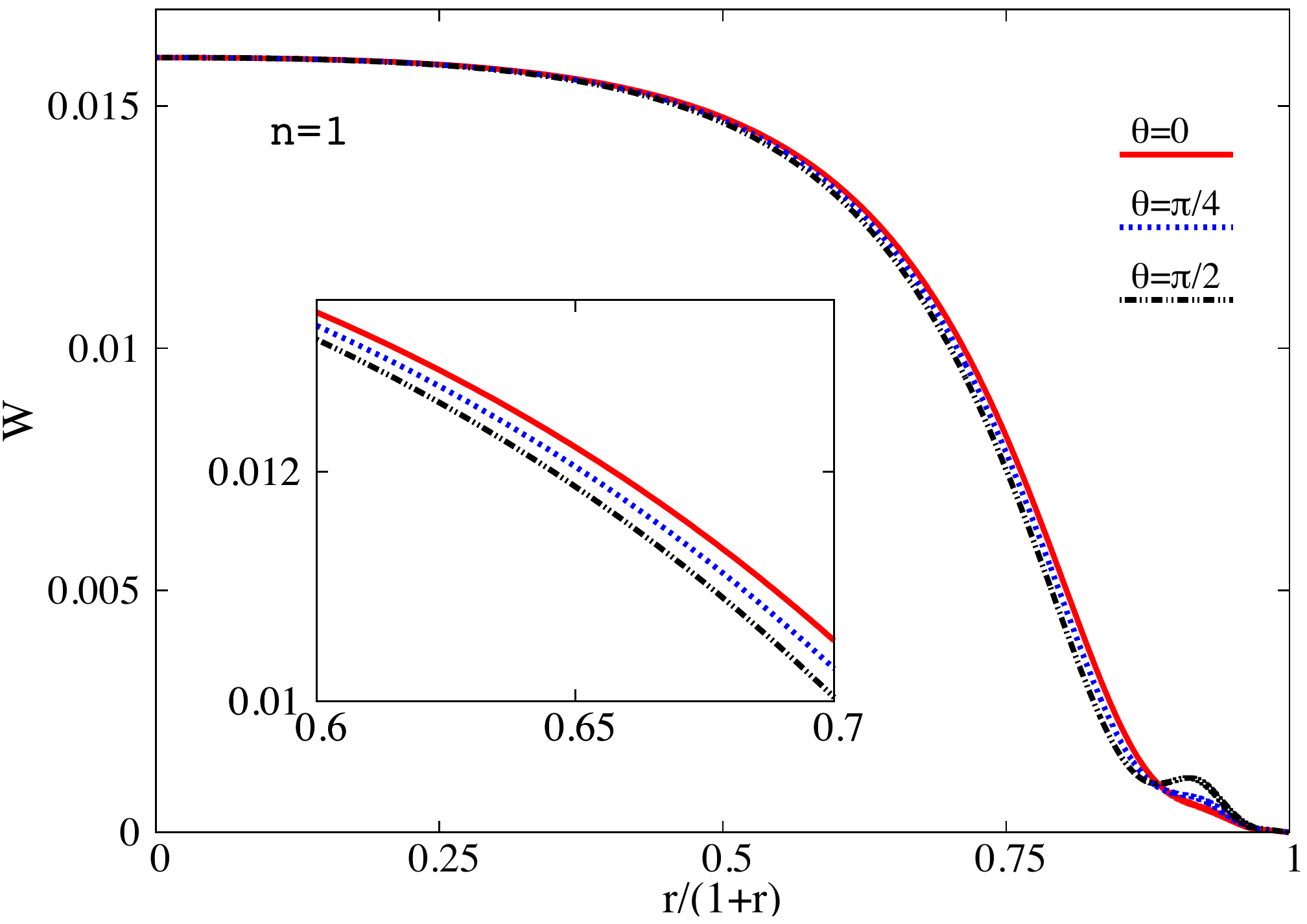}
\caption{Metric functions of two Proca stars: $n=0$ (left panels) and $n=1$ (right panels).}
\label{fig6}
\end{figure}

\begin{figure}
\centering
\includegraphics[width=.485\linewidth]{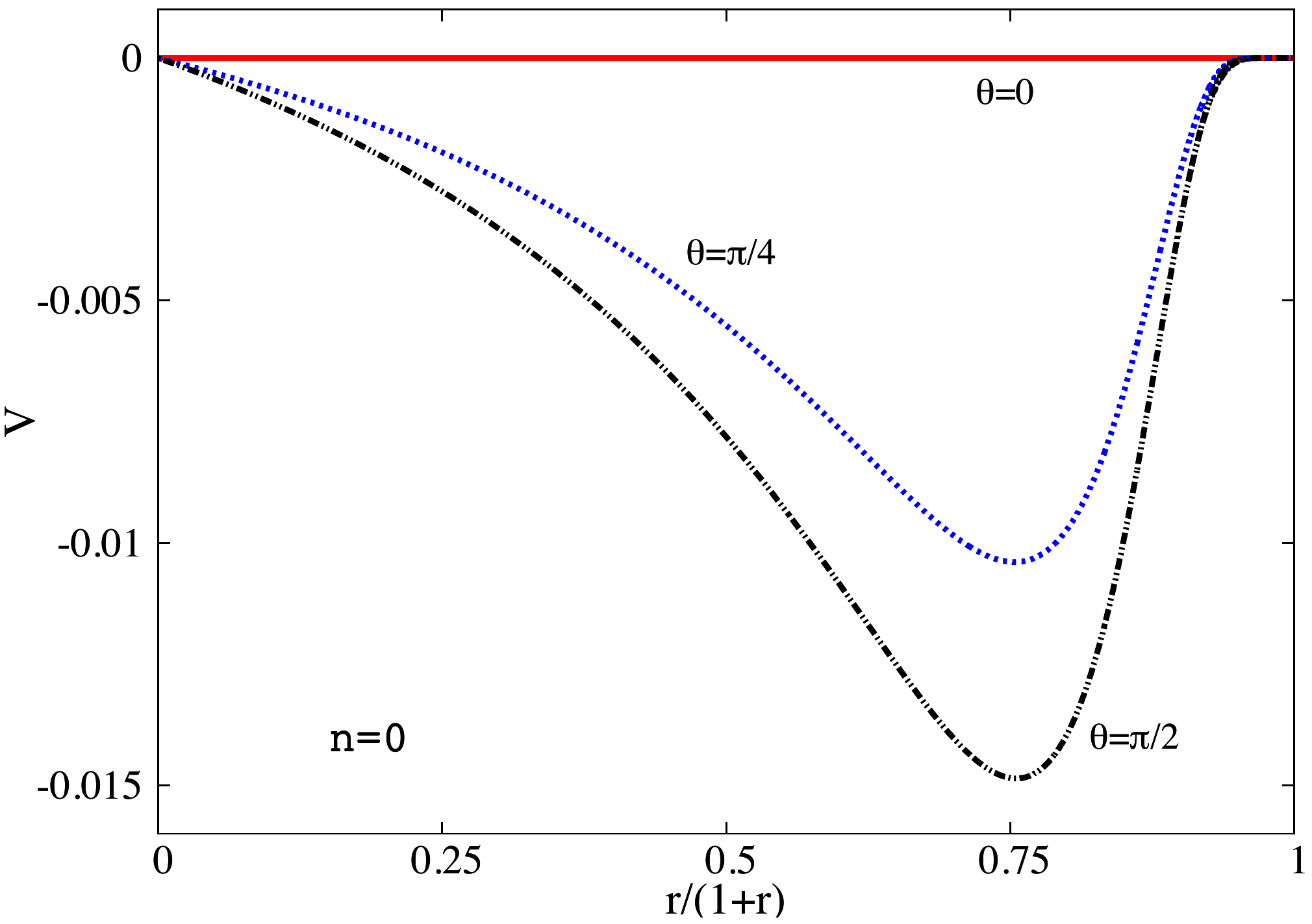}
\includegraphics[width=.485\linewidth]{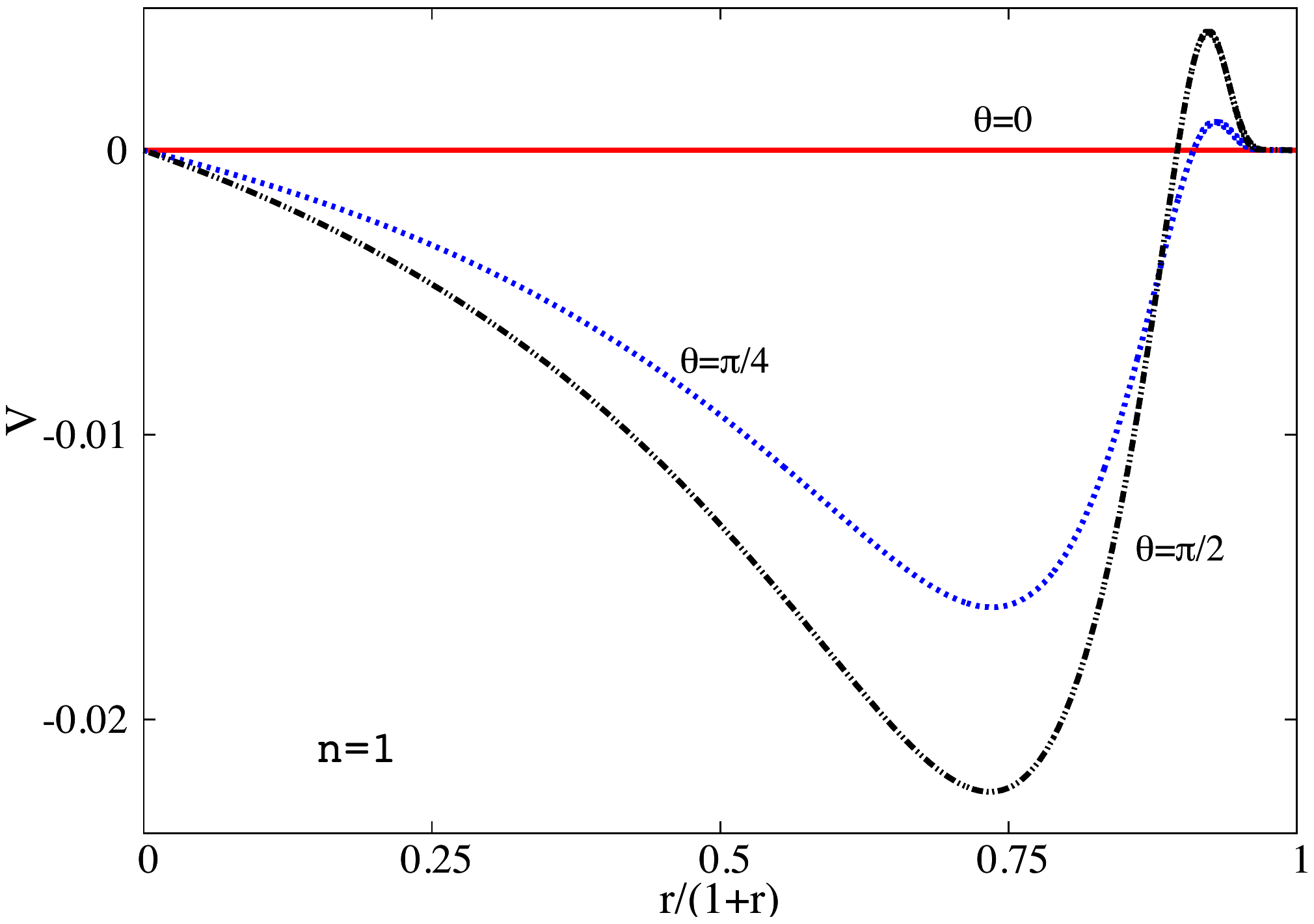}
\includegraphics[width=.485\linewidth]{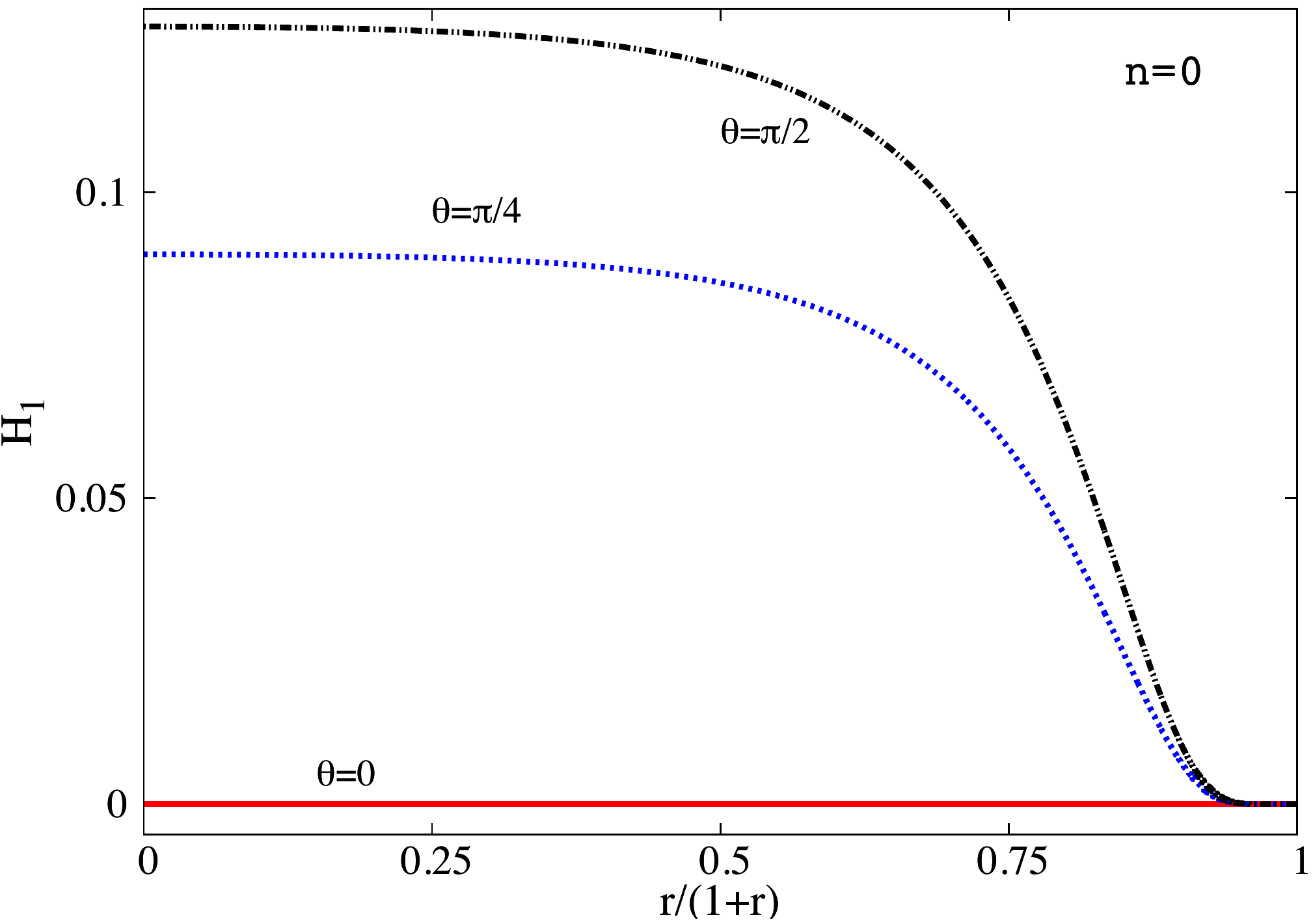}
\includegraphics[width=.485\linewidth]{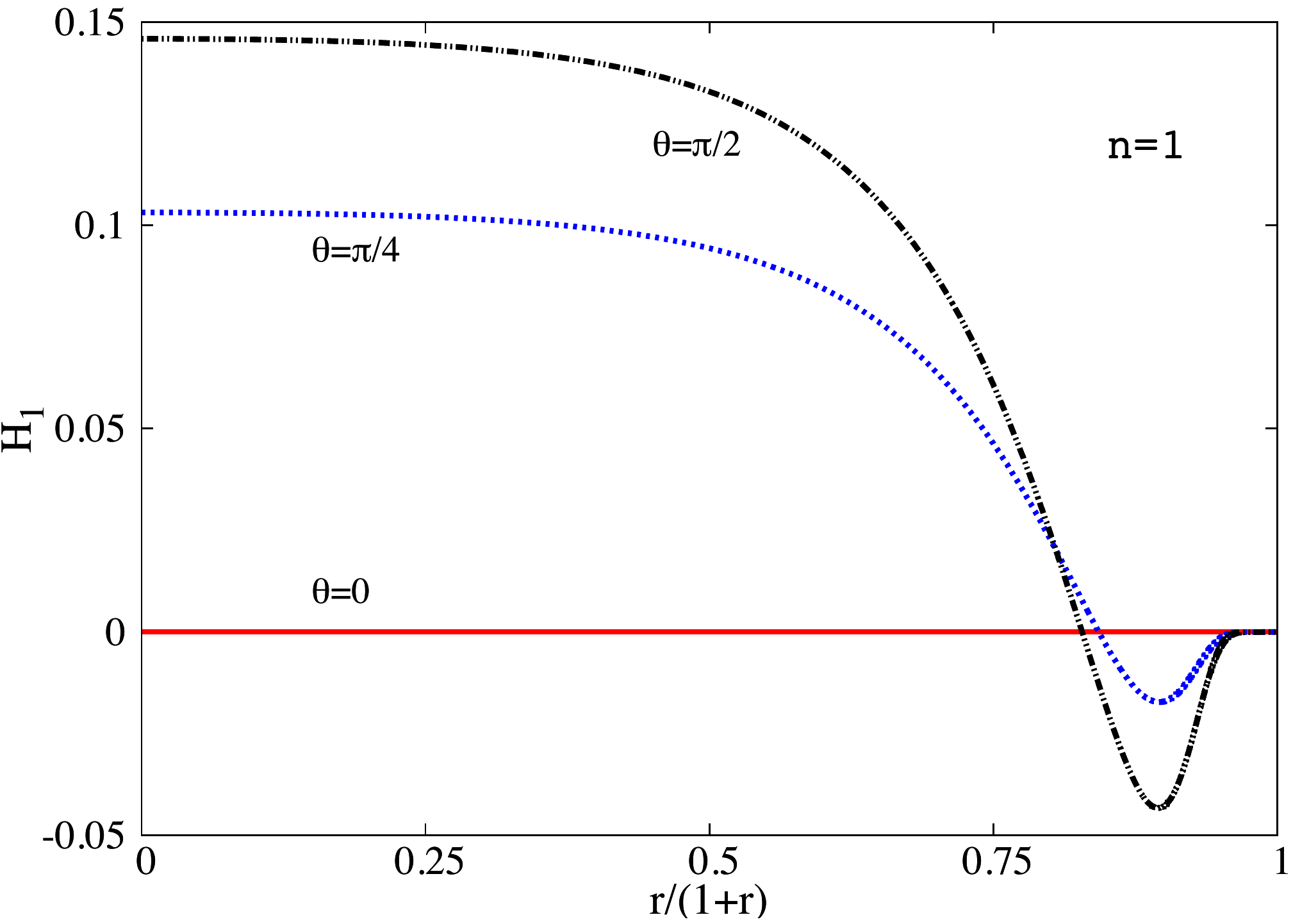}
\includegraphics[width=.485\linewidth]{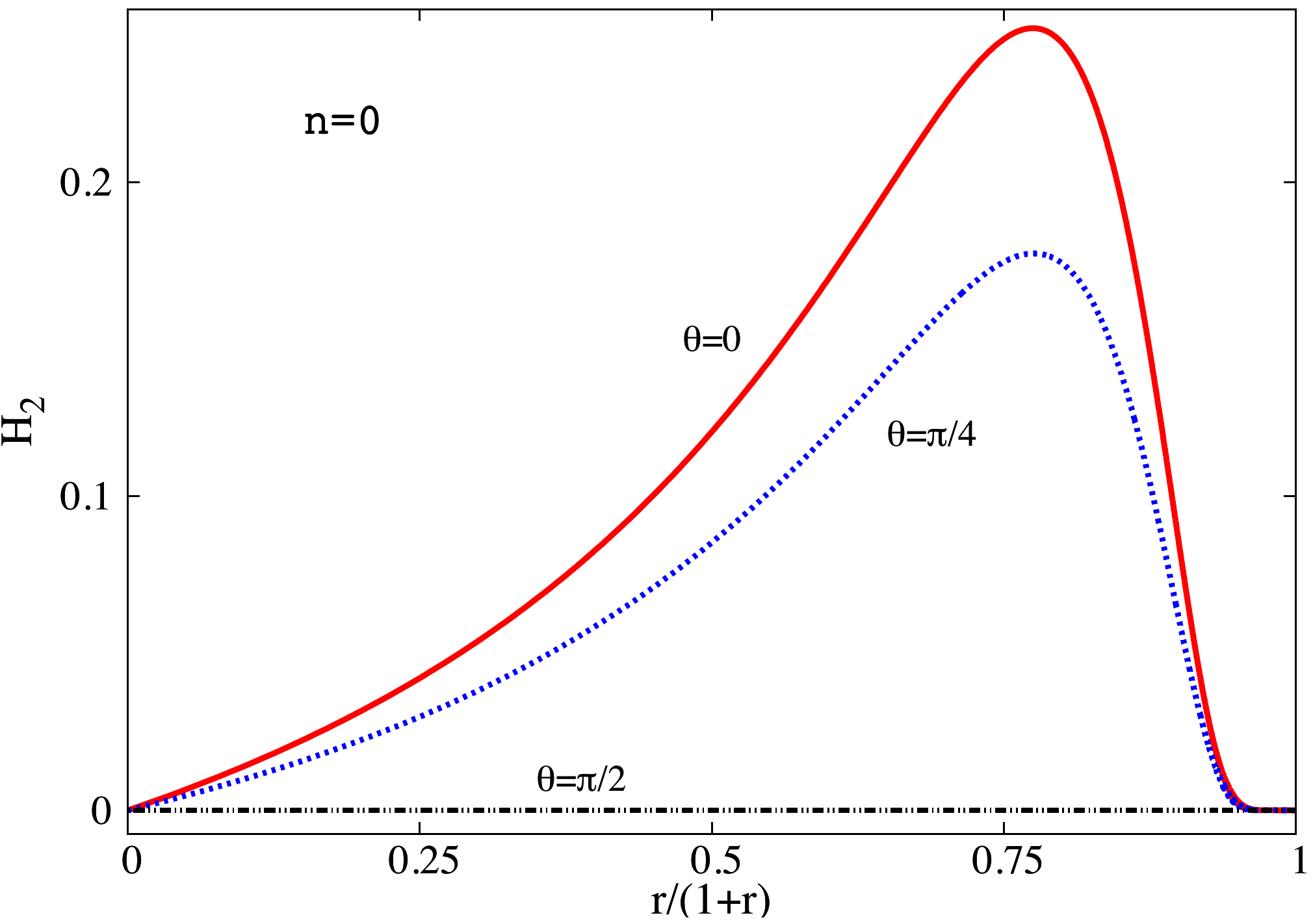}
\includegraphics[width=.485\linewidth]{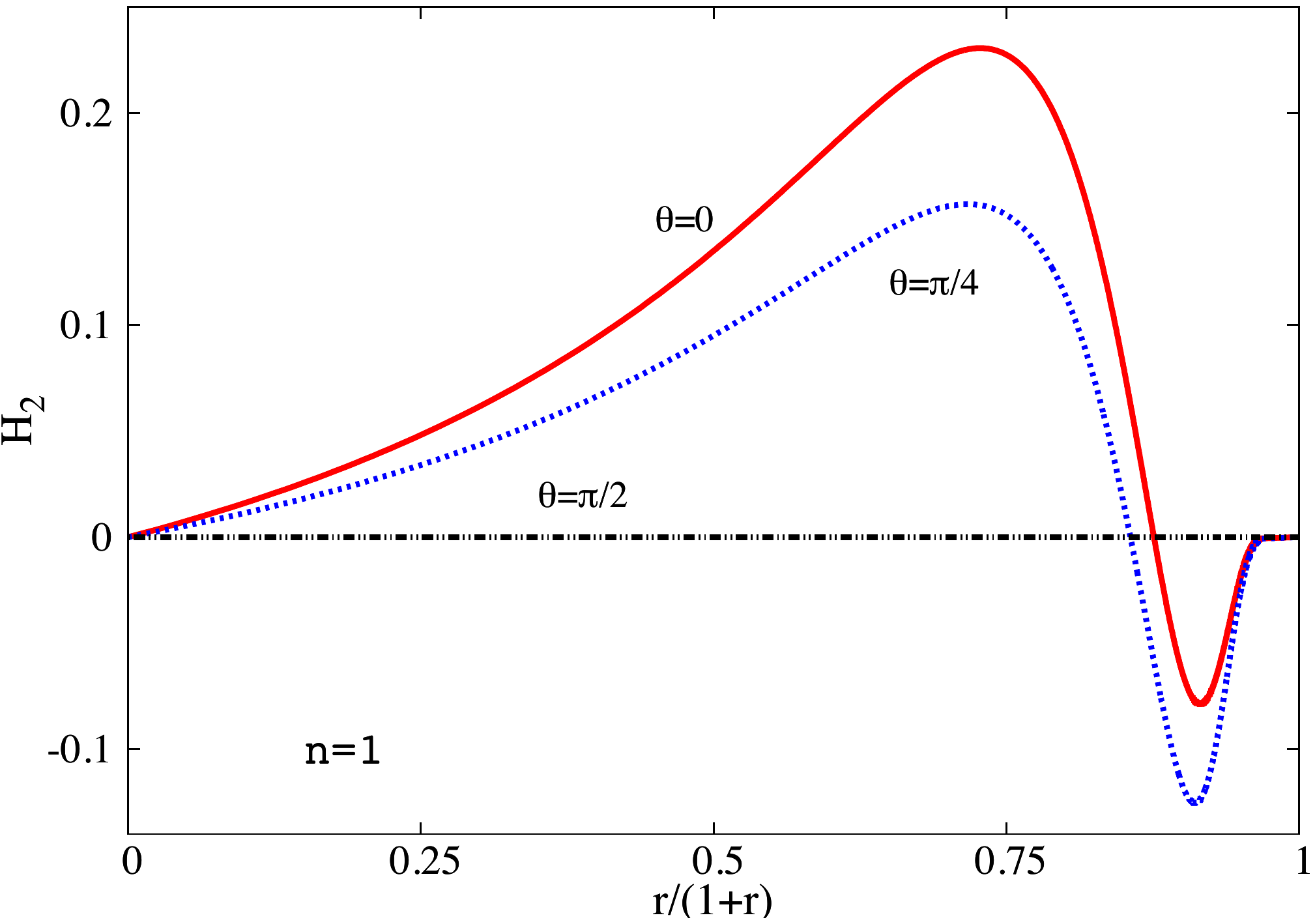}
\includegraphics[width=.485\linewidth]{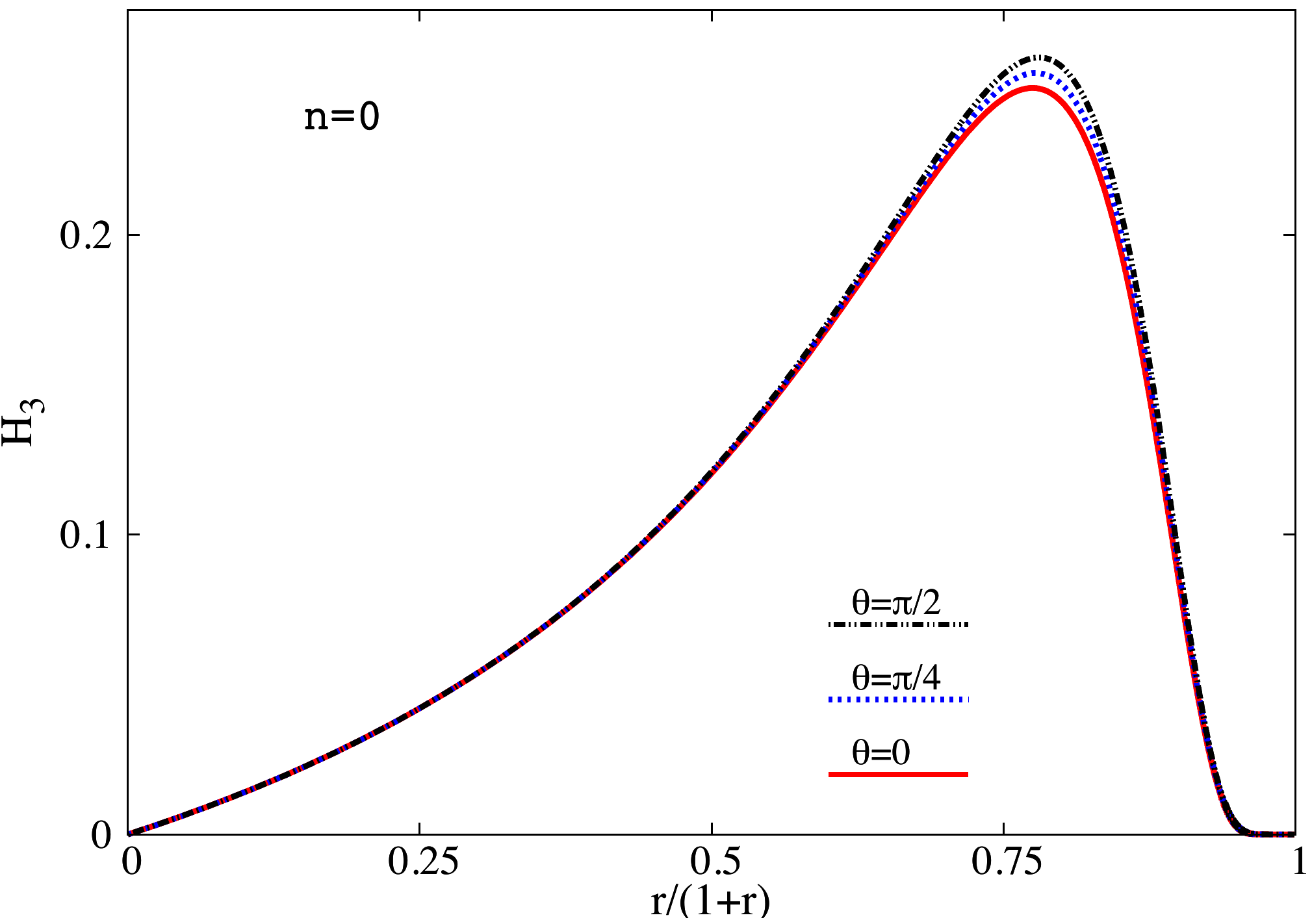}
\includegraphics[width=.485\linewidth]{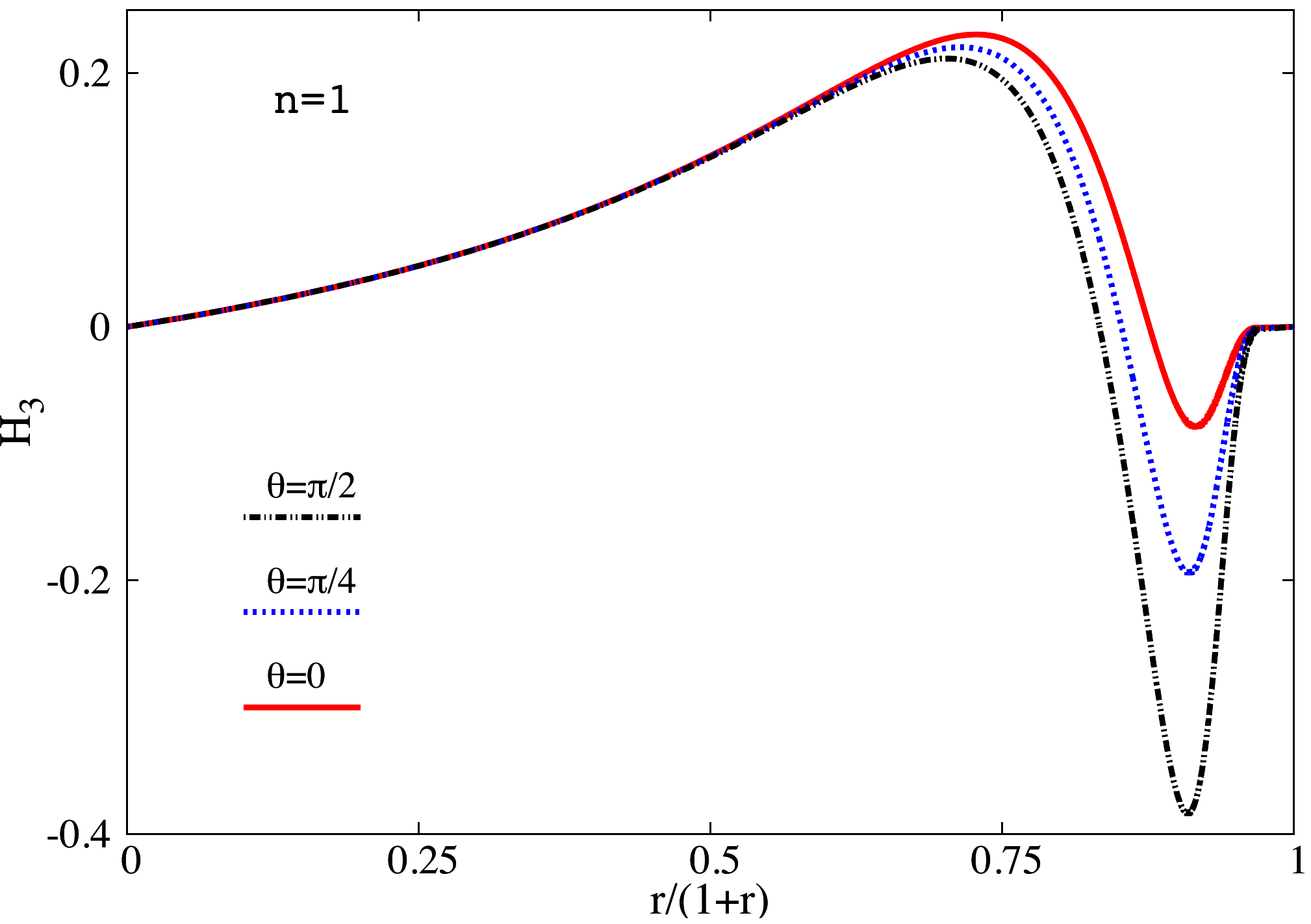}
\caption{Same as in \autoref{fig6}, but for the Proca potential functions.}
\label{fig7}
\end{figure}

\begin{figure}
\centering
\includegraphics[width=.483\linewidth]{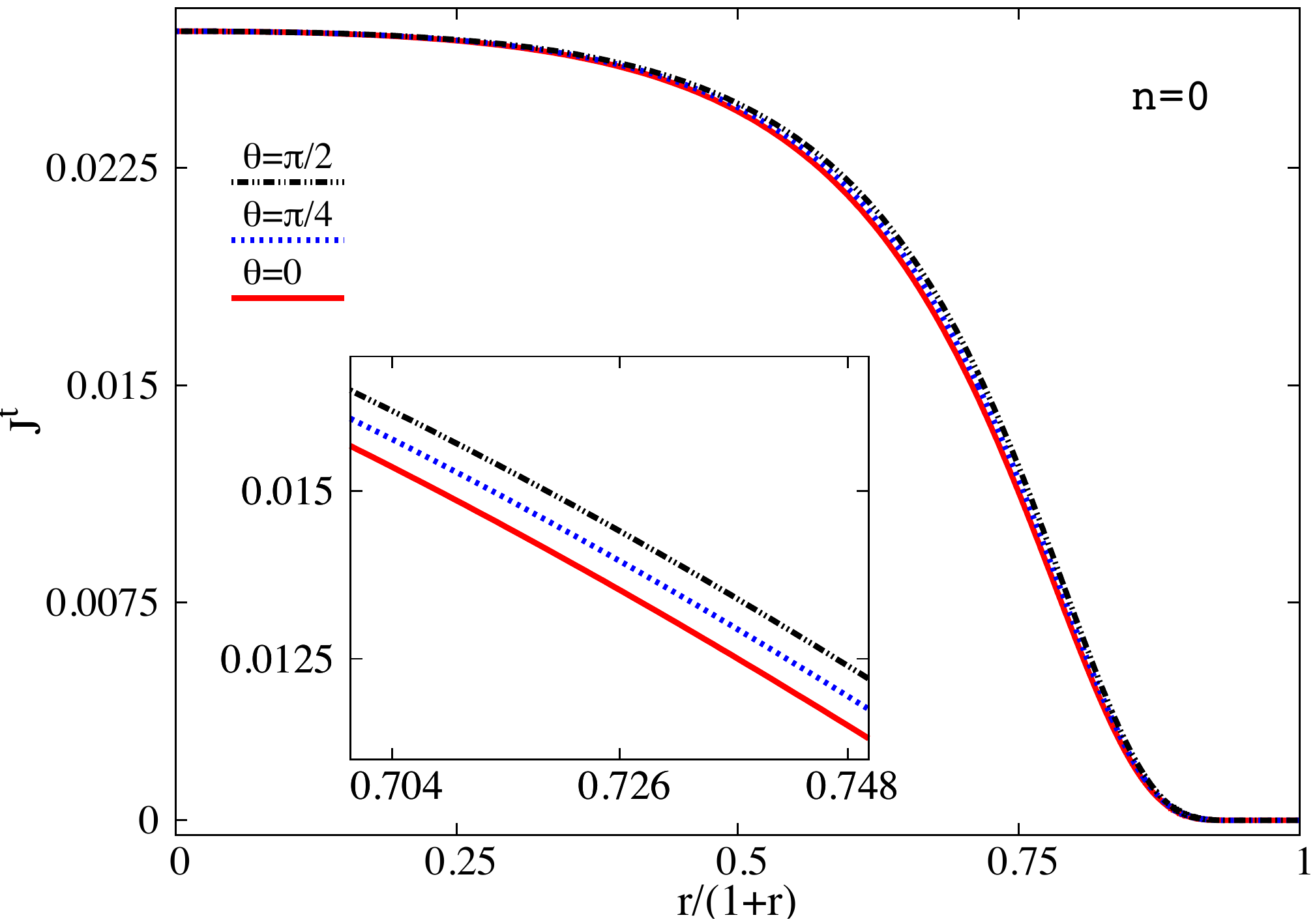}
\includegraphics[width=.483\linewidth]{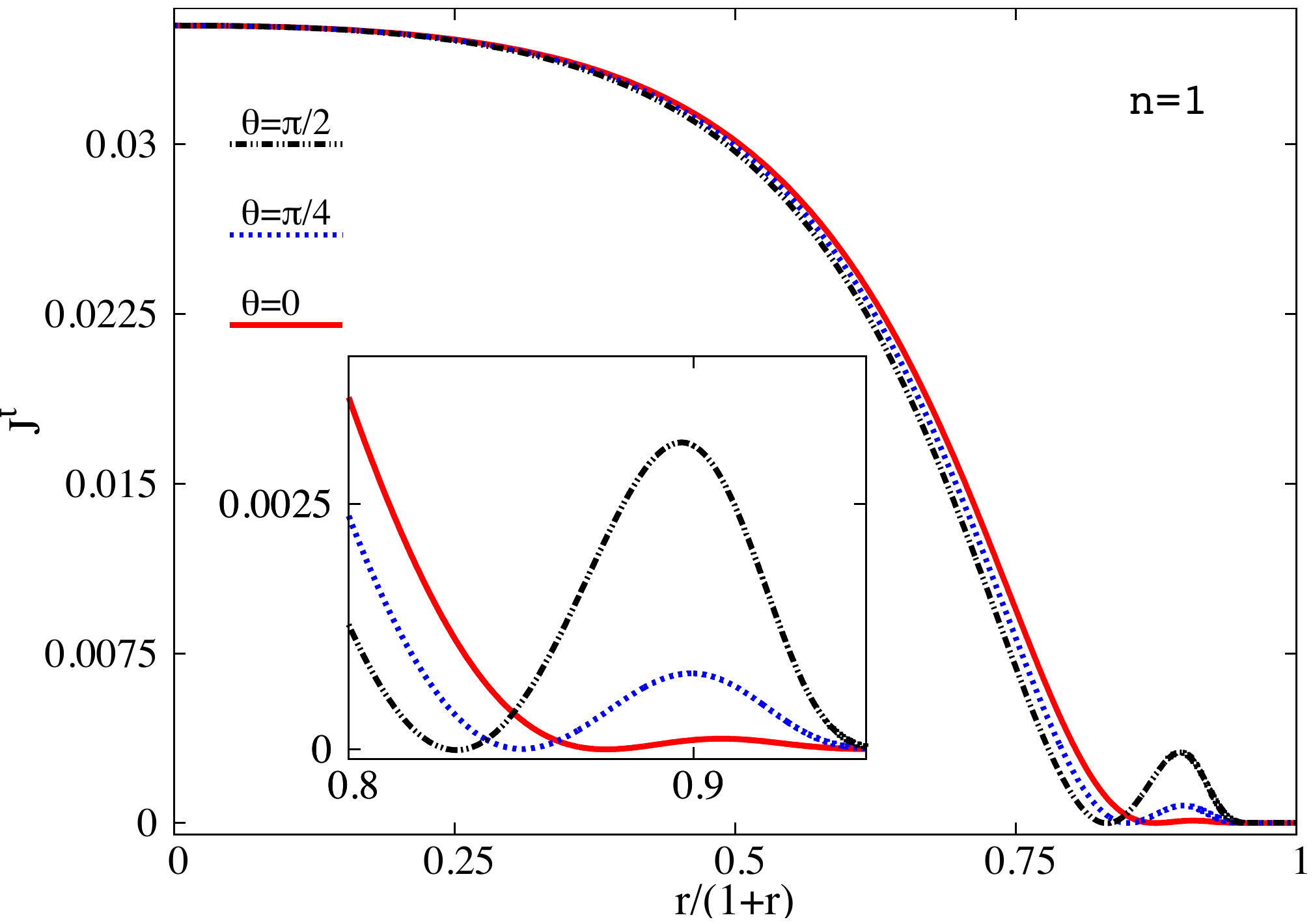}
\includegraphics[width=.483\linewidth]{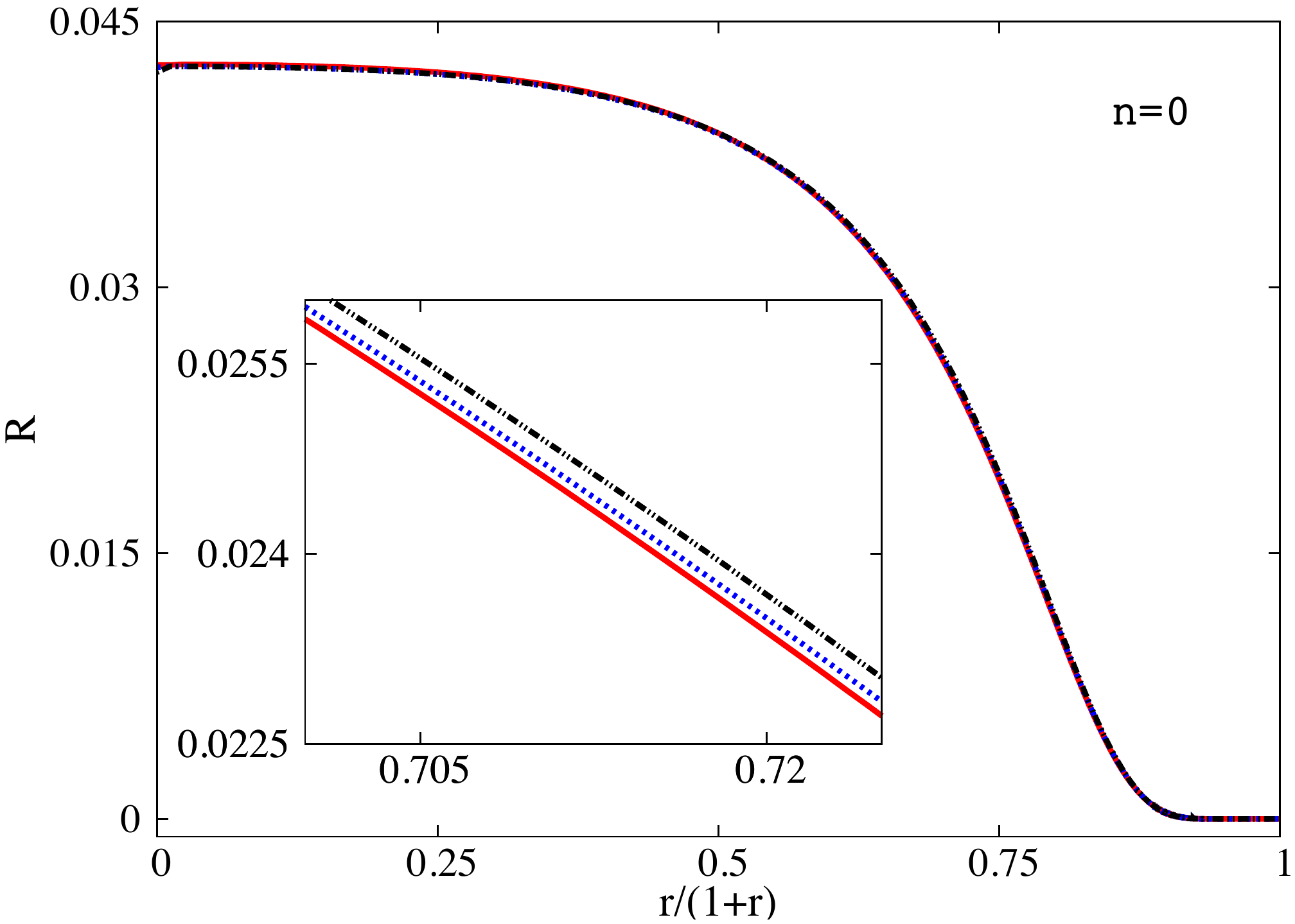}
\includegraphics[width=.483\linewidth]{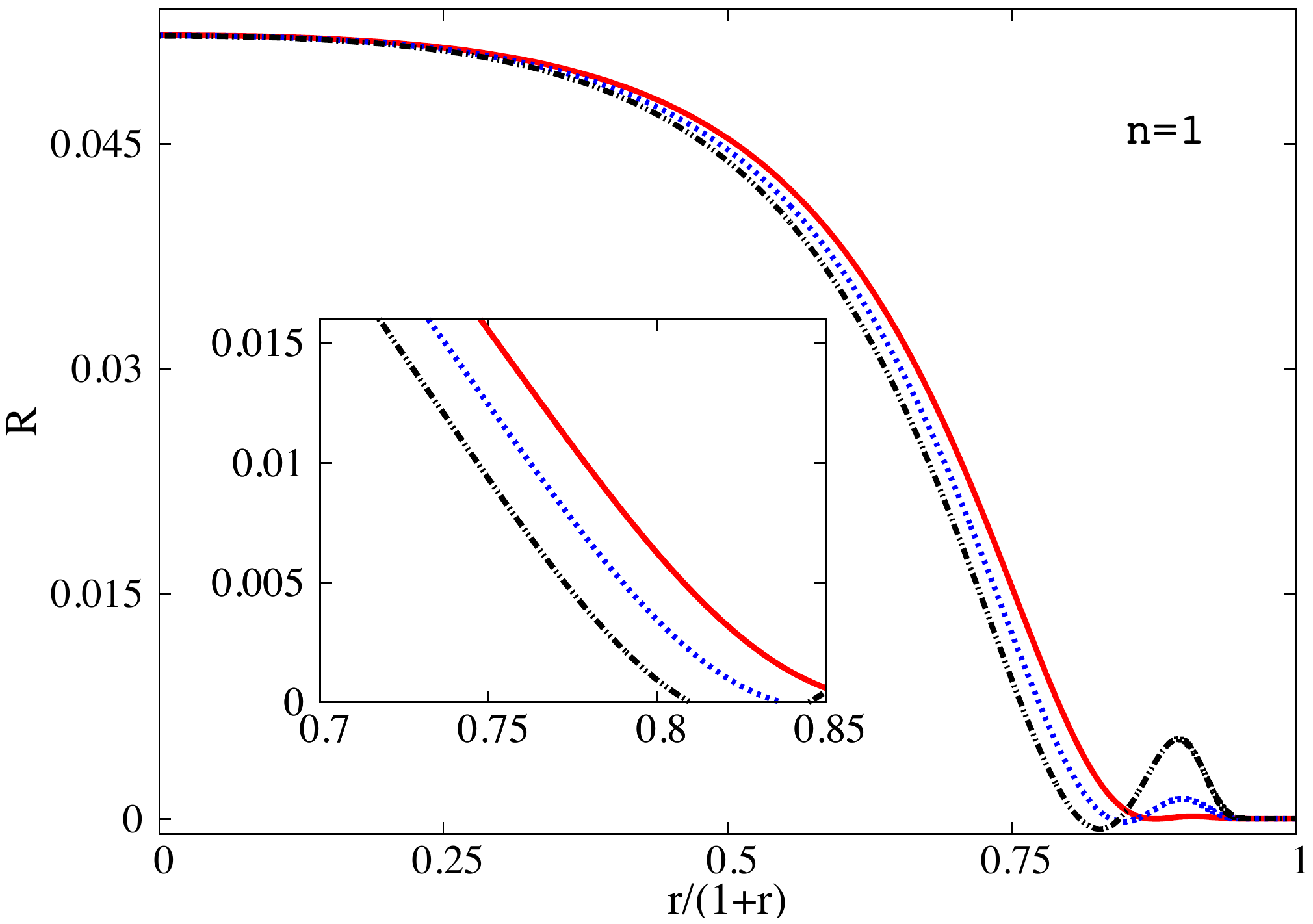}
\includegraphics[width=.483\linewidth]{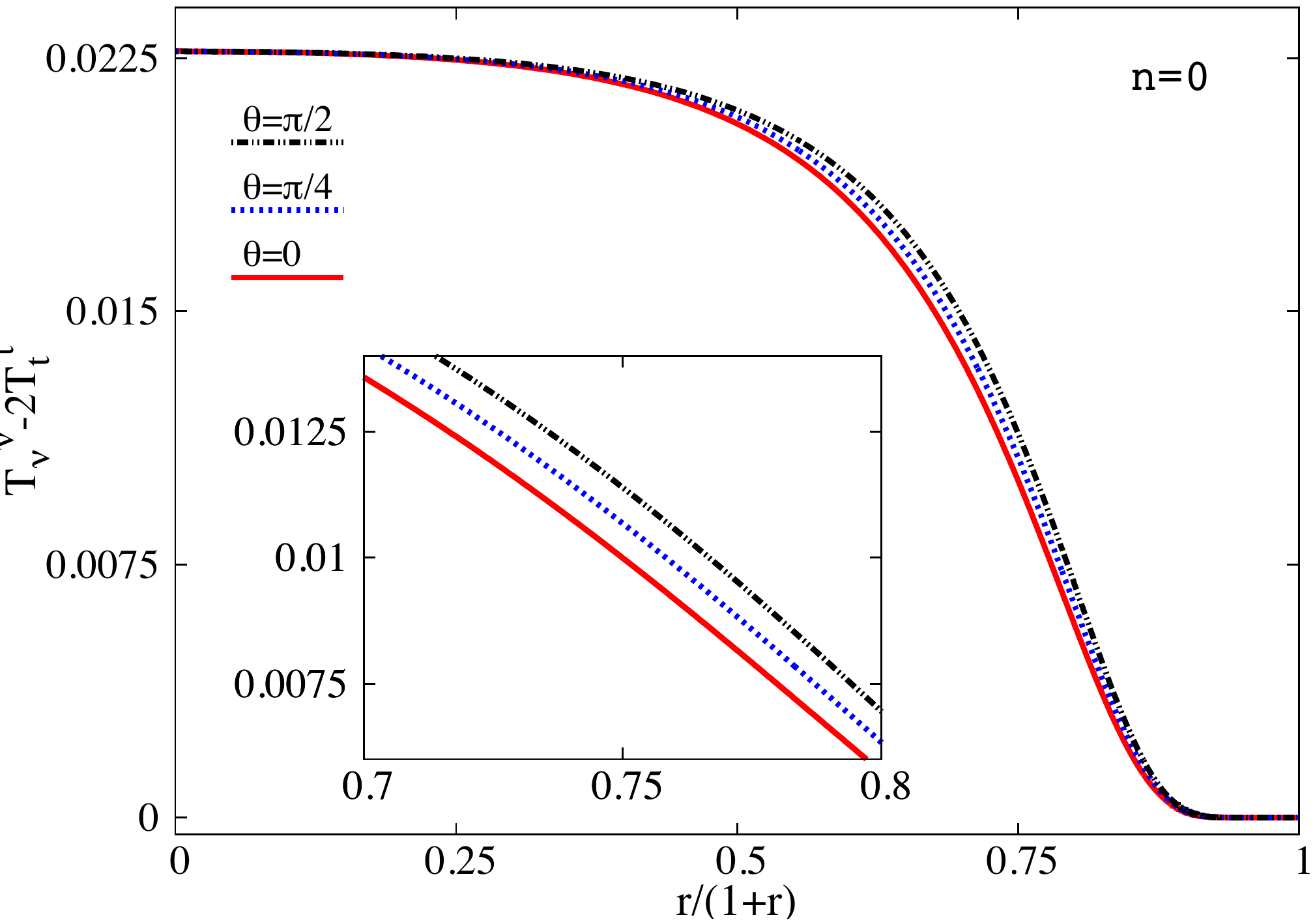}
\includegraphics[width=.483\linewidth]{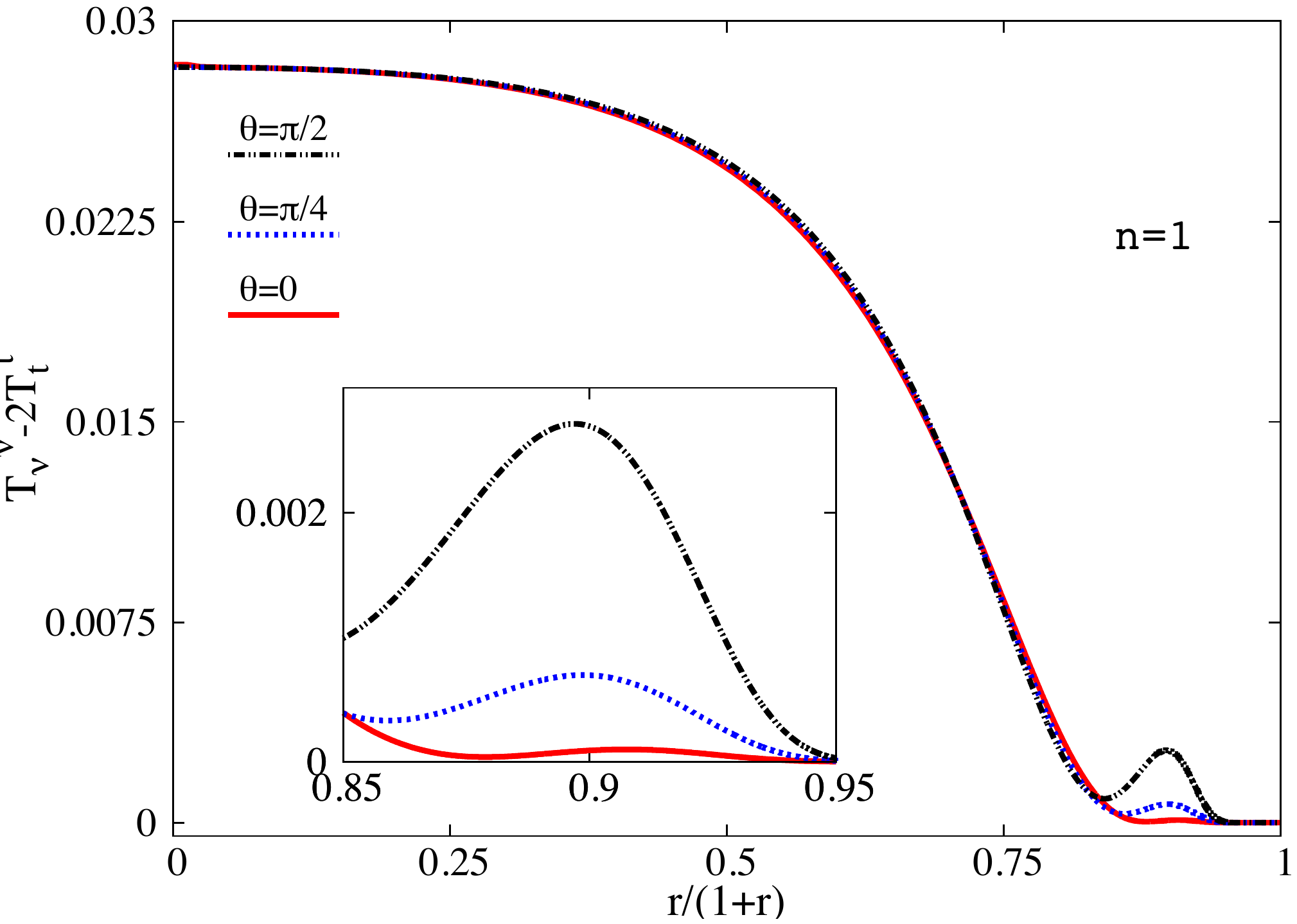}
\includegraphics[width=.483\linewidth]{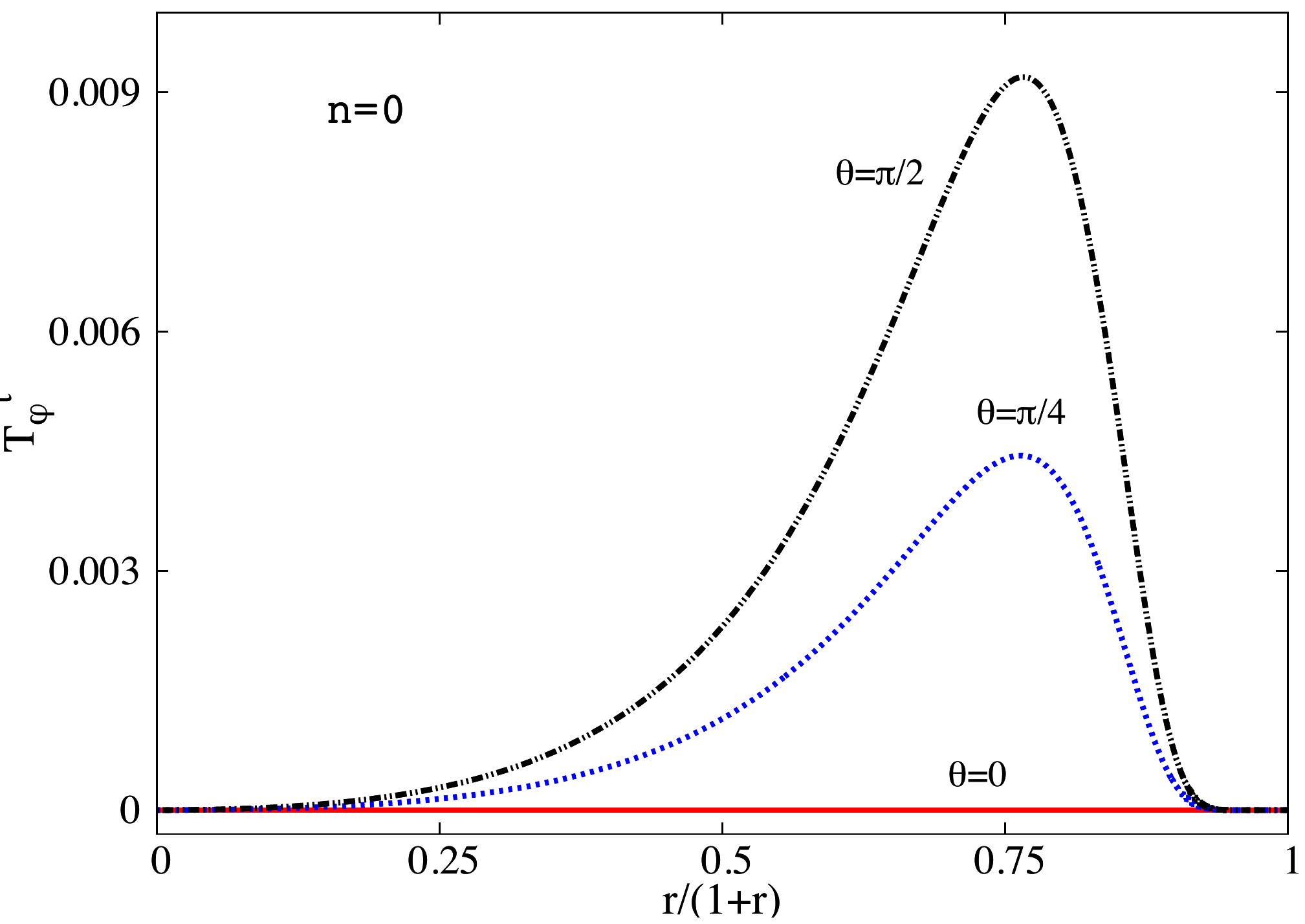}
\includegraphics[width=.483\linewidth]{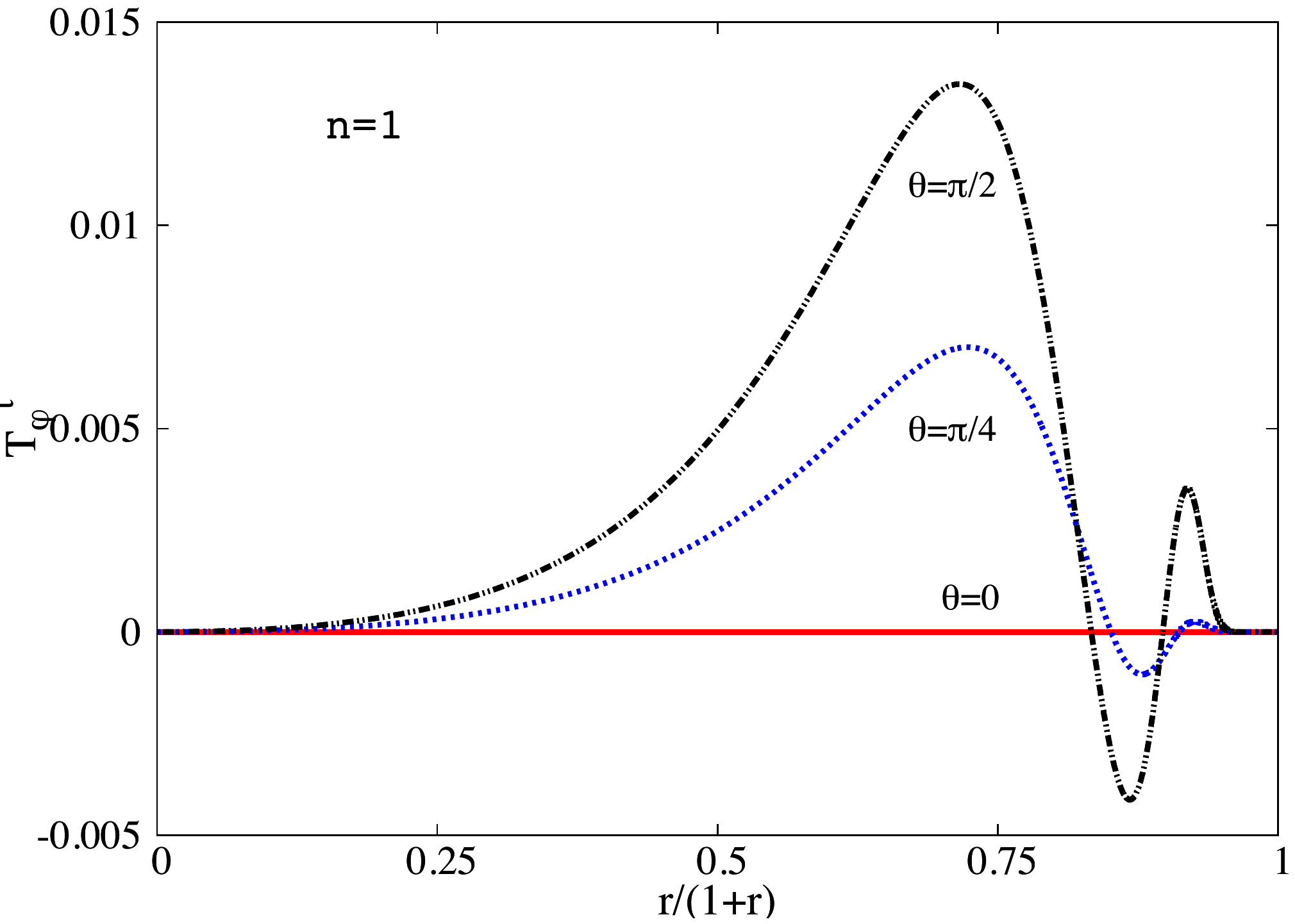}
\caption{Same as in \cref{fig6,fig7}, but for some physical quantities.}
\label{fig8}
\end{figure}

In \autoref{fig6} we compare the metric functions of the two illustrative Proca stars in terms of a compactified radial coordinate, to have an overview of the whole radial domain, and for three different $\theta$-values. Whereas in the fundamental state all metric functions are rather smooth and monotonic, in the first excited state there is some extra structure, mostly noticeable along the equatorial plane ($\theta=\pi/2$). The metric function $W$, in particular, is no longer monotonic. The nodeless $vs.$ nodeful structure of the $n=0$ $vs.$ $n=1$ spinning Proca stars becomes evident in \autoref{fig7}. One observes, in particular, that all four Proca potential functions have the same number of nodes, $n=0$ or $n=1$, for each star. Moreover, the temporal and radial component of the potential have a trivial structure along the $\theta=0$ symmetry axis. Finally, the extra structure of the excited states becomes clear when analysing more invariant quantities, such as the Noether charge density, the Ricci curvature scalar and the Komar energy density, that are exhibited in \autoref{fig8}.

The Noether charge results from the global $U(1)$ symmetry of Eq.~\eqref{epaction}, which is invariant under the global transformation 
$A_\beta\rightarrow e^{i\chi}A_\beta$, where $\chi$ is a constant. Thus, a conserved 4-current exists
\begin{equation}
\label{current}
J^\alpha=\frac{i}{2}\left[\bar{F}^{\alpha \beta}A_\beta-F^{\alpha\beta}\bar{A}_\beta\right] \ , \qquad  \nabla_\alpha J^\alpha=0 \ .
\end{equation}
The Noether charge, $Q$, which is interpreted as the particle number (indeed becomes the particle number upon quantisation), is obtained integrating the time component of this current on a spacelike hypersurface $\Sigma$:
\begin{equation}
Q=\int_\Sigma \D^3x~J^t \ .
\end{equation}
The Nother charge density is thus $J^t$, which is plotted in the top panel of \autoref{fig8}. 

The Komar energy density results from the Komar mass computed at infinity. Using Gauss's law, one relates the latter with the horizon Komar mass $M_\text{H}$ (in the cases which have a horizon) and a volume integral on a spacelike hypersurface between the horizon and infinity. One obtains~\cite{Herdeiro:2015gia} (where $k^\alpha$ is the asymptotic timelike Killing vector field):
\begin{equation}
M=M_\text{H}-2\int_{\Sigma}\D S_{\alpha}\left(T^\alpha_\beta k^\beta-\frac{1}{2}Tk^\alpha\right) \equiv M_\text{H}+M^{(\mathcal{P})} \ ,
\end{equation}
where $M^{(\mathcal{P})} $ is the energy contained in the Proca field (outside a horizon, in case there is one):
\begin{equation}
M^{({\cal P})}\equiv - \int_{\Sigma} \D r \D\theta\D\varphi \sqrt{-g}(2T_t^t-T_\alpha^\alpha)  \ .
\end{equation}
The integrand is the Proca energy density, which is plotted in the second from bottom panels of \autoref{fig8}. A similar analysis can be done for the Komar angular momentum density, showing that the Komar angular momentum density is $T^t_\varphi$~\cite{Herdeiro:2015gia}, plotted in the bottom panels of \autoref{fig8}.

All invariant quantities in~\autoref{fig8} demonstrate that whereas the $n=1$ stars have a Saturn-like morphology -- which was observed in~\cite{Herdeiro:2015gia} --, with the energy density or the particle number having a global maximum at the centre and a local maximum at some radial distance,  the fundamental states are spheroidal. This contrasts with the toroidal shape of the fundamental spinning scalar boson stars~\cite{Schunck:1996he}. This morphological difference was argued to be related to the different dynamical stability of the fundamental states of scalar/vector spinning bosonic stars~\cite{Sanchis-Gual:2019ljs}.

%%%%%%%%%%%%%%%

We now turn to hairy BHs. The single most important observation concerning hairy BHs in this model is that they can be quite Kerr-like or strongly non-Kerr-like. This follows from the fact that the hairy BHs interpolate between the Kerr family and a solitonic limit (Proca stars) whose properties and phenomenology can be quite different from Kerr. So, here we shall focus on two illustrative solutions that exemplify this range of possibilities. 

First, we consider an example of a fairly Kerr-like BH with Proca hair, with $n=0$. It is chosen in the region where these BHs matched the endpoint of the dynamical of evolutions reported in~\cite{East:2017ovw} -- see~\cite{Herdeiro:2017phl}. Moreover, within this region, it is chosen to be as hairy as those evolutions suggest a hairy BH can be, when forming dynamically from the superradiant instability of Kerr BHs. This hairy BH, labelled HBH$_{\bf 1}$, has\footnote{Do not confuse the dimensionless spin $j$ in this Section with the total angular momentum $j$ of \autoref{sec2}.} [all quantities in Eqs.~\eqref{hbh1} and \eqref{hbh2}  are given  in units of $\mu$, which was omitted]
\begin{equation}
\left[M,\frac{M_\text{H}}{M}; \ J, \frac{J_\text{H}}{J};\  j,j_\text{H}; \ \Omega_\text{H},r_\text{H}\right]_{\bf 1}=(0.239 ,0.905 ; \ 0.055, 0.607 ; \ 0.98 , 0.726   ; \  0.97  ,  0.3) \ ,
\label{hbh1}
\end{equation}
where $j\equiv J/M^2$ and   $j_\text{H}\equiv J_\text{H}/M_\text{H}^2$ are the dimensionless spin in terms of global and horizon quantities, respectively. 
Thus, HBH$_{\bf 1}$ has 9.5\% of its energy and 39.3\% of its spin outside the horizon. These were roughly the maximal values of extraction via superradiance observed in~\cite{East:2017ovw}. Also note that both $j$ and $j_\text{H}$ are smaller than unity; thus the hairy BH obeys the Kerr bound, both in terms of horizon and asymptotic quantities. It is known that spinning BHs with synchronised hair can violate the Kerr bound -- see $e.g.$~\cite{Herdeiro:2014goa,Herdeiro:2015moa}.

Second, we consider an example of a fairly non-Kerr-like BH with Proca hair, with $n=0$. This hairy BH, labelled HBH$_{\bf 2}$, has
\begin{equation}
\left[M,\frac{M_\text{H}}{M}; \ J, \frac{J_\text{H}}{J};\  j,j_\text{H}; \ \Omega_\text{H},r_\text{H}\right]_{\bf 2}=(0.501 ,0.231 ; \ 0.392, 0.022 ; \  1.56 ,  0.642 ; \ 0.93 ,  0.2) .
\label{hbh2}
\end{equation}
Thus, HBH$_{\bf 2}$ has 76.9\% of its energy and 97.8\% of its spin outside the horizon. Moreover, this BH violates the Kerr bound in terms of asymptotic quantities, since $j=1.56>1$, but not in terms of horizon quantities. In this sense it behaves more like a star. Both these solutions are marked with triangles, and labelled with the corresponding numbers, in \autoref{fig3}.

\begin{figure}
\centering
\includegraphics[width=.485\linewidth]{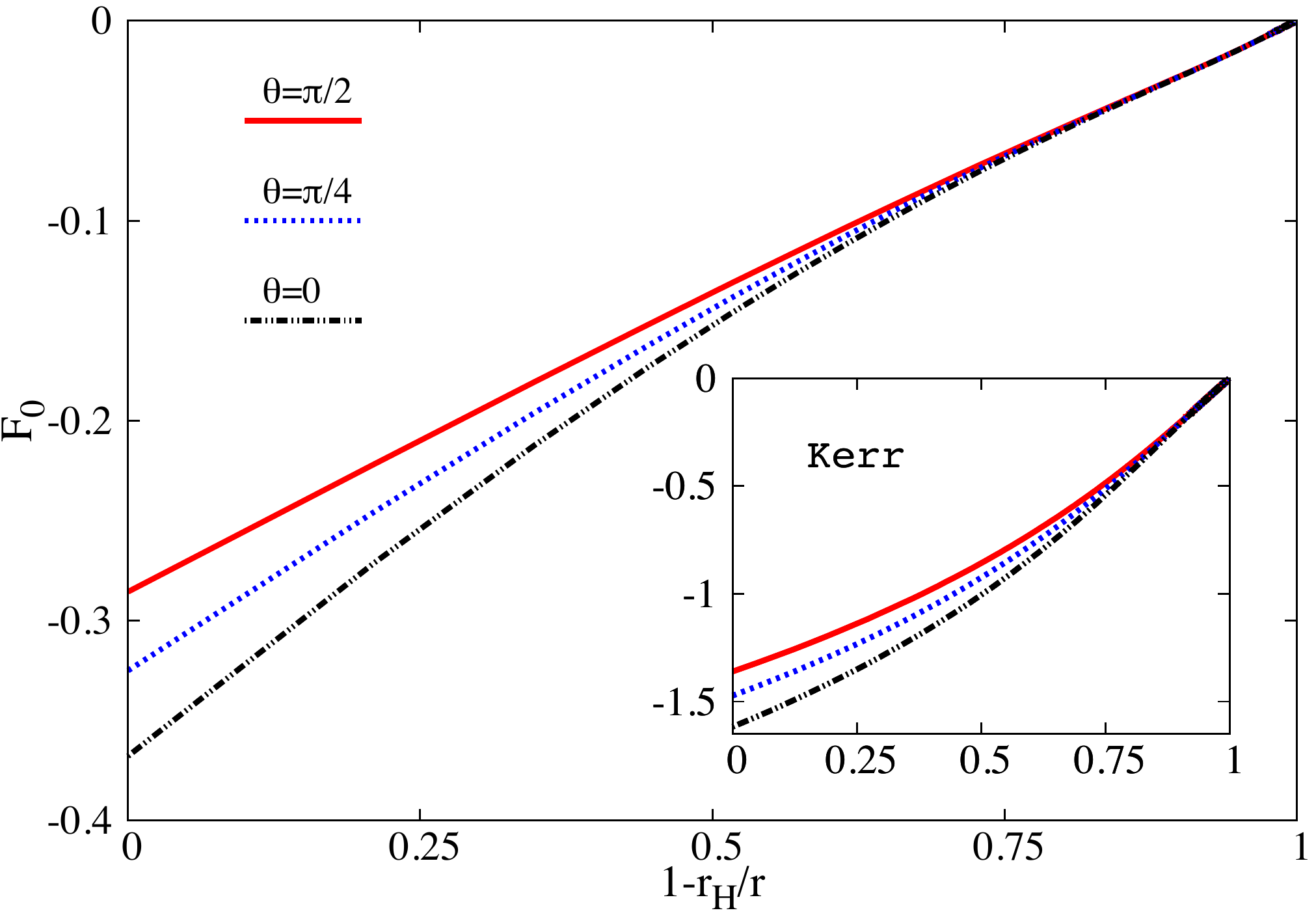}
\includegraphics[width=.485\linewidth]{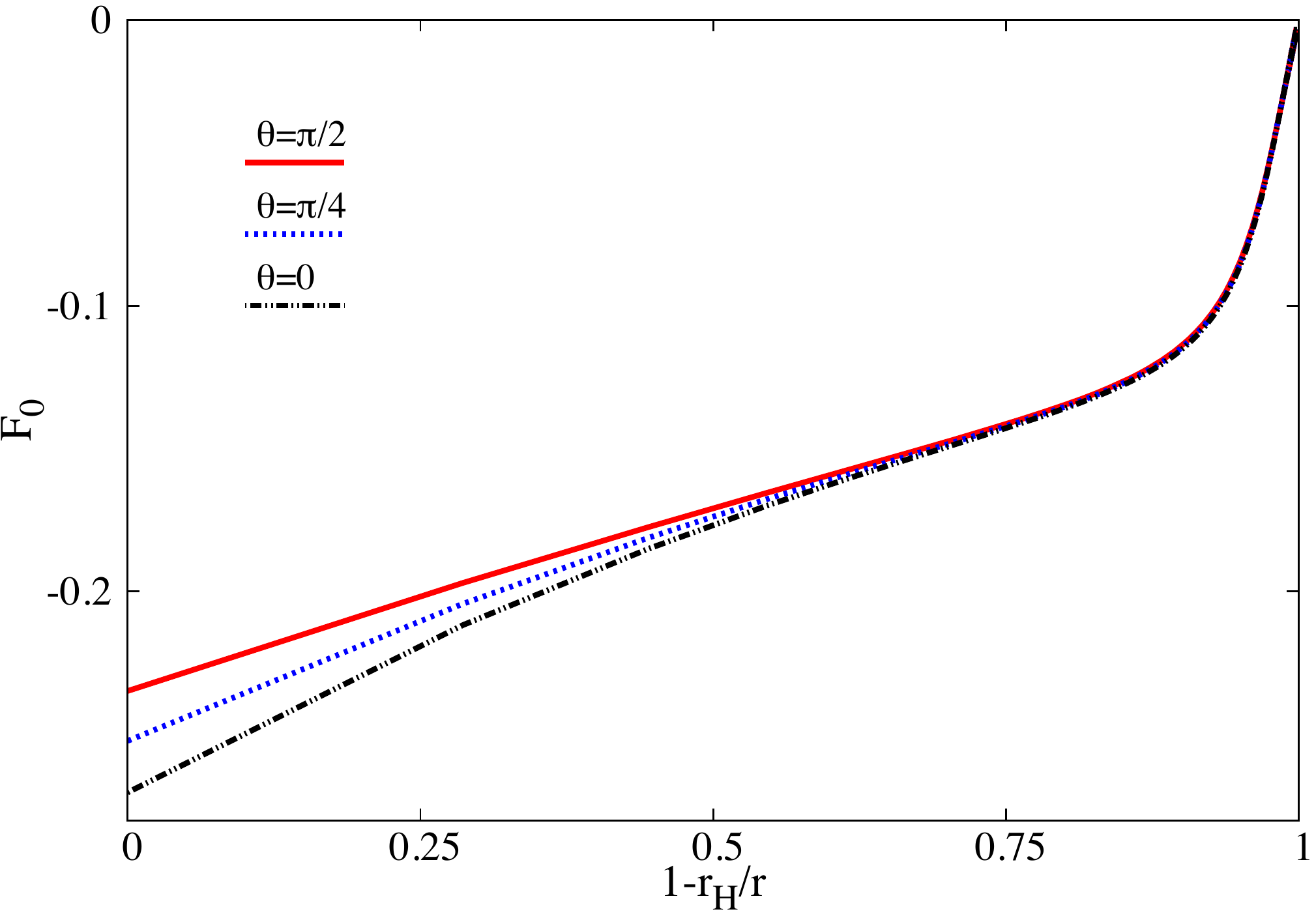}
\includegraphics[width=.485\linewidth]{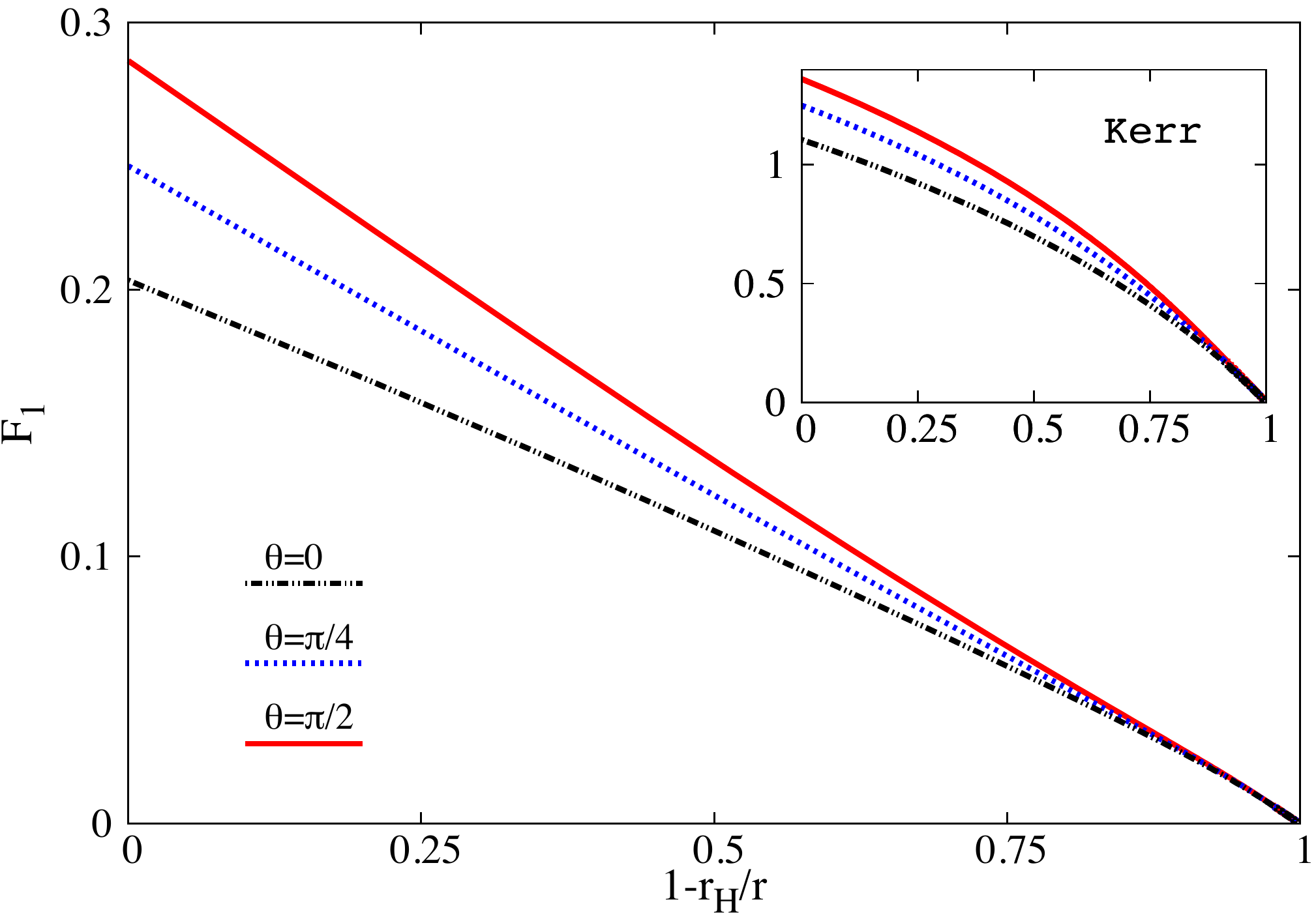}
\includegraphics[width=.485\linewidth]{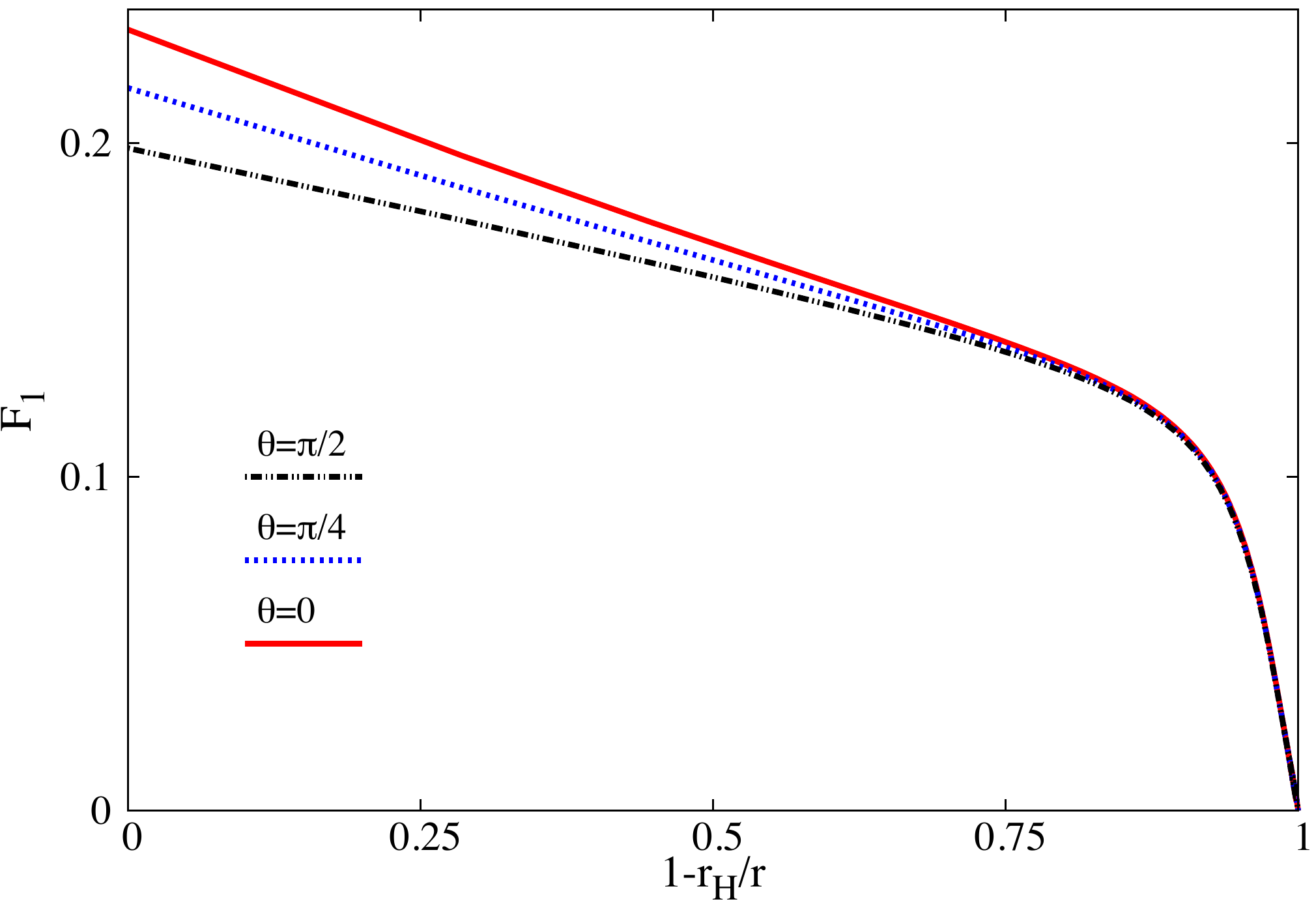}
\includegraphics[width=.485\linewidth]{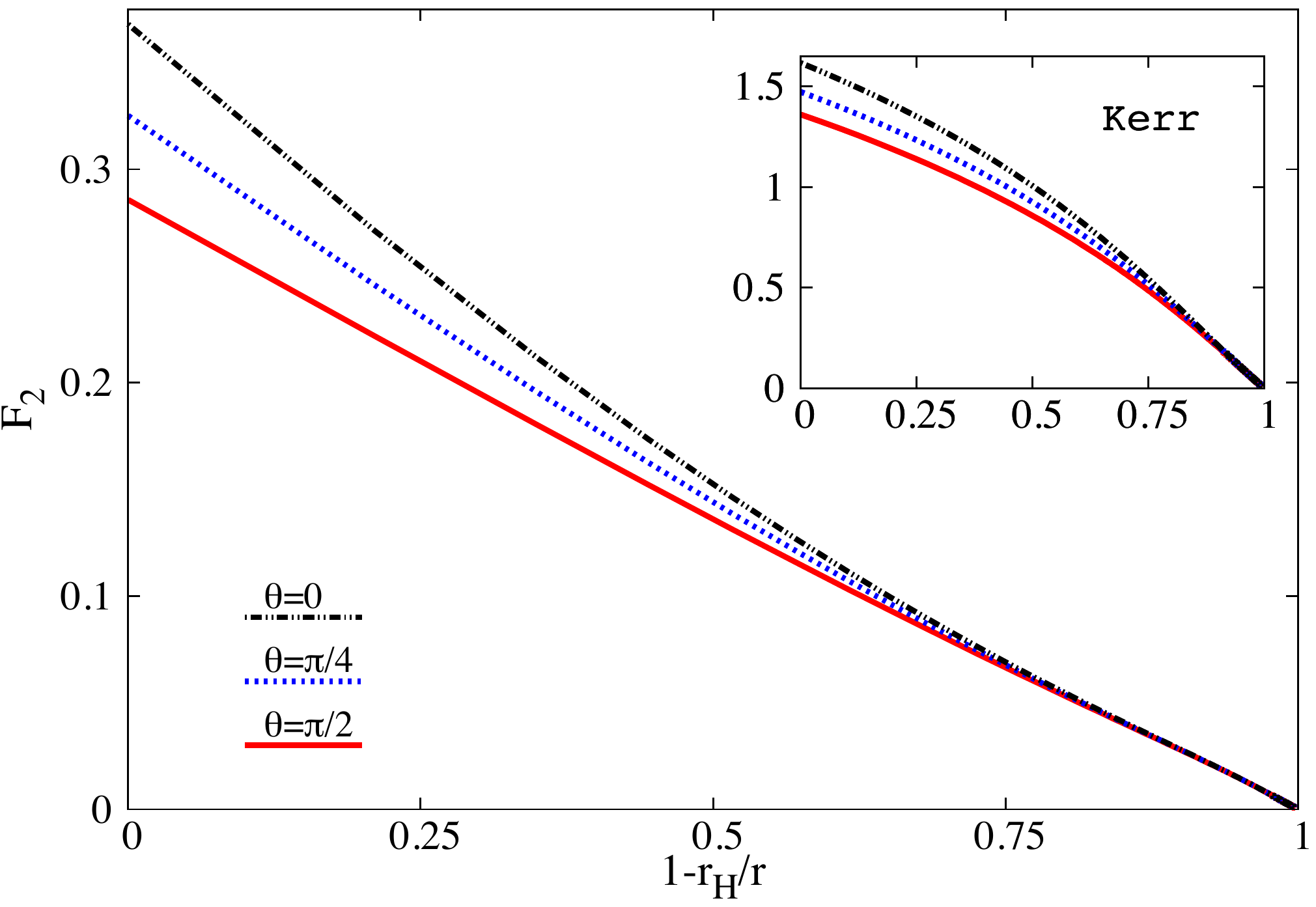}
\includegraphics[width=.485\linewidth]{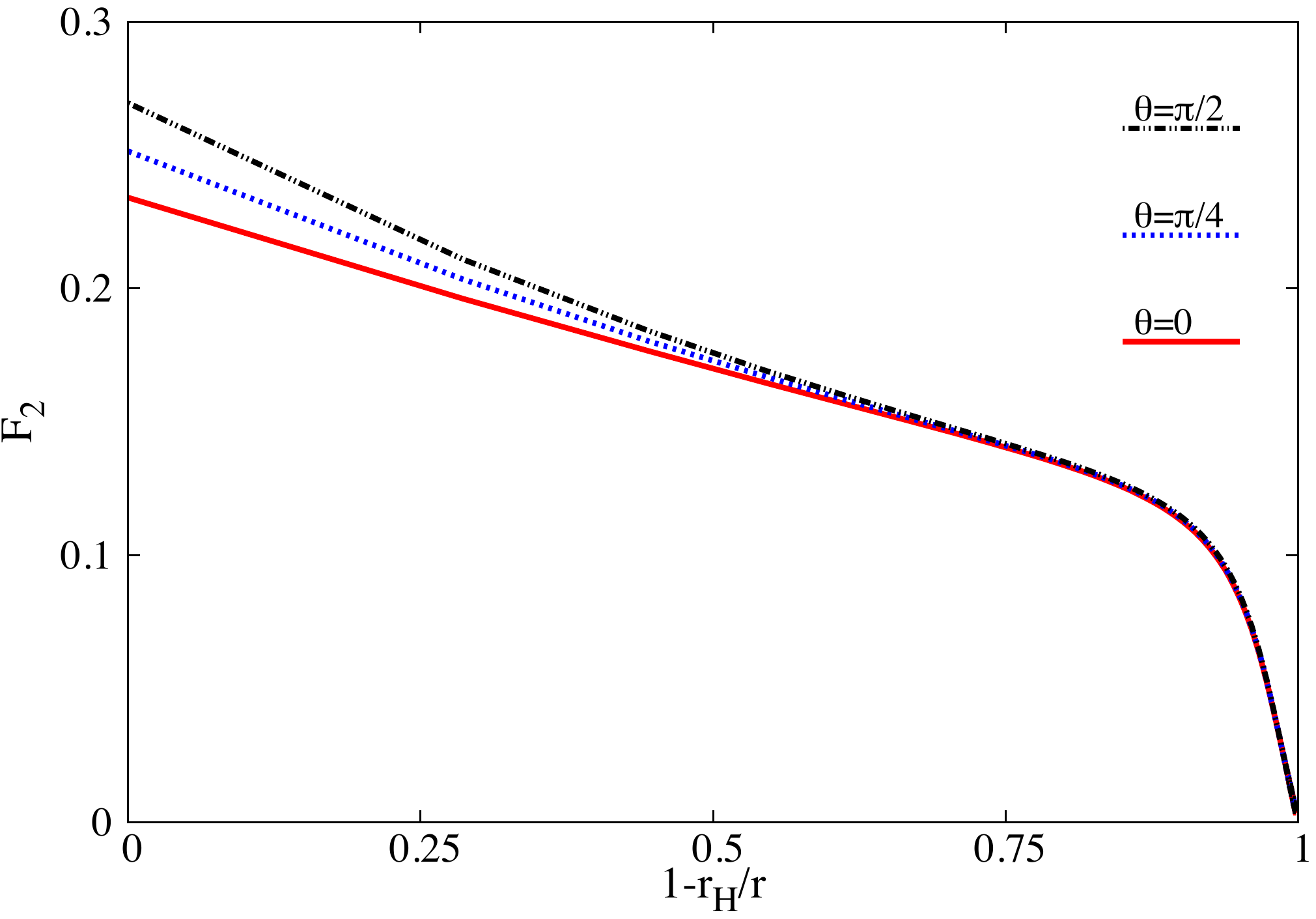}
\includegraphics[width=.485\linewidth]{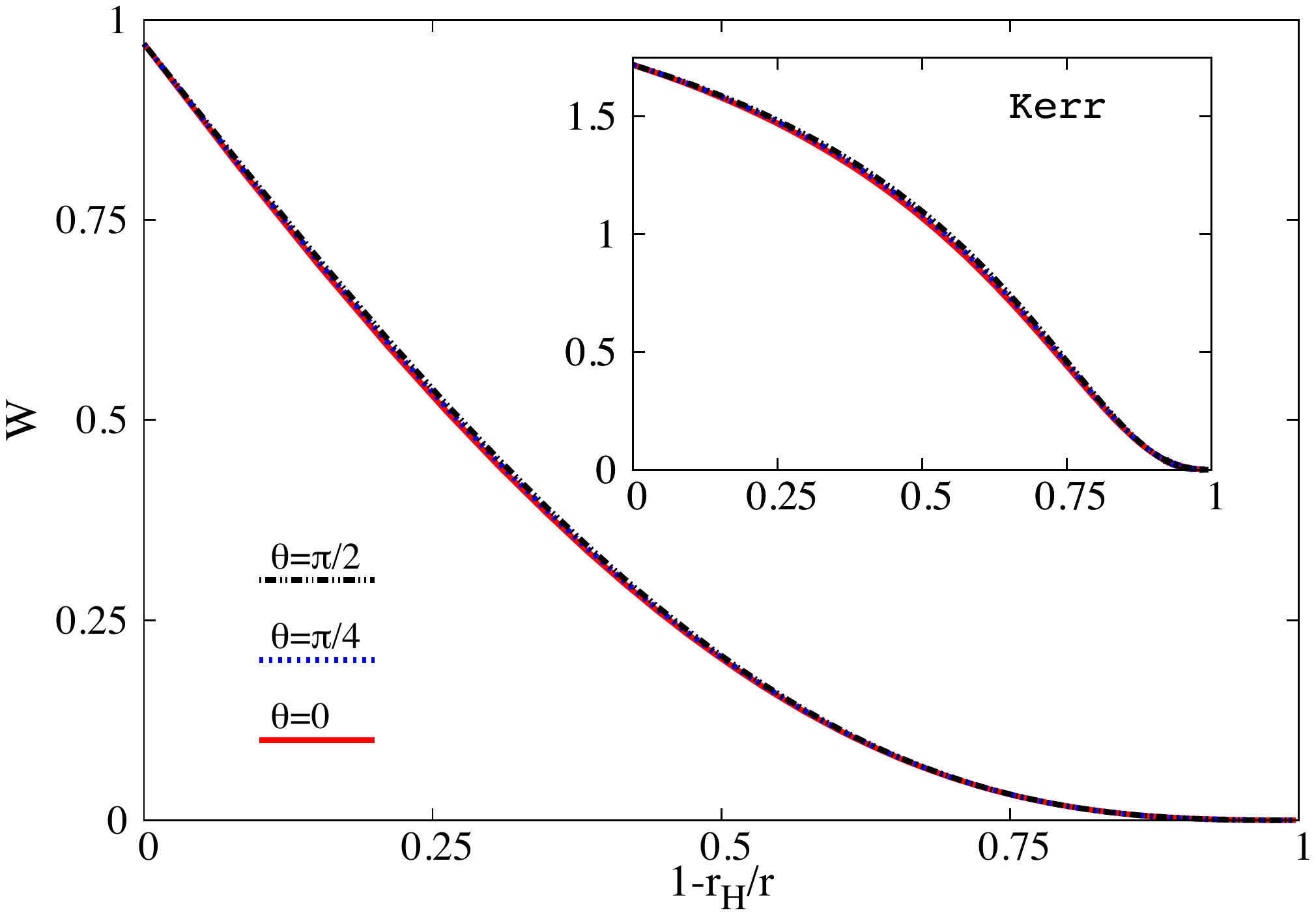}
\includegraphics[width=.485\linewidth]{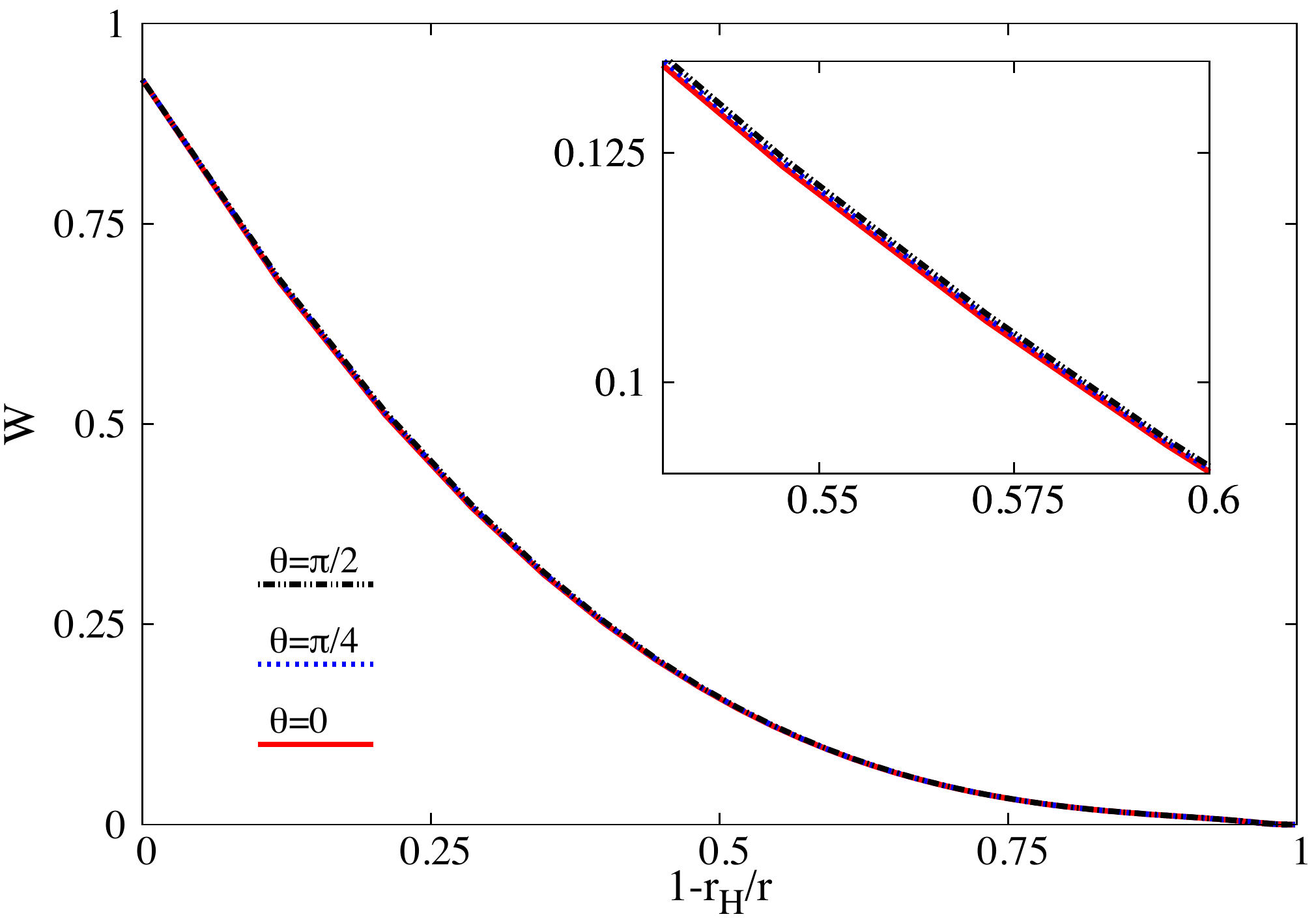}
\caption{Metric functions of HBH$_{\bf 1}$ (left panel) and HBH$_{\bf 2}$ (right panel).}
\label{fig9}
\end{figure}

In \autoref{fig9} we exhibit the metric functions outside the horizon for the two hairy BHs. In the case of HBH$_{\bf 1}$, the insets show a \textit{comparable Kerr BH}, that is with the same total mass $M$ and angular momentum $J$. In order to make the latter comparison, the Kerr metric is expressed in the gauge~\eqref{mansatz} -- see Appendix A in~\cite{Herdeiro:2015gia}. No comparable Kerr BH exists for HBH$_{\bf 2}$, as the latter violates the Kerr bound. Again we use a compactified radial coordinate.  The metric functions of HBH$_{\bf 1}$ already show some qualitative differences relatively to those of the comparable Kerr BH. The latter has a considerably higher $\Omega_\text{H}/\mu\simeq 1.72$; indeed HBH$_{\bf 1}$ and the comparable Kerr roughly correspond to the two points in Fig. 5 of~\cite{Herdeiro:2017phl}, representing the longest migration. On the other hand, the metric functions of HBH$_{\bf 2}$ are a hybrid between the Kerr metric functions and those of a Proca star -- see \autoref{fig6} (left panels). Indeed, as can be seen from the physical parameters in Eq.~\eqref{hbh2},  HBH$_{\bf 2}$ has over three quarters of the total mass and almost the totality of the angular momentum stored in the Proca field outside the horizon. Thus, it is more accurately described as a spinning Proca star with a BH horizon at its centre, than as a BH horizon surrounded by a Proca cloud. The latter is an appropriate description for HBH$_{\bf 1}$.

\begin{figure}
\centering
\includegraphics[width=.485\linewidth]{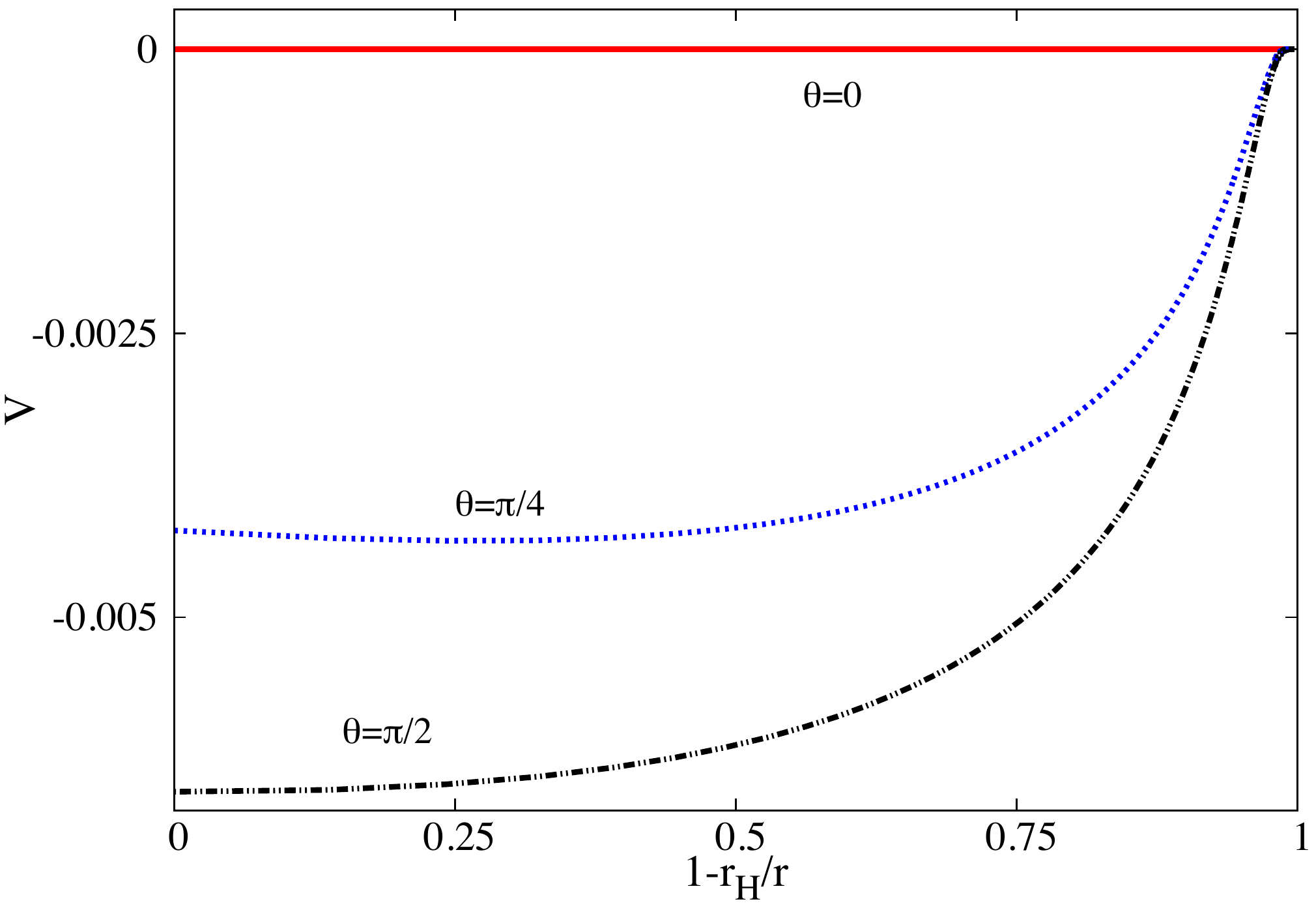}
\includegraphics[width=.485\linewidth]{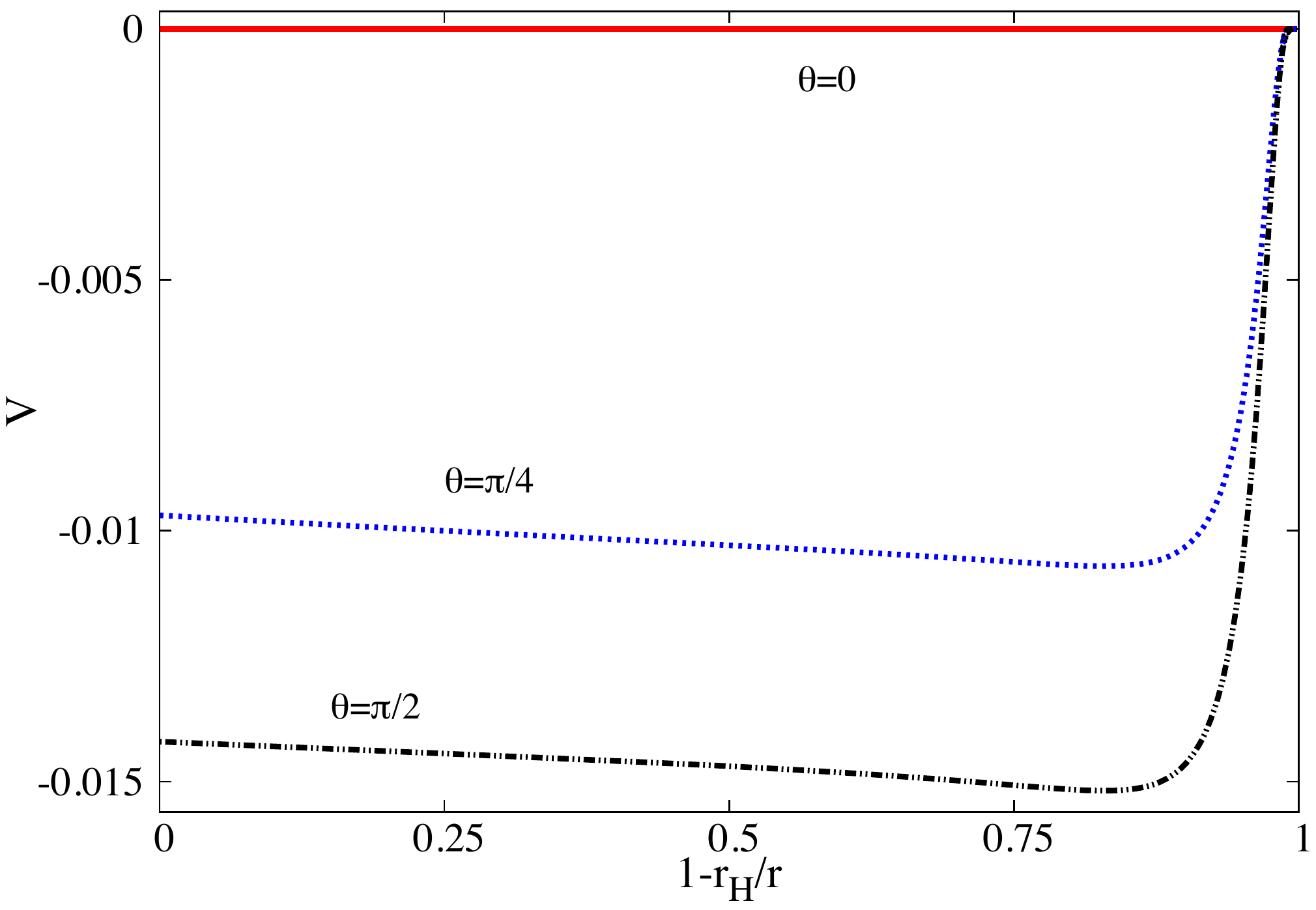}
\includegraphics[width=.485\linewidth]{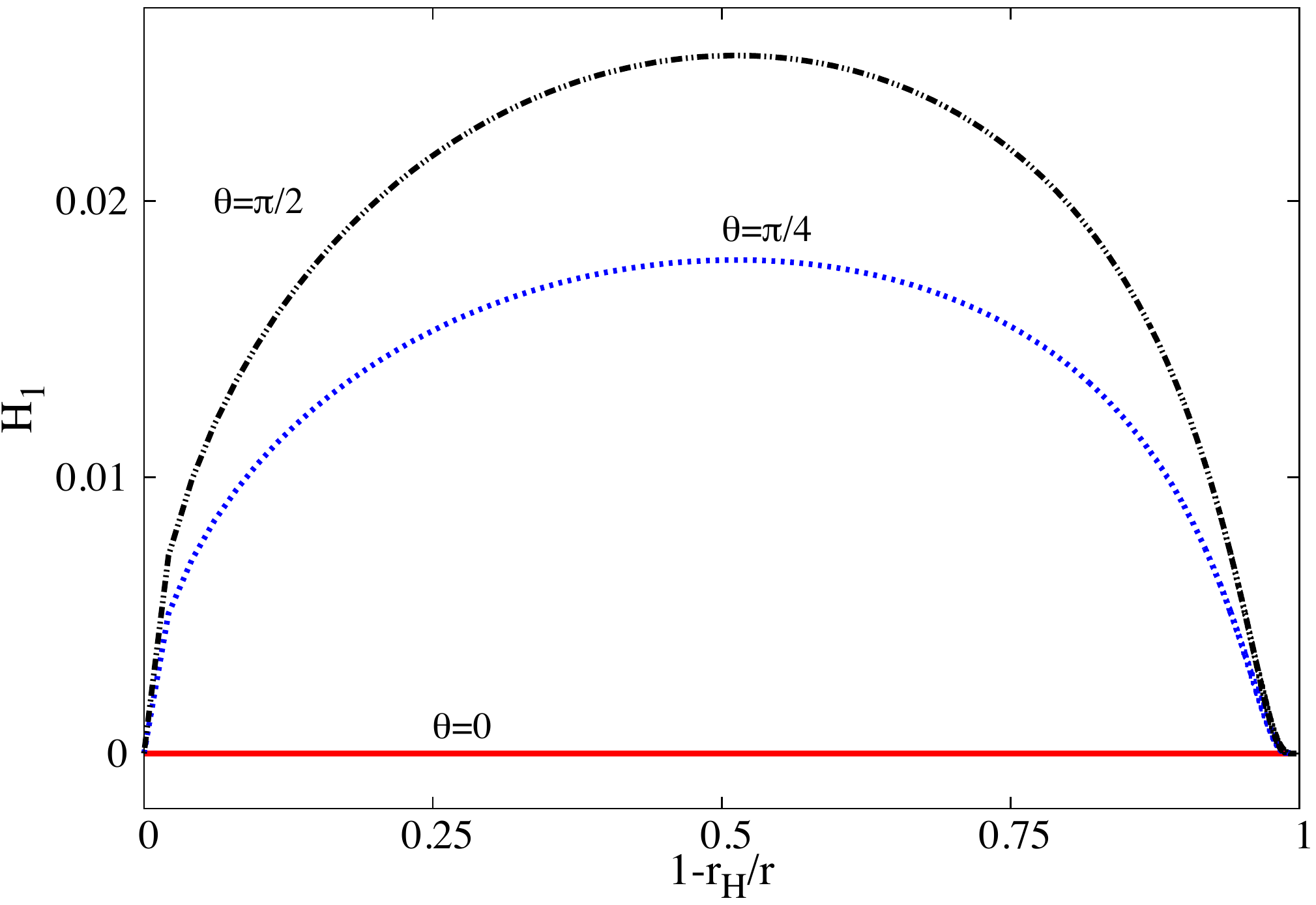}
\includegraphics[width=.485\linewidth]{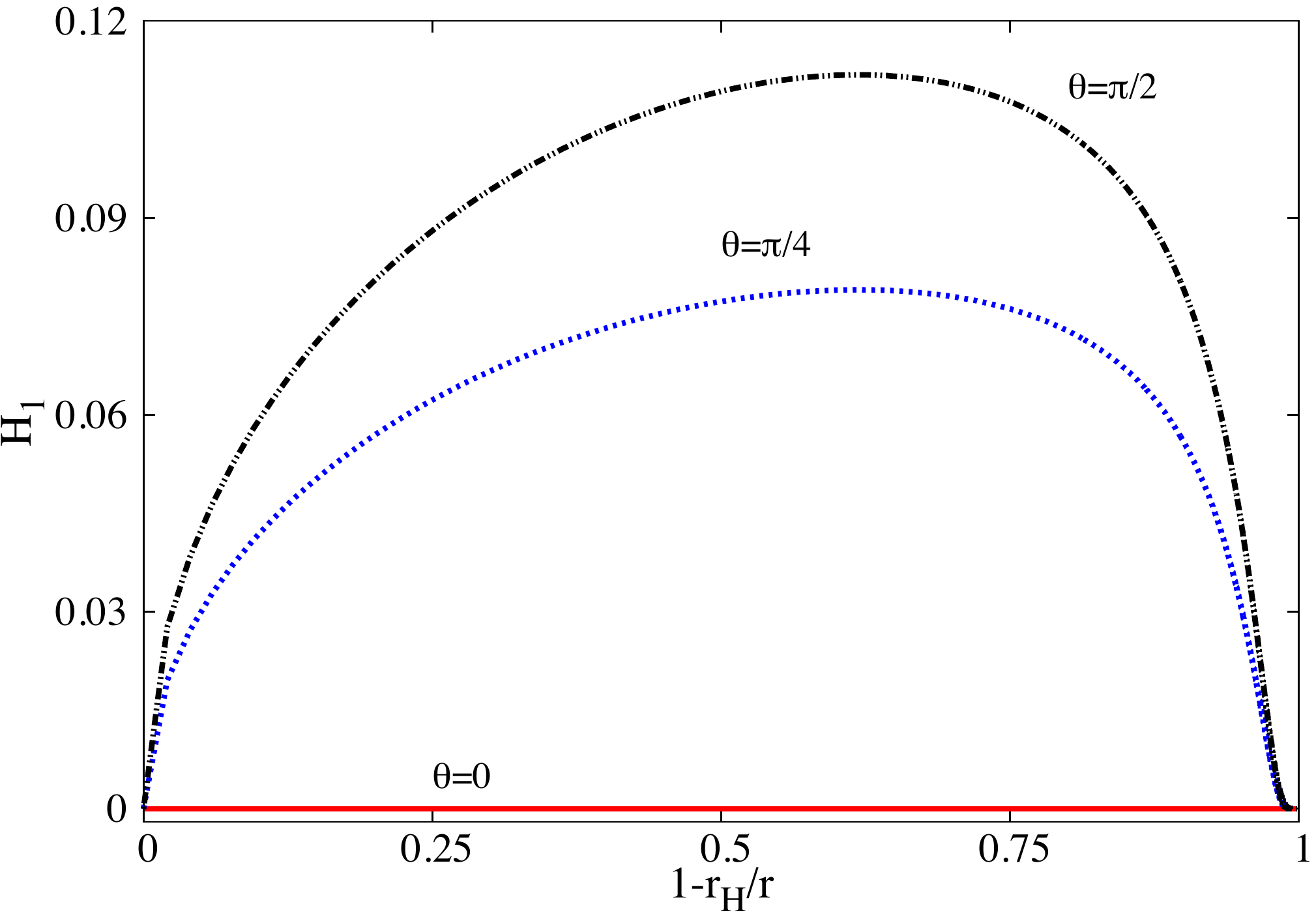}
\includegraphics[width=.485\linewidth]{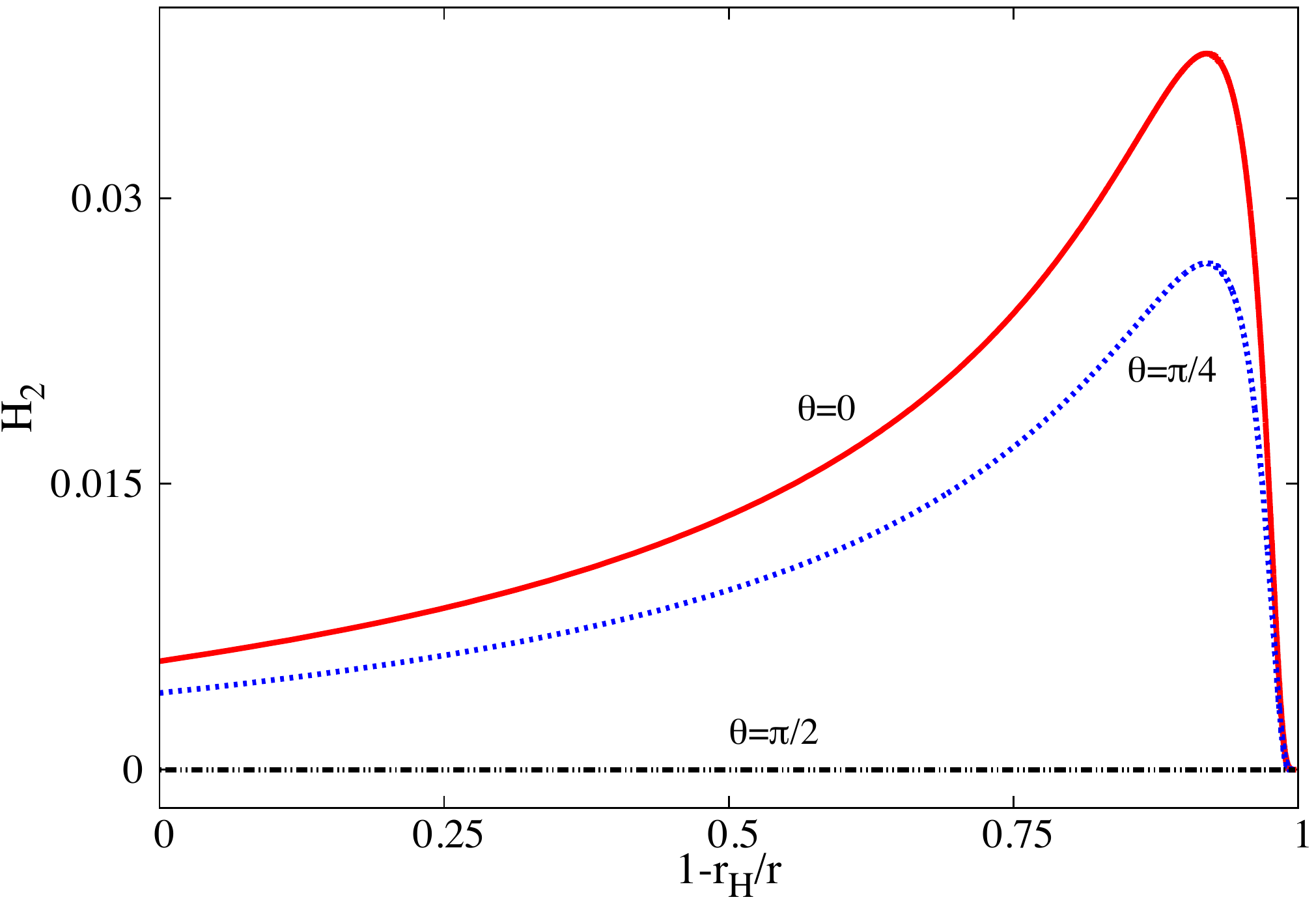}
\includegraphics[width=.485\linewidth]{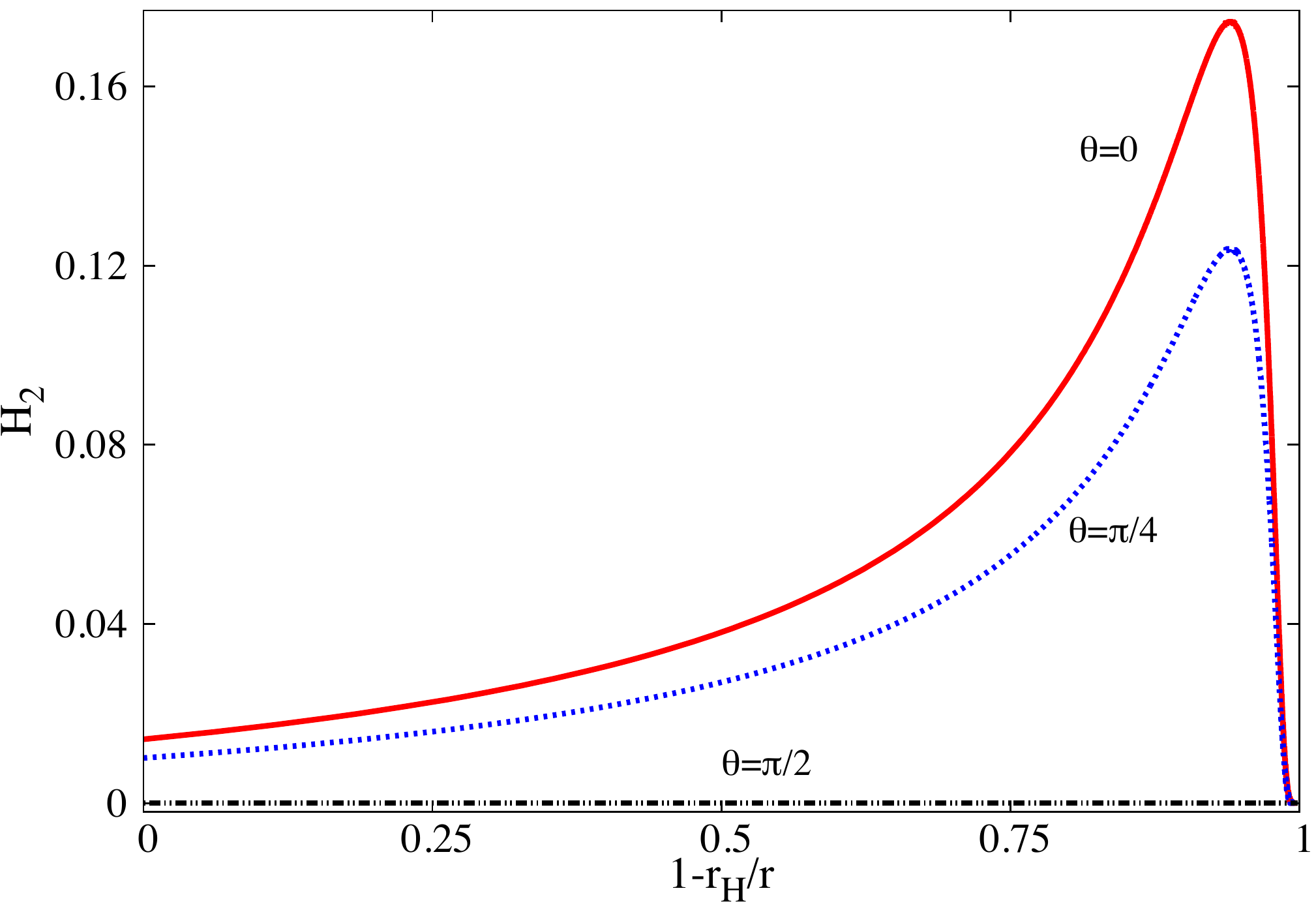}
\includegraphics[width=.485\linewidth]{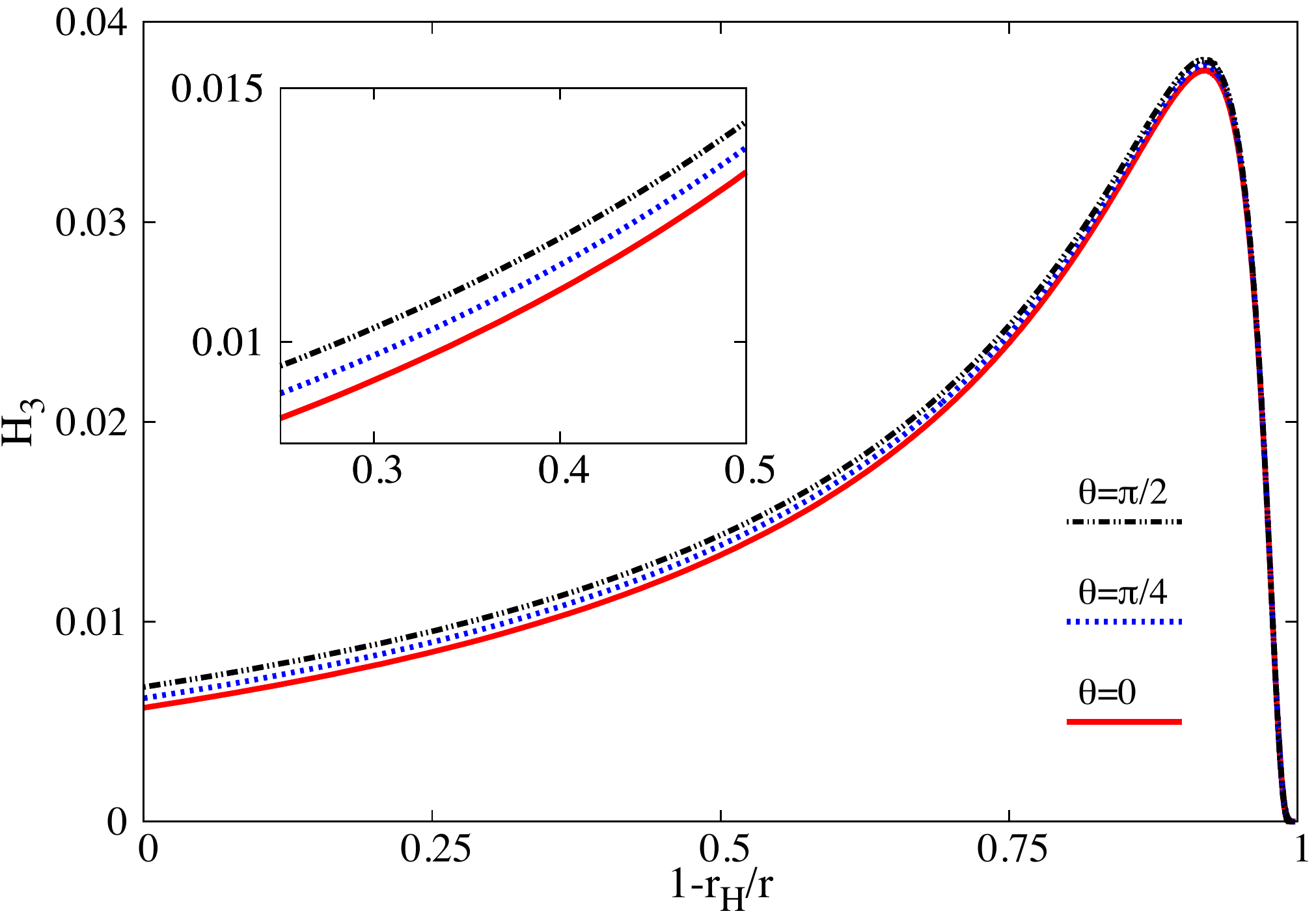}
\includegraphics[width=.485\linewidth]{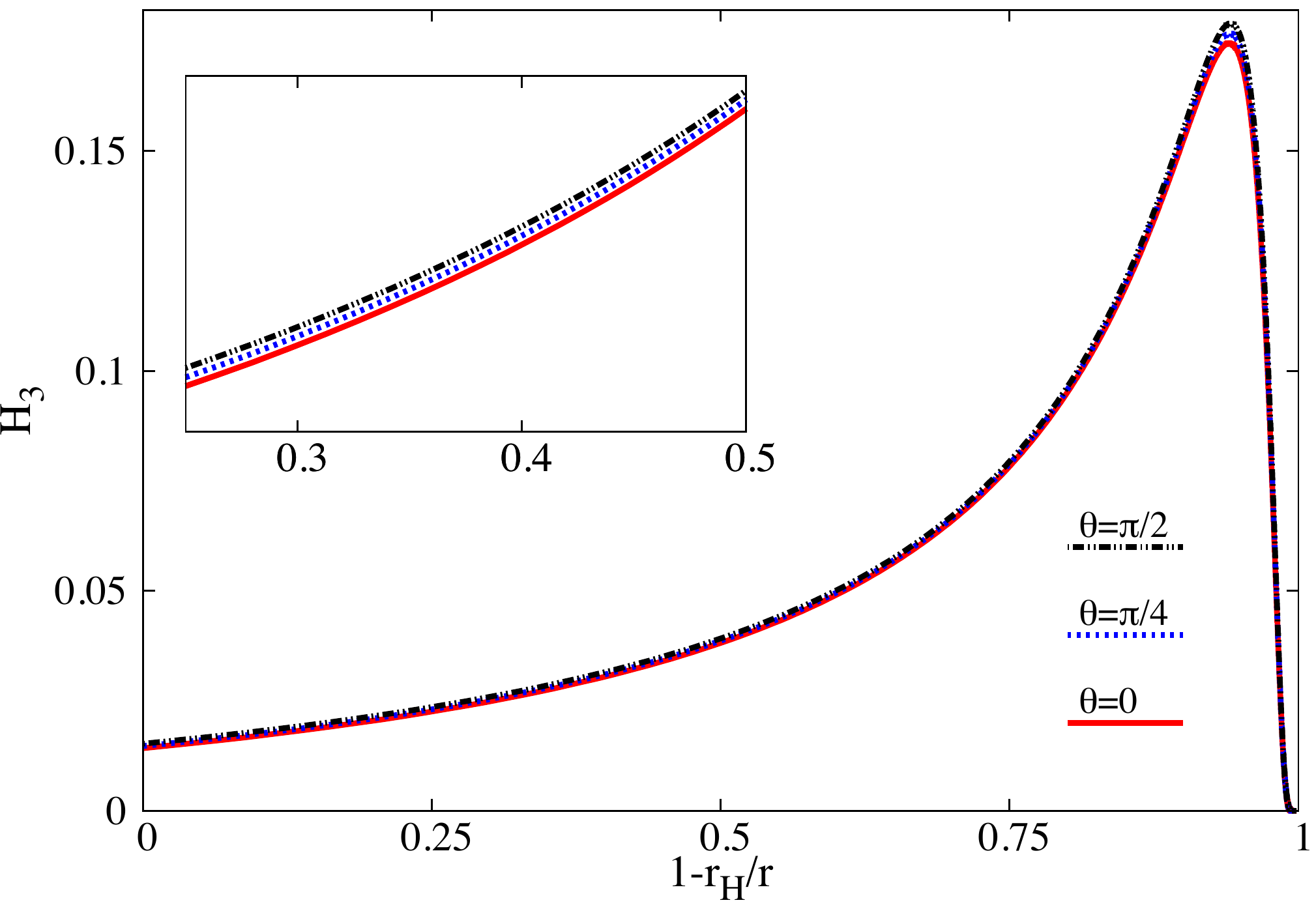}
\caption{Proca potential functions of HBH$_{\bf 1}$ (left panel) and HBH$_{\bf 2}$ (right panel).}
\label{fig10}
\end{figure}

In \autoref{fig10} the Proca potentials are shown for the two hairy BHs. One can appreciate the difference in boundary conditions as compared to the Proca stars in \autoref{fig7}. $V, H_2, H_3$ are non-zero on the BH horizon and zero at the origin, for stars; $H_1$ is the opposite. On the other hand, the most apparent differences between HBH$_{\bf 1}$ and HBH$_{\bf 2}$ are the larger magnitude of the Proca potential functions for the latter, together with a steeper behaviour. This is intuitive from the fact the second BH has a much larger fraction of its energy in the Proca field.

\begin{figure}
\centering
\includegraphics[width=.485\linewidth]{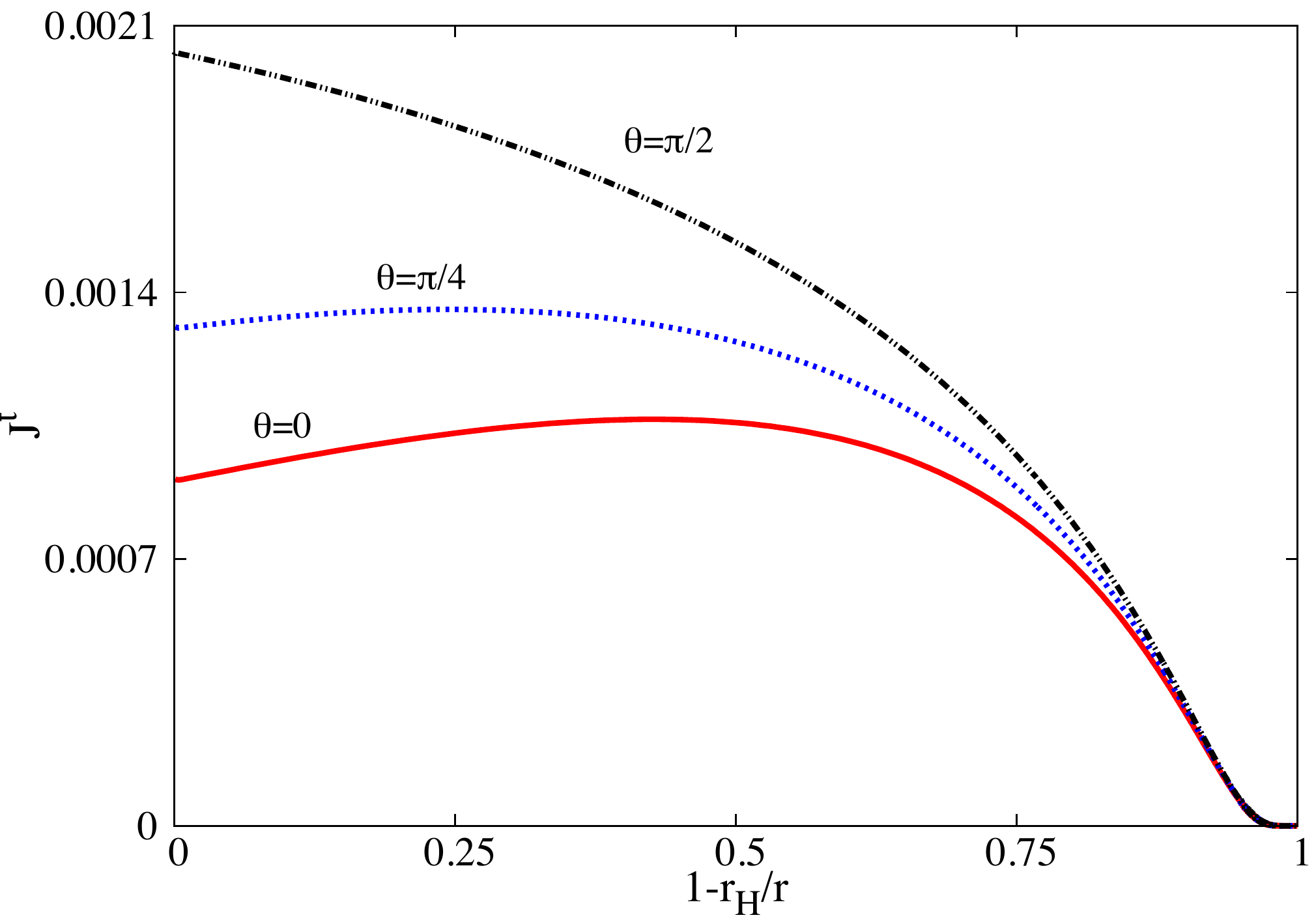}
\includegraphics[width=.485\linewidth]{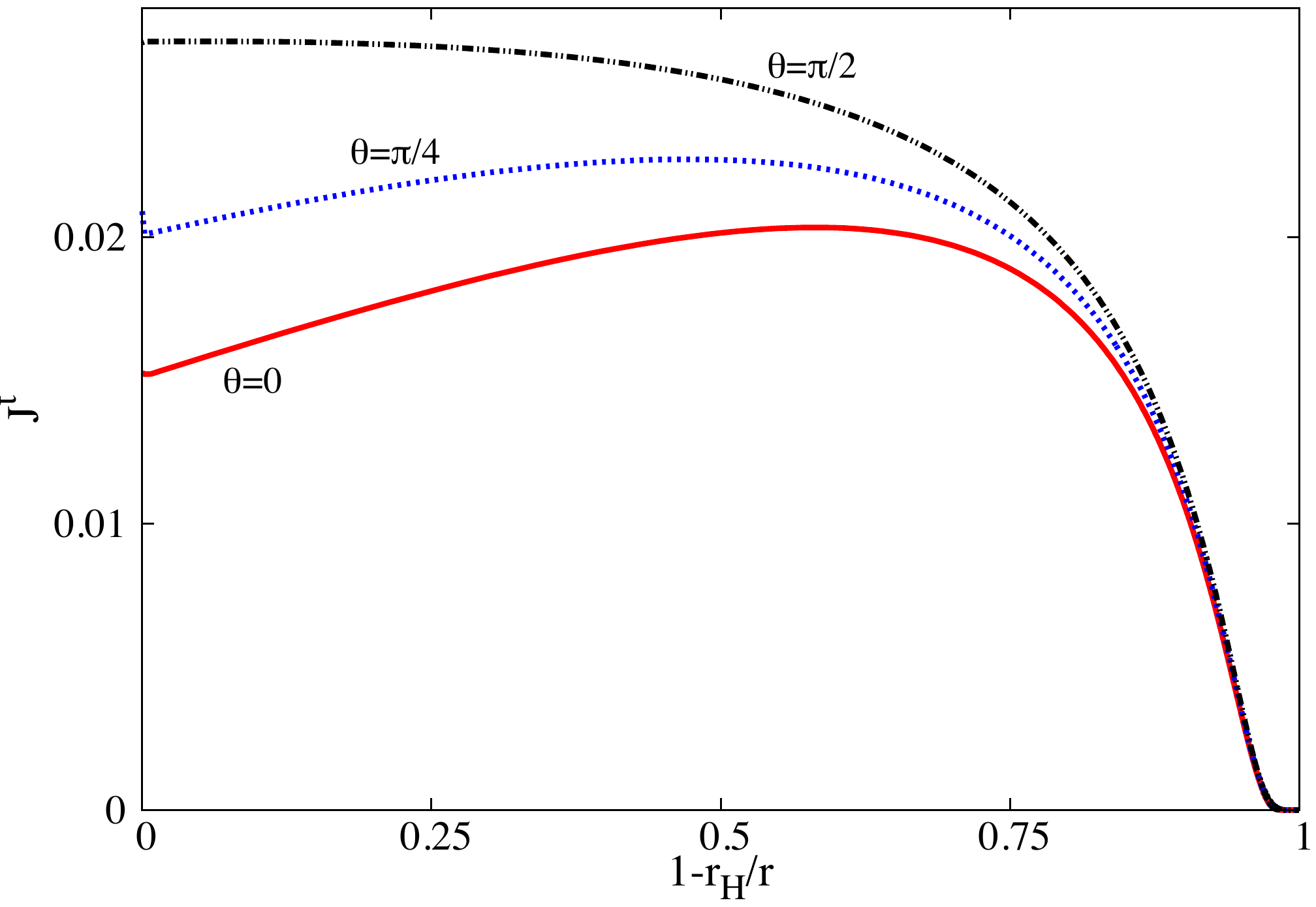}
\includegraphics[width=.484\linewidth]{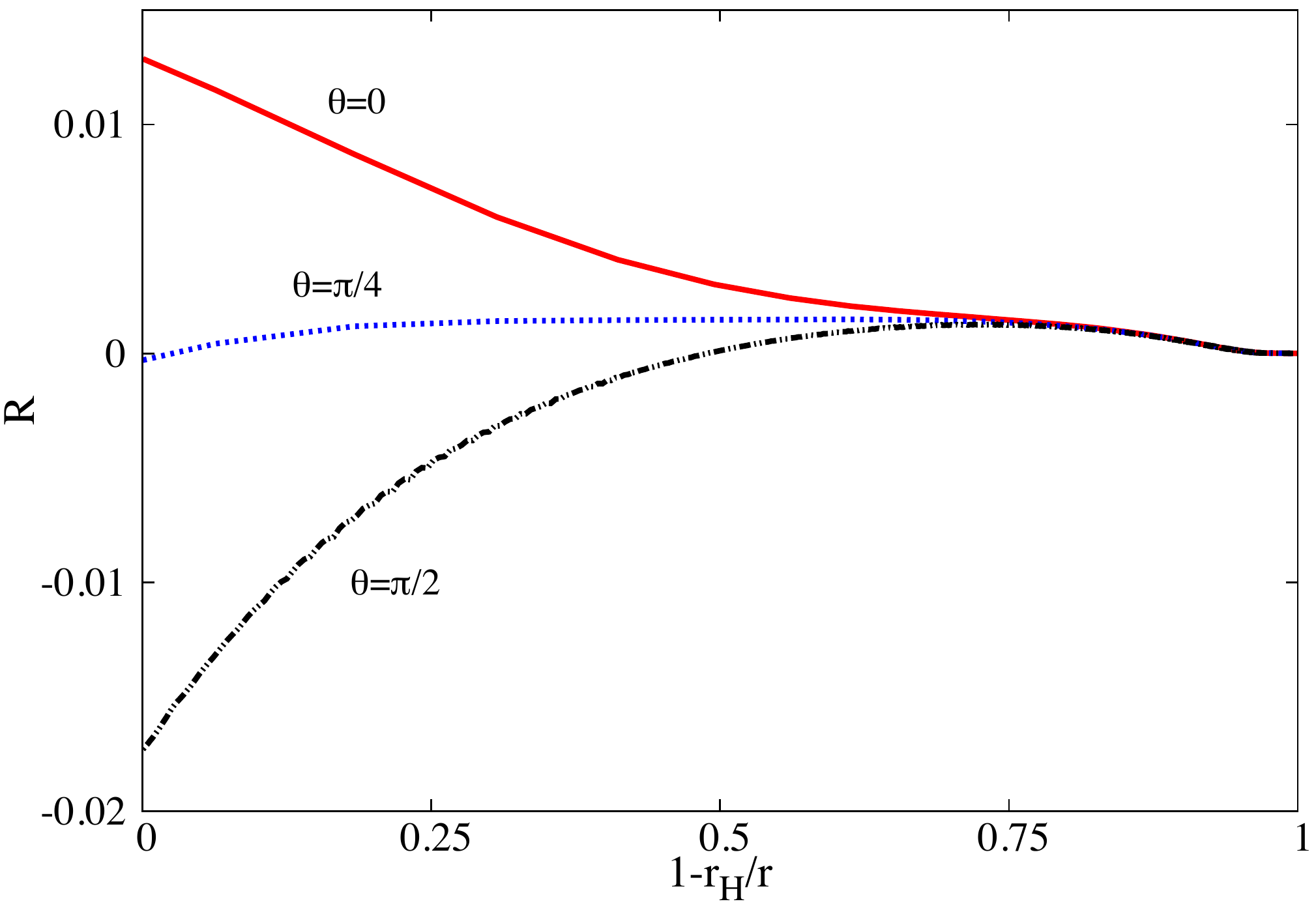}
\includegraphics[width=.485\linewidth]{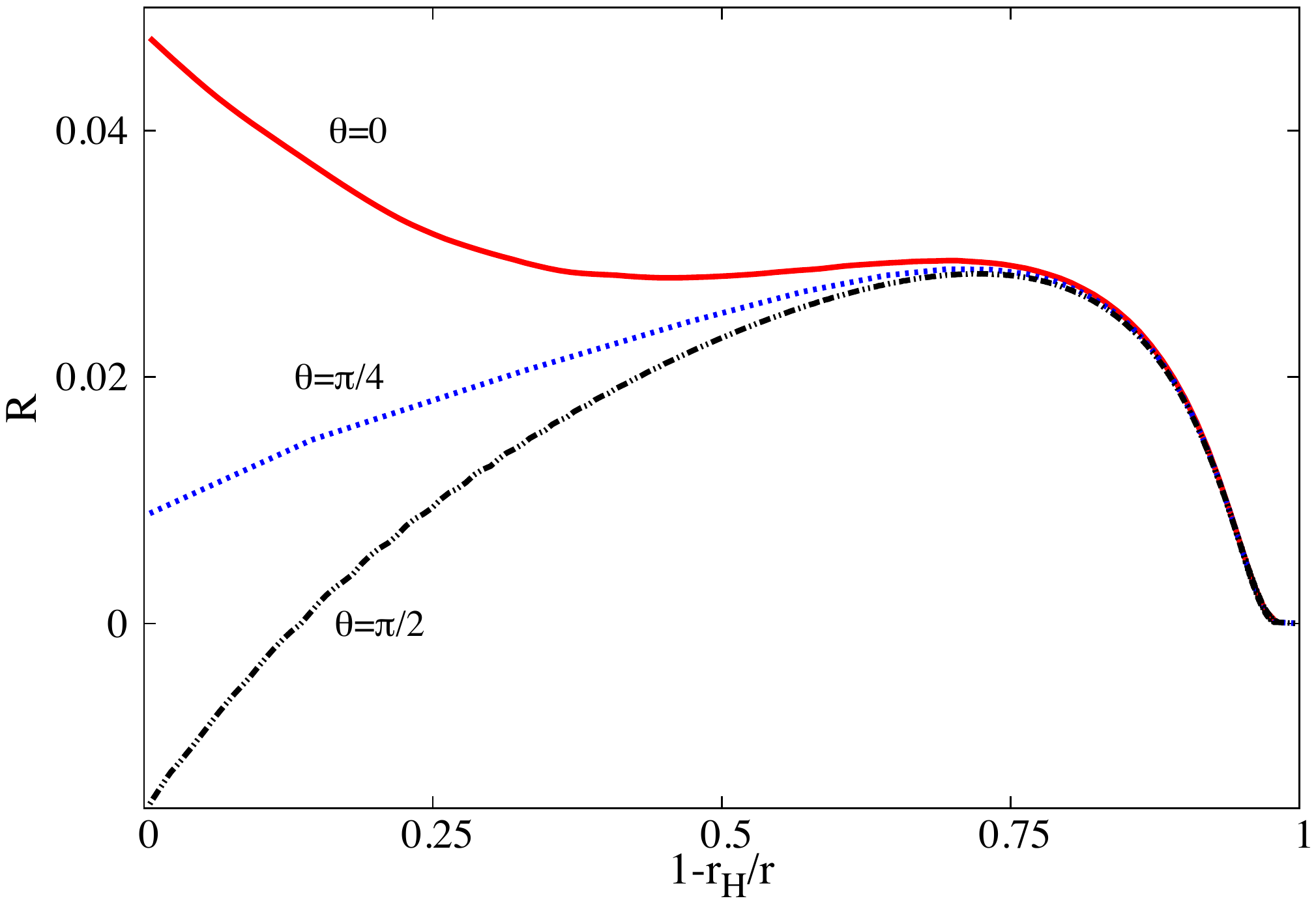}
\includegraphics[width=.485\linewidth]{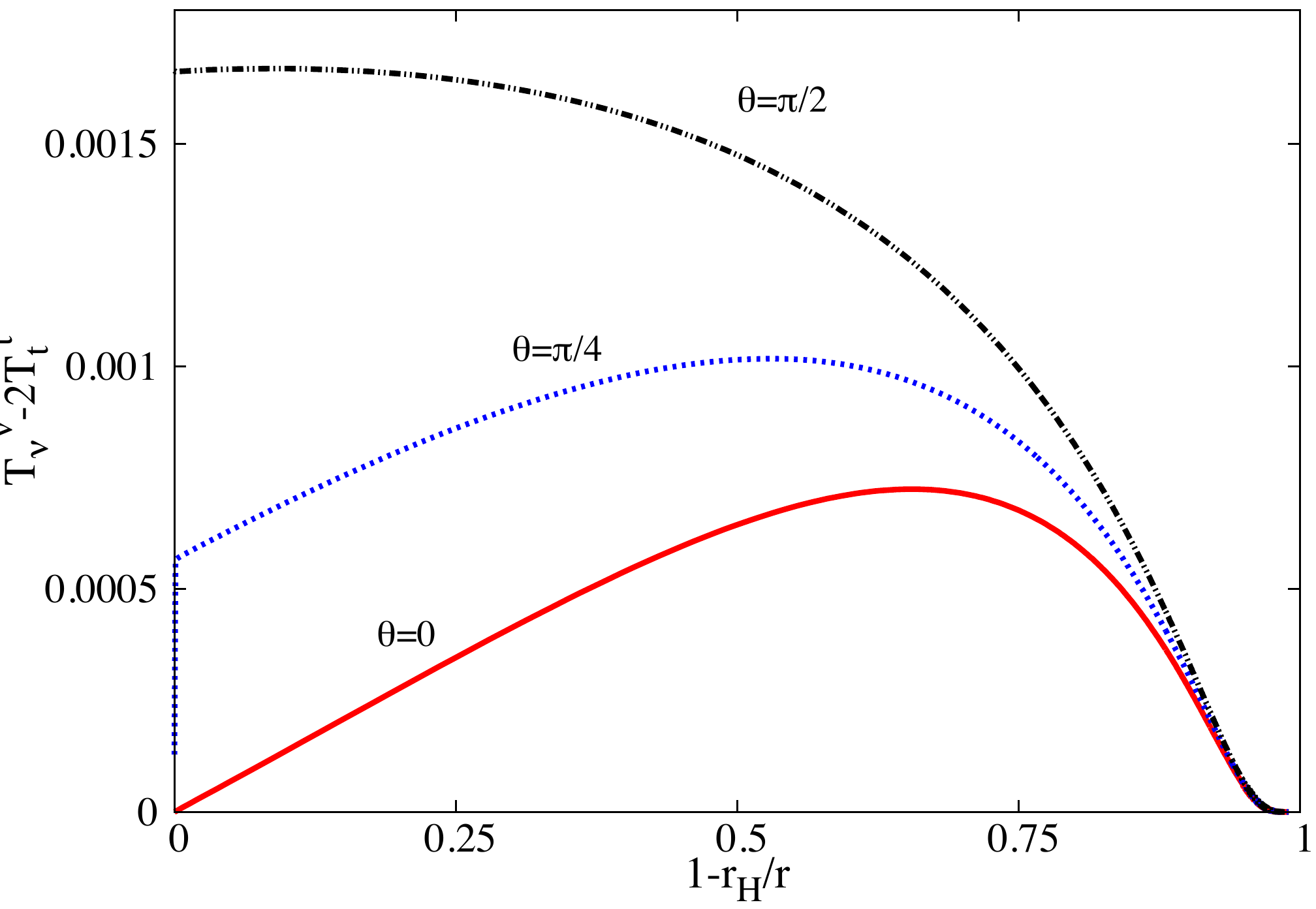}
\includegraphics[width=.485\linewidth]{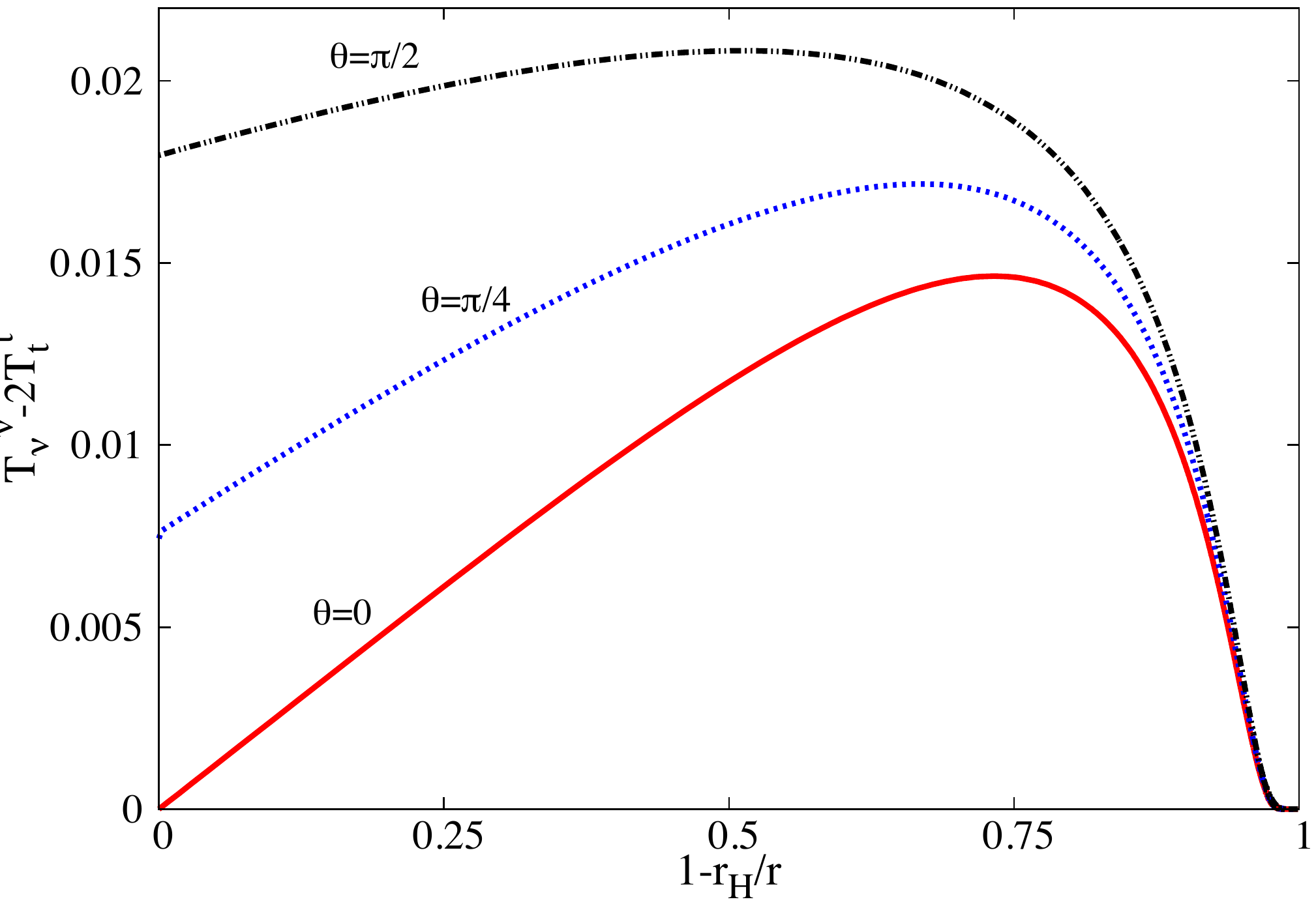}
\includegraphics[width=.485\linewidth]{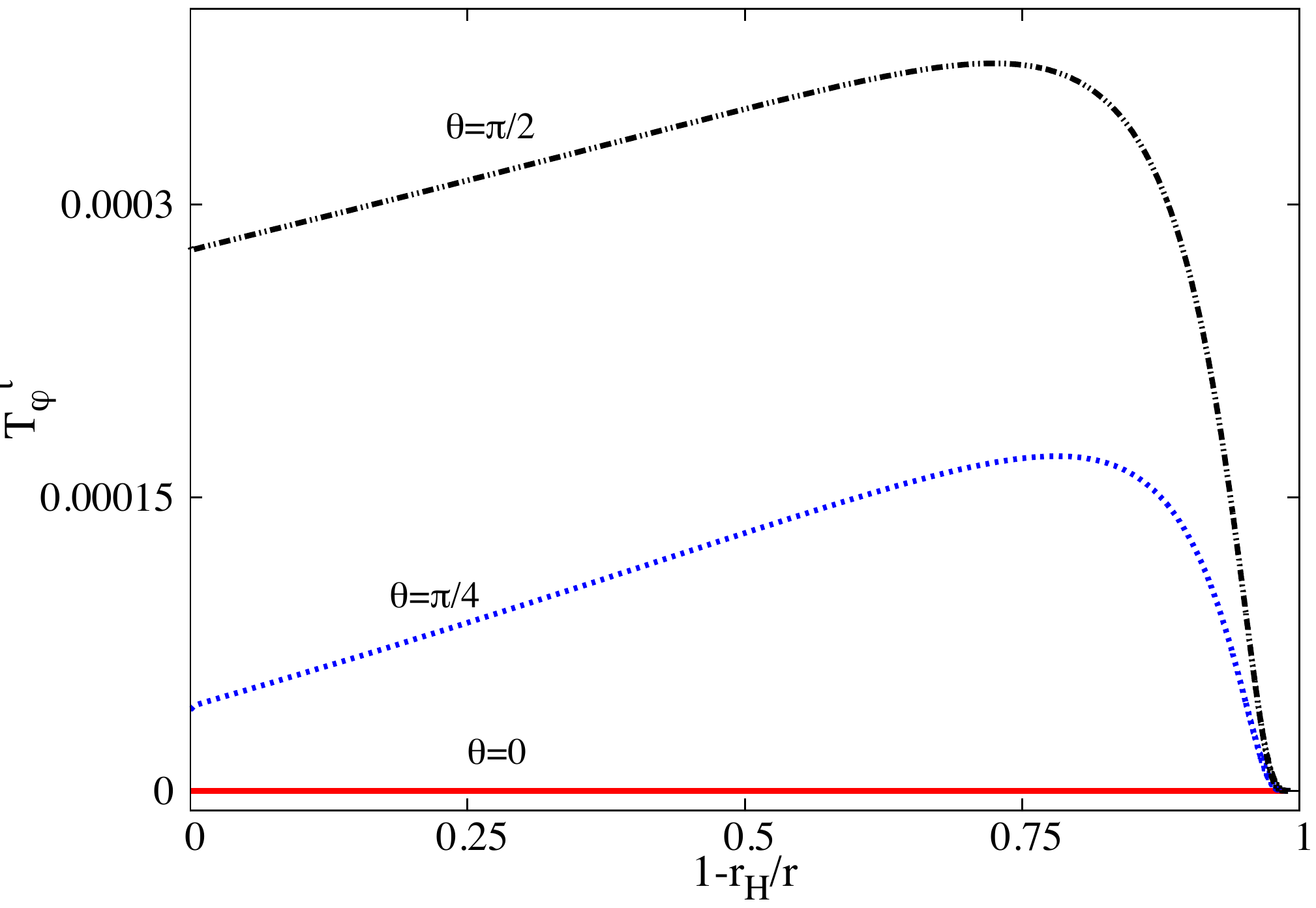}
\includegraphics[width=.485\linewidth]{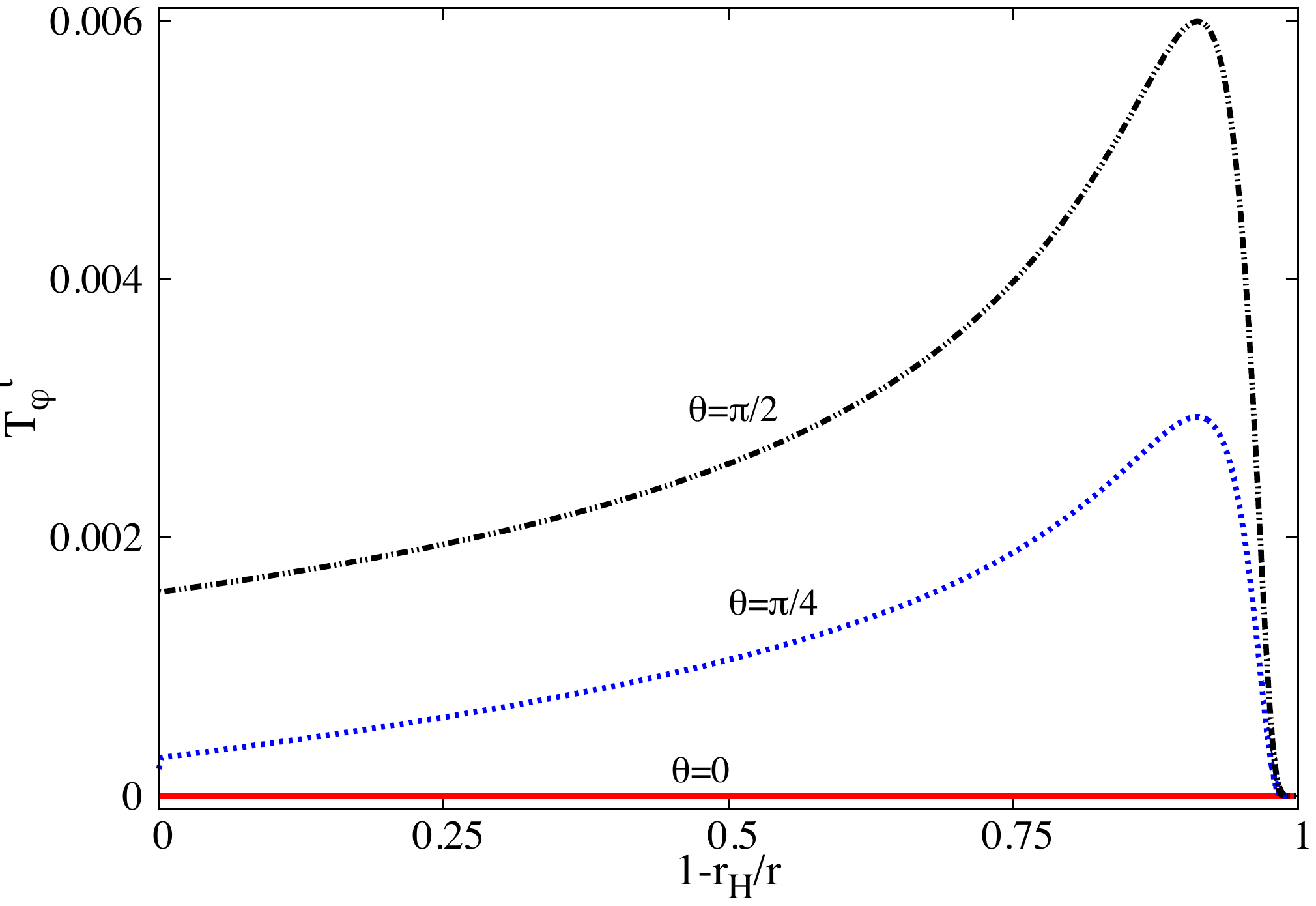}
\caption{Some physical quantities for HBH$_{\bf 1}$ (left panel) and HBH$_{\bf 2}$ (right panel).}
\label{fig11}
\end{figure}

Finally, in \autoref{fig11} we represent some physical quantities of the two hairy BH solutions. The Noether charge density is one order of magnitude larger for HBH$_{\bf 2}$ and with a steeper profile. This impacts on the Ricci scalar curvature, known to manifest the spacetime deformation due to matter, which has a clear lump outside the horizon for the hairiest solution. The Komar energy and angular momentum densities are also larger in magnitude and with sharper profiles, becoming asymptotically more similar to those of the Proca star exhibited in \autoref{fig8} (left panels).

%%%%%%%%%%%%%%%%%%%%%%%%%%%  
\section{Conclusion\label{sec:Conclusion}}
%%%%%%%%%%%%%%%%%%%%%%%%%%%

In this paper we have analysed linear vector clouds of a massive Proca field around a Kerr BH and the BHs with synchronised Proca hair that can be considered as the non-linear realisations of these clouds. Our analysis has been inspired by a series of fairly recent developments that motivates revisiting the Einstein-(complex)-Proca model and its BH and solitonic solutions. Notice, however, that the linear analysis in \autoref{sec2} does not depend on the fact that the Proca field is complex, unlike the analysis in \autoref{sec3}, where the existence of the stationary solutions describing hairy BHs and Proca stars relies on the field being complex.

Concerning the linear analysis of~\autoref{sec2},  the key physical property of these bound-state configurations is the synchronization of their phase angular velocity with the event horizon angular velocity. Furthermore, they resemble the hydrogen's atomic orbitals and can be described in terms of $\{n,\ell,j,m_j\}$. The quantum numbers label the existence lines of stationary vector clouds in the two-dimensional parameter space of Kerr BHs. These curves mark the bifurcation of the Kerr family towards the new family of BHs with Proca hair and constitute one of the boundaries of the domain of existence of the latter~\cite{Herdeiro:2016tmi}.

As for massive scalar bosons~\cite{Benone:2014ssa}, the analysis of the vector clouds shows that, for a fixed value of the azimuthal total angular momentum $m_j$, the cloud's energy, which is proportional to its phase angular velocity, is mainly determined by the node number $n$, and the orbital angular momentum $\ell$. The bound-state $\ket{0,0,1,1}$ has the lowest possible energy and higher values of $n$ and/or $\ell$ correspond to higher-energy states. Thus, the existence line of this bound state is wherein the fundamental states of the hairy BHs bifurcate from. Moreover, despite not having a relevant impact on the cloud energy, the total angular momentum $j$ allows for the existence of $\ell=0$ bound states, which is rooted in the non-vanishing intrinsic angular momentum of the bosons. 

The existence lines obtained numerically were compared with analytical approximations recently reported in the literature~\cite{Baumann:2019eav}. In general, the agreement is excellent for all values of the Kerr BH's rotation parameter, except when $j=m_j$ and $\ell<j$. In this case, a discrepancy arises for near-extremal Kerr BHs;  the reason behind this observation remains to be clarified. 

The analysis' starting point was the FKKS ansatz~\cite{Frolov:2018ezx} for the separation of the Proca equation. This ansatz prevents the need for approximations or time-consuming numerical algorithms when studying massive vector bosons in Kerr-NUT-(A)dS spacetimes and has already been used to address quasi-bound states in the Kerr and Kerr-Newman backgrounds~\cite{Frolov:2018ezx,Dolan:2018dqv,Baumann:2019eav,Siemonsen:2019ebd,Cayuso:2019ieu}. This ansatz, however, does decouple and separate the torsion-modified Proca equation (known as Troca equation) in the Chong-Cveti\v{c}-L\"{u}-Pope spacetime of $D=5$ minimal gauged supergravity~\cite{Cayuso:2019ieu} and in the Kerr-Sen spacetime of low-energy heterotic string theory~\cite{Cayuso:2019ieu}. Note that All these works focused on the dynamics of massive vector bosons in the frequency domain, in which the particles are described as monochromatic waves. Future research should then dive into a yet-to-be-explored time-domain analysis of these simplified equations of motion. Of particular interest would be to perform long-time evolutions of massive vector Gaussian wave packets in superradiance-prone spacetimes.  

Concerning the non-linear analysis in \autoref{sec3}, here we have exhibited the domain of existence of the fundamental states of BHs with synchronised Proca hair and compared some of their properties with the first excited states, discussed in~\cite{Herdeiro:2016tmi}, and the cousin hairy BHs obtained in the scalar case~\cite{Herdeiro:2014goa,Herdeiro:2015gia}. Then, we have analysed some illustrative solutions of both the solitonic limit (Proca stars) and hairy BHs. We emphasise that all the solutions considered here have azimuthal harmonic index $m~(\equiv m_j)=1$. Higher $m$ solutions also exist, corresponding to another sort of excitation.\footnote{Higher $m$ increases the number of nodes in the azimuthal direction.  An in-depth study of the higher $m$ solutions in the scalar case is found in~\cite{Delgado:2019prc}. Some particular higher $m$ solutions in the Proca case can be found in~\cite{Sanchis-Gual:2019ljs}. We remark that $n=0$
spinning Proca stars with $m>1$ possess  surfaces of  constant energy density with a toroidal
morphology.}  There are two main ideas to retain from our results. 

\begin{figure}[h!]
\centering
\includegraphics[width=1\linewidth]{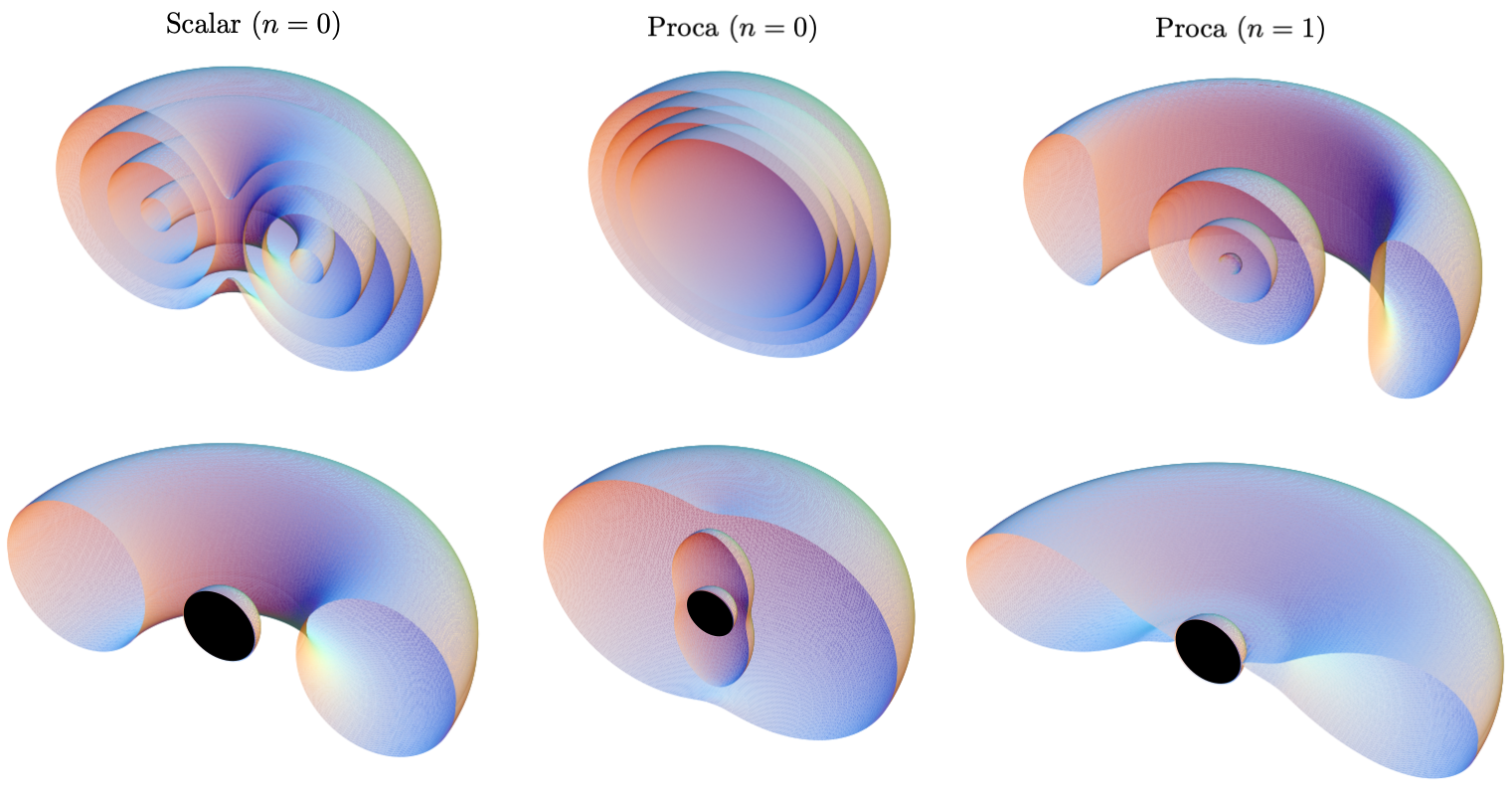}
\caption{Morphology of the surfaces of constant energy density of scalar ($n=0$) and vector/Proca ($n=0$ and $n=1$) spinning bosonic stars (top panels) and BHs with synchronised hair of the corresponding type (bottom panels). For the BHs,  the black disk represents the horizon. All these solutions have $m=1$. The hairy BHs bifurcate from clouds with $\ell=0$. Besides further radial excitations (with higher $n$), these families of solutions possess azimuthal excitations (with higher $m$) and also further polar-angular excitations (corresponding, in the linear limit, to higher $\ell$).}
\label{fig12}
\end{figure}

Firstly, there are morphological differences between the cases compared herein, which may have various implications. This is summarised in \autoref{fig12}, where surfaces of constant scalar or Proca density for stars and hairy BHs are exhibited for the three families of solutions we have compared. Spinning scalar bosonic stars ($n=0$) are toroidal; spinning vector stars with $n=0$ are spheroidal and with $n=1$ have a Saturn-like morphology. The corresponding hairy BHs are a non-linear bound state of such a star with a horizon, deforming the star's morphology.

Secondly, in all these families of hairy BHs, in particular in the $n=0$ BHs with Proca hair, there are Kerr-like solutions, but also
rather non-Kerr-like examples. At the moment, at least one formation channel for the hairy Kerr-like solutions is known -- superradiance;  HBH$_{\bf 1}$ discussed in \autoref{sec33} belongs to this set. The solution shows already some interesting deviations from Kerr and it will be very interesting to analyse how these deviations impact on astrophysical observables. In this respect, one could reconsider some of the analysis done for the scalar case or for the excited BHs with Proca hair, namely of shadows~\cite{Cunha:2015yba,Cunha:2019ikd}, X-ray spectroscopy~\cite{Ni:2016rhz,Zhou:2017glv} or quasi-periodic oscillations~\cite{Franchini:2016yvq}. Of course, one of the most interesting open questions concerns the dynamical properties of these BHs, including quasi-normal modes. The recently established dynamical robustness of spinning Proca stars~\cite{Sanchis-Gual:2019ljs} has paved the way to perform dynamical evolutions of these BHs, from which one could, in particular, extract waveforms for binary evolutions. Work in this direction is underway.

%%%%%%%%%%%%%%%%%%%%%%%%%%%  
\section*{Acknowledgements}
%%%%%%%%%%%%%%%%%%%%%%%%%%%
We would like to thank N. Sanchis-Gual for reading a draft of this paper. The authors thank Conselho Nacional de Desenvolvimento Cient\'ifico e Tecnol\'ogico (CNPq) and Coordena\c{c}\~ao de Aperfei\c{c}oamento de Pessoal de N\'{\i}vel Superior (Capes) - Finance Code 001, in Brazil, for partial financial support.
This work is supported  by the Center for Astrophysics and Gravitation (CENTRA) and by the Center for Research and Development in Mathematics and Applications (CIDMA) through the Portuguese Foundation for Science and Technology (FCT - Funda\c{c}\~ao para a Ci\^encia e a Tecnologia), references UIDB/04106/2020, UIDB/00099/2020 and UIDP/04106/2020. We acknowledge support  from the projects PTDC/FIS-OUT/28407/2017 and CERN/FIS-PAR/0027/2019 and from national funds (OE), through FCT, I.P., in the scope of the framework contract foreseen in the numbers 4, 5 and 6 of the article 23, of the Decree-Law 57/2016, of August 29,
changed by Law 57/2017, of July 19.   This work has further been supported by  the  European  Union's  Horizon  2020  research  and  innovation  (RISE) programme H2020-MSCA-RISE-2017 Grant No.~FunFiCO-777740. The authors would like to acknowledge networking support by the COST Action CA16104.

\bibliography{references}

\appendix
%%%%%%%%%%%%%%%%%%%%%%%%%%%
\section{Vector spherical harmonics}
\label{ap1}
\begin{figure}[t]
\centering
   \subfigure{\includegraphics[scale=0.4]{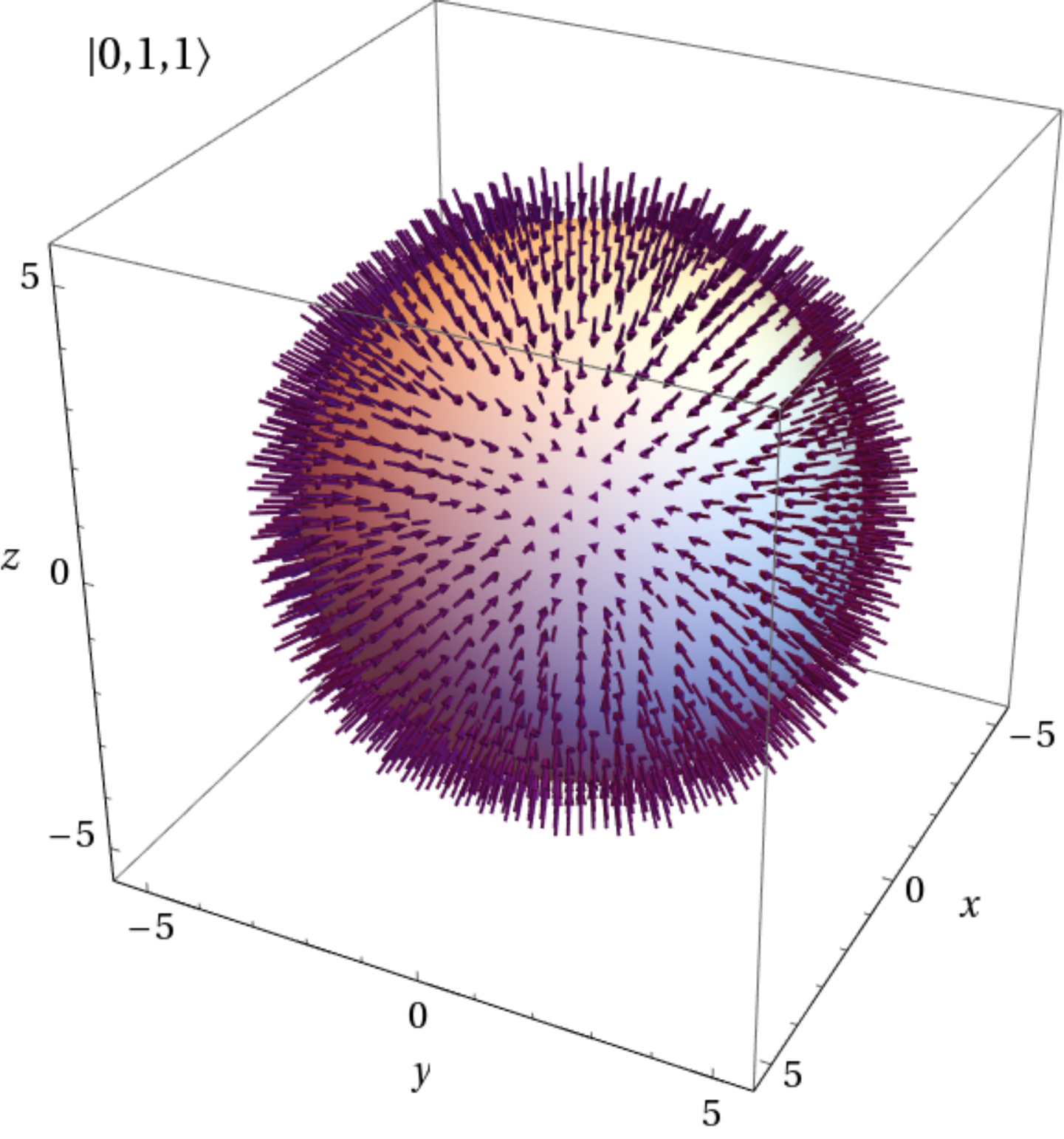}}
   \quad\quad\quad
   \subfigure{\includegraphics[scale=0.4]{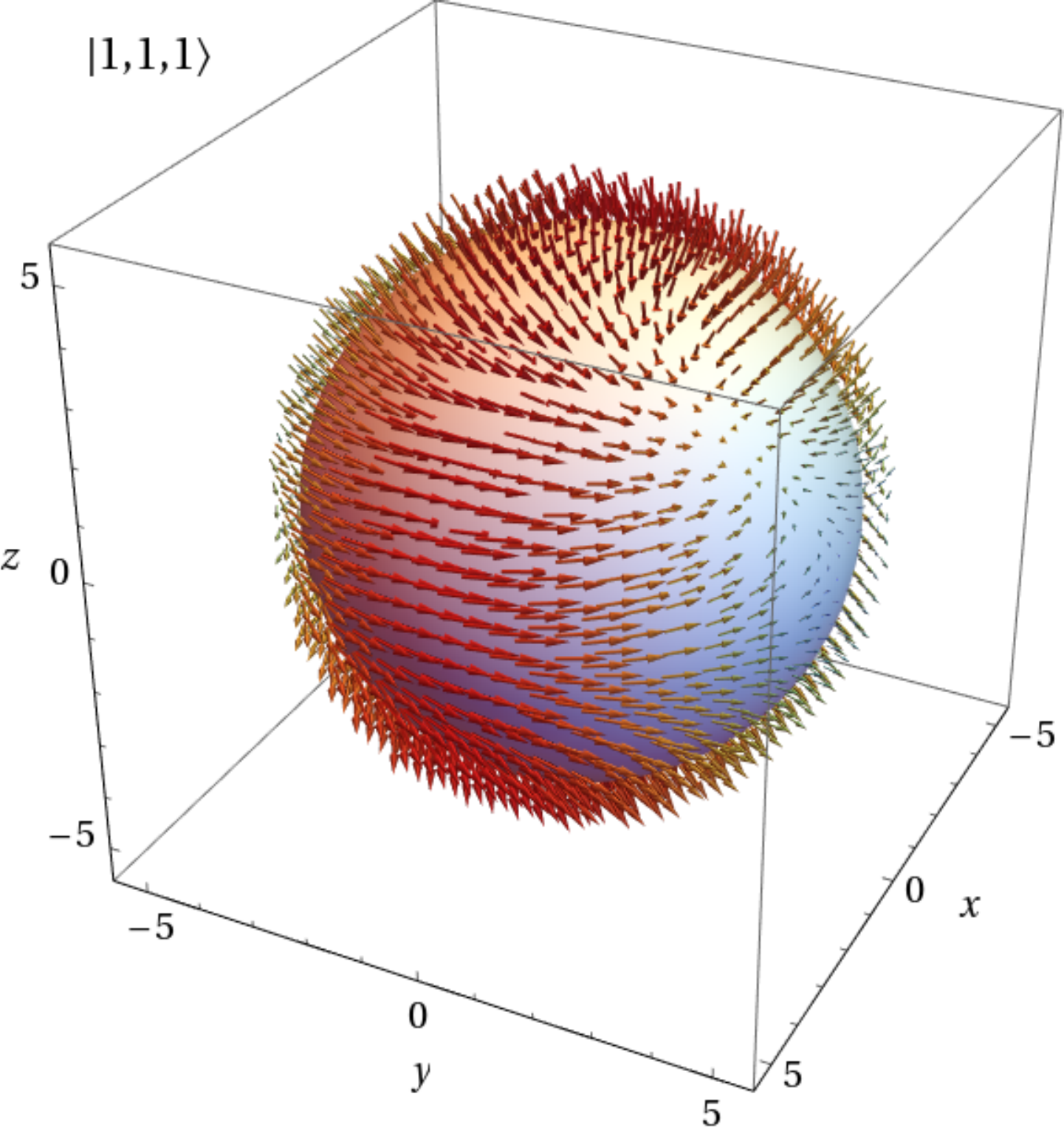}}\\
      \subfigure{\includegraphics[scale=0.4]{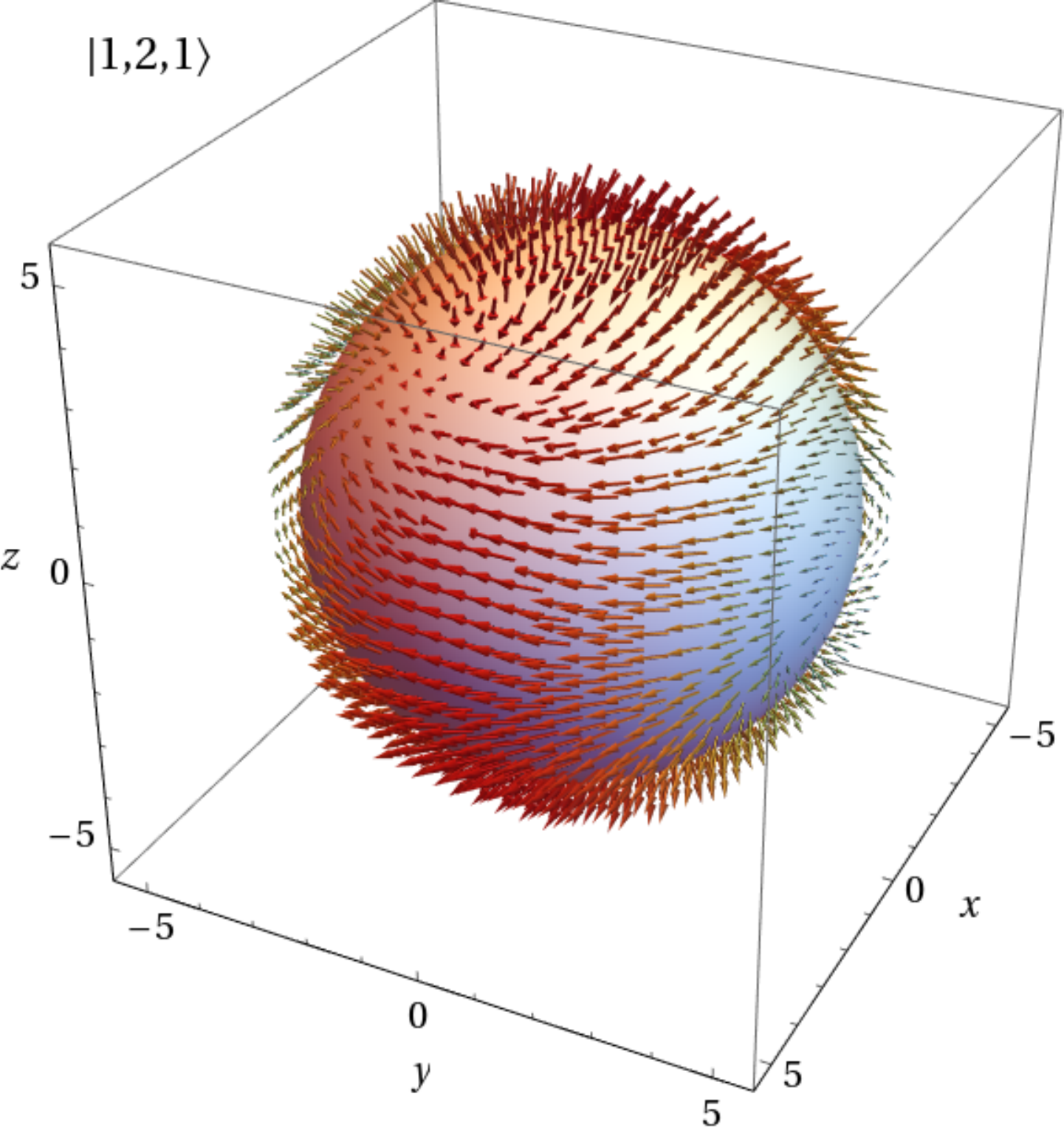}}
   \quad\quad\quad
   \subfigure{\includegraphics[scale=0.4]{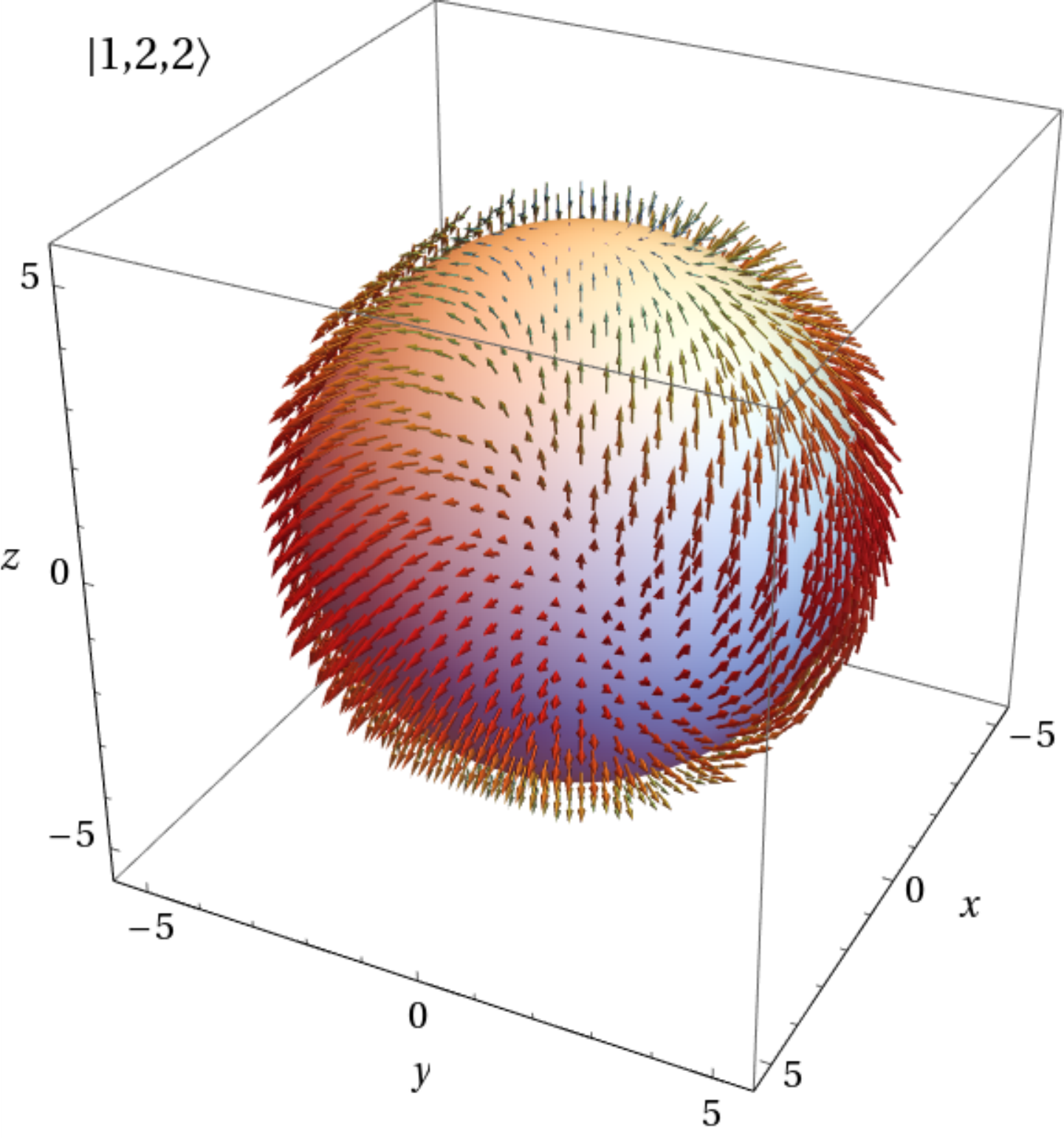}}\\
     \subfigure{\includegraphics[scale=0.4]{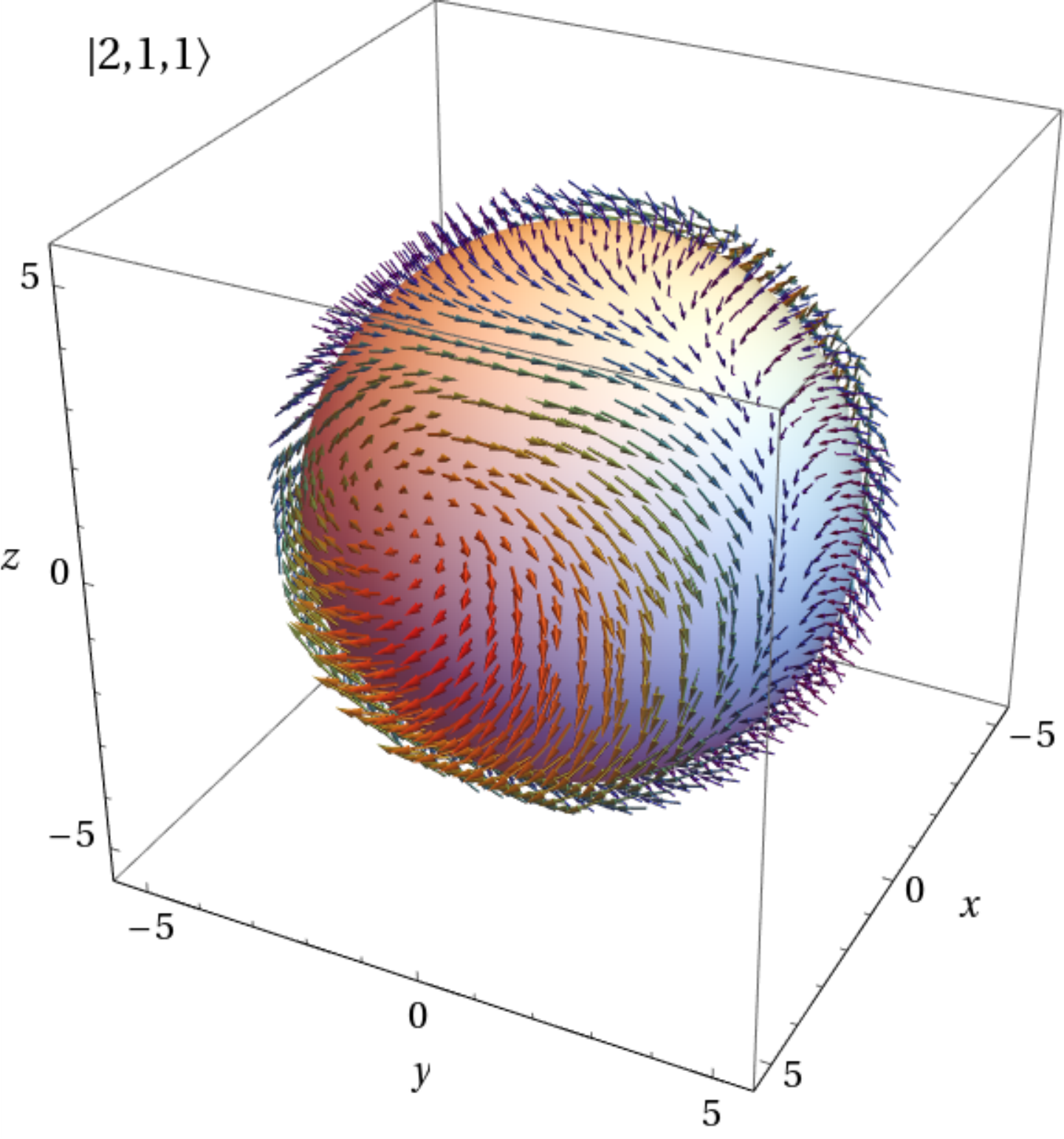}}
   \quad\quad\quad
   \subfigure{\includegraphics[scale=0.4]{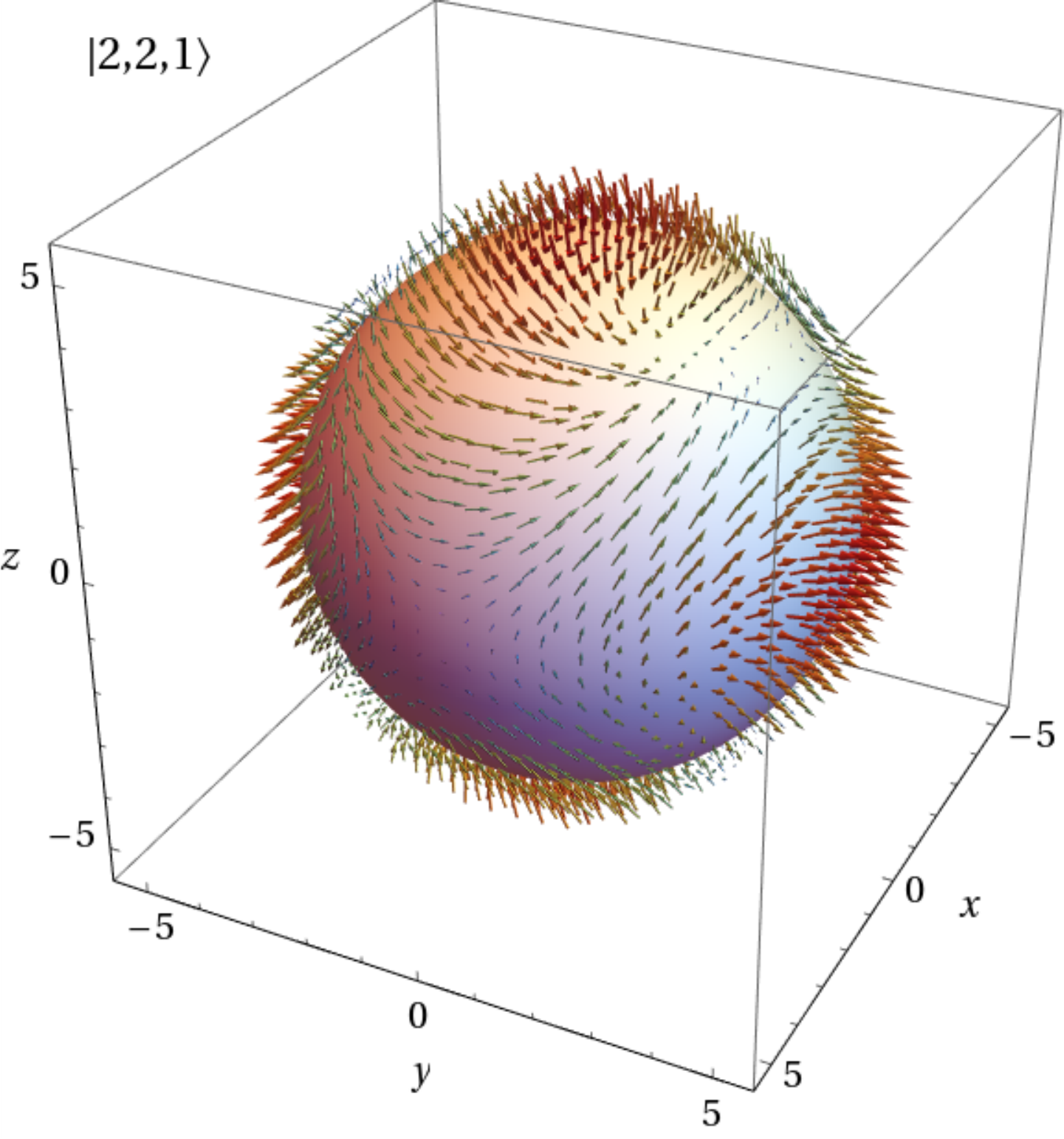}}\\
	\caption{Real part of the first few (`pure-orbital') vector spherical harmonics $\boldsymbol{Y}^\ell_{j,m_j}$ in the three-dimensional real coordinate space ($x=r\sin\theta\cos\varphi$, $y=r\sin\theta\sin\varphi$, $z=r\cos\theta$).
} 
\label{fig13}
\end{figure}

%%%%%%%%%%%%%%%%%%%%%%%%%%%
The components of the first few (`pure-orbital') vector spherical harmonics $\boldsymbol{Y}^\ell_{j,m_j}$ in terms of the spherical unit  vectors  $\{\hat{\boldsymbol{e}}_{(r)},\hat{\boldsymbol{e}}_{(\theta)},\hat{\boldsymbol{e}}_{(\varphi)}\}$ are
\begin{align*}
\boldsymbol{Y}^{0}_{1,1}&=-\frac{1}{2\sqrt{2\pi}}
\begin{pmatrix}
    1\\
    i\\
    0
\end{pmatrix},
&&\boldsymbol{Y}^{1}_{1,1}=-\frac{1}{4}\sqrt{\frac{\pi}{3}}
\begin{pmatrix}
    \cos\theta\\
    i\cos\theta\\
    -e^{i\varphi}\sin\theta
\end{pmatrix},\\
\boldsymbol{Y}^{1}_{2,1}&=-\frac{1}{4}\sqrt{\frac{3}{\pi}}
\begin{pmatrix}
    \cos\theta\\
    i\cos\theta\\
    e^{i\varphi}\sin\theta
\end{pmatrix},
&&\boldsymbol{Y}^{1}_{2,2}=\frac{1}{4}\sqrt{\frac{3}{\pi}}e^{i\varphi}\sin{\theta}
\begin{pmatrix}
    1\\
    i\\
    0
\end{pmatrix},\\
\boldsymbol{Y}^{2}_{1,1}&=\frac{1}{8\sqrt{\pi}}
\begin{pmatrix}
    3e^{2i\varphi}\sin^2\theta-3\cos^2\theta+1\\
    i(-3e^{2i\varphi}\sin^2\theta-3\cos^2\theta+1)\\
    6e^{i\varphi}\sin\theta\cos\theta
\end{pmatrix},
&&\boldsymbol{Y}^{2}_{2,1}=-\frac{1}{8}\sqrt{\frac{5}{\pi}}
\begin{pmatrix}
    e^{2i\varphi}\sin^2\theta+3\cos^2\theta-1\\
    i(-e^{2i\varphi}\sin^2\theta+3\cos^2\theta-1)\\
    -2e^{i\varphi}\sin\theta\cos\theta
\end{pmatrix},\\
\boldsymbol{Y}^{2}_{2,2}&=\frac{1}{4}\sqrt{\frac{5}{\pi}}e^{i\varphi}\sin\theta
\begin{pmatrix}
    \cos\theta\\
    i\cos\theta\\
    -e^{i\varphi}\sin\theta
\end{pmatrix},
&&\boldsymbol{Y}^{2}_{3,1}=\frac{1}{8\sqrt{\pi}}
\begin{pmatrix}
    e^{2i\varphi}\sin^2\theta-6\cos^2\theta+2\\
    i(-e^{2i\varphi}\sin^2\theta-6\cos^2\theta+2)\\
    -8e^{i\varphi}\sin\theta\cos\theta
\end{pmatrix},\\
\boldsymbol{Y}^{2}_{3,2}&=-\frac{1}{2}\sqrt{\frac{5}{2\pi}}e^{i\varphi}\sin\theta
\begin{pmatrix}
    \cos\theta\\
    i\cos\theta\\
    e^{i\varphi}\sin\theta
\end{pmatrix},
&&\boldsymbol{Y}^{2}_{3,3}=-\frac{1}{8}\sqrt{\frac{15}{\pi}}e^{2i\varphi}\sin^2\theta
\begin{pmatrix}
    1\\
    i\\
    0
\end{pmatrix}.
\end{align*}
Their real part are displayed in \autoref{fig13}.

\end{document}